\documentclass[preprint2]{aastex6}

\newcommand{\Vmac}{V_{\rm mac}}
\newcommand{\eps}[1]{\log\varepsilon_{\rm #1}}
\newcommand{\kms}{km\,s$^{-1}$}
\newcommand{\iso}[2]{\mbox{$^{#1}{\rm #2}$}}
\newcommand{\kH}{$S_{\rm H}$}
\newcommand{\Teff}{T_{\rm eff}}
\newcommand{\eexc}{E_{\rm exc}}
\newcommand{\vt}{\xi_t}

\newcommand{\alp}{$\alpha$}
\newcommand{\alpfe}{[$\alpha$/Fe]}
\def\gtsim {>\kern-0.9em\lower1.1ex\hbox{$\sim$}~}
\def\ltsim {<\kern-0.9em\lower1.1ex\hbox{$\sim$}~}

\slugcomment{Accepted by ApJ for publication}

\shorttitle{Non-LTE study of F and G dwarfs. II. Abundance patterns from Li to Eu} \shortauthors{Zhao et al.}

\begin{document}

\shorttitle{Non-LTE study of F and G dwarfs. II. Abundance patterns from Li to Eu} \shortauthors{Zhao et al.}

\title{Systematic non-LTE study of the $-2.6 \le$ [Fe/H] $\le 0.2$ F and G dwarfs \\
 in the solar neighbourhood. \\
II. Abundance patterns from Li to Eu \thanks{Based on
    observations collected at the UCO/Lick observatory, USA and Canada-France-Hawaii Telescope}}

\author{G. Zhao\altaffilmark{1,3},
        L. Mashonkina\altaffilmark{2},
        H. L. Yan\altaffilmark{1},
        S. Alexeeva\altaffilmark{2},
        C. Kobayashi\altaffilmark{4},
        Yu. Pakhomov\altaffilmark{2},
        J. R. Shi\altaffilmark{1,3},
        T. Sitnova\altaffilmark{2},
        K. F. Tan\altaffilmark{1},
        H. W. Zhang\altaffilmark{5,6},
        J. B. Zhang\altaffilmark{1,3},
        Z. M. Zhou\altaffilmark{1,3},
        M. Bolte\altaffilmark{7},
        Y. Q. Chen\altaffilmark{1},
        X. Li\altaffilmark{1},
        F. Liu\altaffilmark{1,8},
        M. Zhai\altaffilmark{1},
        }

\affil{\\
\altaffilmark{1}Key Laboratory of Optical Astronomy, National Astronomical Observatories, Chinese Academy of Sciences, Beijing 100012, China}
\email{gzhao@nao.cas.cn}
\affil{\altaffilmark{2}Institute of Astronomy, Russian Academy of Sciences, RU-119017 Moscow, Russia}
\email{lima@inasan.ru}
\affil{\altaffilmark{3}School of Astronomy and Space Science, University of Chinese Academy of Sciences, Beijing 100049, China}
\affil{\altaffilmark{4}School of Physics, Astronomy and Mathematics, Centre for Astrophysics Research, University of Hertfordshire, \\
College Lane, Hatfield AL10 9AB, UK}
\affil{\altaffilmark{5}Department of Astronomy, School of Physics, Peking University, Beijing 100871, China}
\affil{\altaffilmark{6}Kavli Institute for Astronomy and Astrophysics, Peking University, Beijing 100871, China}
\affil{\altaffilmark{7}UCO/Lick Observatory, University of California, 1156 High Street, Santa Cruz, CA 95064, USA}
\affil{\altaffilmark{8}Research School of Astronomy and Astrophysics, Australian National University, Canberra, ACT 2611, Australia}

\begin{abstract}
For the first time, we present an extensive study of stars with individual non-local thermodynamic equilibrium (NLTE) abundances for 17 chemical elements from Li to Eu in a sample of stars uniformly distributed over the $-2.62 \le$ [Fe/H] $\le +0.24$ metallicity range that is suitable for the Galactic chemical evolution research. The star sample has been kinematically selected to trace the Galactic thin and thick disks and halo. We find new and improve earlier results as follows. (i) The element-to-iron ratios for Mg, Si, Ca, and Ti form a MP plateau at a similar height of 0.3~dex, and the knee occurs at common [Fe/H] $\simeq -0.8$. The knee at the same metallicity is observed for [O/Fe], and the MP plateau is formed at [O/Fe] = 0.61. (ii) The upward trend of [C/O] with decreasing metallicity exists at [Fe/H] $< -1.2$, supporting the earlier finding of Akerman et al. (iii) An underabundance of Na relative to Mg in the [Fe/H] $< -1$ stars is nearly constant, with the mean [Na/Mg] $\simeq -0.5$. (iv) The K/Sc, Ca/Sc, and Ti/Sc ratios form well-defined trends, suggesting a common site of the K-Ti production. (v) Sr follows the Fe abundance down to [Fe/H] $\simeq -2.5$, while Zr is enhanced in MP stars. (vi) The comparisons of our results with some widely used Galactic evolution models are given. The use of the NLTE element abundances raises credit to the interpretation of the data in the context of the chemical evolution of the Galaxy.

\end{abstract}

\keywords{Line: formation -- Stars: abundances  -- Stars: atmospheres -- Stars: late-type -- Galaxy: evolution}

\section{Introduction}
The understanding of the formation and evolution of the Galaxy mainly relies on the stellar spectroscopic analysis, which gives the chemical compositions of un-evolved cool stars. Pioneering studies in this field \citep[e.g.][]{1962ApJS....6..407W, 1990A&A...238..242Z, 1990A&AS...86...85Z} have found that \alpfe\footnote{In the classical notation, where [X/H] = $\log(N_{\rm X}/N_{\rm H})_{star} - \log(N_{\rm X}/N_{\rm H})_{Sun}$.} versus [Fe/H] shows a plateau below [Fe/H] $= -1$, and then there is a steady decline to \alpfe\ $\sim 0$ at the solar metallicity, which is actually a composite of the ratios transiting from the halo, to the thick- and thin-disk populations. That is, the ``chemical tagging'' can be used to recover the star formation and evolution history of the Galaxy. However, they are found to span a large range in the overall abundance, of more than 8 orders of magnitude. Stellar abundance trends establish important observational constraints on current models of nucleosynthesis and the chemical evolution of the Galaxy \citep[e.g.][]{1978A&A....67...23S, 1995AJ....109.2736M, 2004A&A...416.1117C, 2006ApJ...653.1145K, 2014A&A...562A..71B}.

In order to interpret these abundance ratios, chemical evolution model is introduced to understand some basic qualitative concepts. The general trends of observational abundances of stars can be readily explained by the simple model of Galactic chemical evolution (hereafter GCE) with outflow, as suggested by \citet{1995ApJS...98..617T}. However, a number of new results have put more complex constraints to the physical scenarios of the Galaxy. In particular, the building of the Galactic halo in the framework of hierarchical galaxy formation and the significant role of radial migration in the Galactic disk should be properly considered in the GCE models. Several attempts to account for these processes have been undertaken in the past years through numerical simulations \citep[e.g.][]{2007MNRAS.381..647S, 2008ApJ...684L..79R}.

The majority of current chemical evolution models are based on high resolution observations of stars in the solar neighborhood. However, the fractions of population measured in the solar vicinity are not representative enough for the entire Galaxy. In particular, the outer halo of the Galaxy may be dominated by the merging imprints of nearby dwarf galaxies, where the chemical evolution proceeds at a lower rate than that of the inner region of the Galaxy. Moreover, even for stars in the solar neighborhood, they have distinct abundance distributions \citep[see, e.g.][]{1997A&A...326..751N, 2010A&A...511L..10N}, indicating the complex chemical evolution of the Galaxy at different locations. To decipher the spectral fingerprints in terms of abundances requires realistic models for the stellar atmospheres and the line-formation processes. Still today, the vast majority of abundance analysis of late-type stars relies on the assumption of local thermodynamic equilibrium (LTE). It is expected that this approach quite often gives misleading results, and for many elements such systematic errors may be very severe \citep[see Fig.7 of][as an example]{2006A&A...451.1065G}. The principles of NLTE line formation and codes capable of such calculations have been around for a long time but have only been explored in a more systematic fashion for a wide range of stellar parameters over the past decade or so \citep[e.g.][]{1998A&A...333..219Z, 2000A&A...362.1077Z, Mashonkina2001sr, 2002PASJ...54..275T, mg_c6, 2006A&A...458..899F, 2006A&A...456..313M, mash_ca, Mashonkina2008, 2006A&A...457..645Z,2008A&A...492..823B, Zhang2008_sc, 2009A&A...503..541L, 2009A&A...503..533S,2010A&A...509A..88A,2010MNRAS.401.1334B,2010A&A...522A...9B, 2011A&A...528A...9S,2012A&A...541A.143S,2015ApJ...802...36Y}.

Before the observations are used to give constraint to the GCE models, we need to take into account the analysis errors of the stellar abundances, such as the departures from LTE and uncertainties in the atmospheric parameters. In particular, most high-resolution spectroscopic studies are based on the LTE assumption, which may not be valid for spectral lines. It is suspected that both the unexpected behaviour of the scatter of the abundance ratios and the different behaviour between dwarfs and giants found in the LTE studies could be due at least partly to the neglect of the departures from LTE \citep{2010A&A...509A..88A}. Actually, NLTE analysis is important in the sense that it can improve the accuracy of stellar abundances, while LTE can achieve a very high precision with a large systematic deviation. In order to link the observations of abundance ratios with GCE models, we need more accuracy than the precision.  In particular, a modelling technique allowing for departures from LTE can be used to accurately predict iron abundances and spectroscopic stellar parameters for a set of benchmark late-type stars.

The \alpfe\ ratios can be more accurately derived by performing the NLTE analysis of both \alp -elements (Mg, Si, Ca, Ti) and iron abundances. It is well known that [O/Fe] and \alpfe\ ratios are the most important indicators for distinguishing of different chemical enrichment histories among the populations in the Galaxy or between the Galaxy and nearby dwarf galaxies. For various chemical species, a change of the element abundance during the Galactic evolution may consist of only a few tenths of one magnitude. To make stellar abundance trends visible if they exist, one needs to determine the differential abundances with the accuracy of 0.1 dex or better.

A systematic NLTE abundance analysis in F \& G dwarfs and subgiants seems worthwhile and timely. These stars are still the most commonly used beacons for studies of Galactic chemical evolution due to their sheer numbers and long lifetimes. An additional reason for limiting the discussions to F \& G dwarfs is that these should be similar to the Sun, which can therefore be used as a test-bench of the modeling. However, due to the increasing numerical complexity, compared with the LTE case, NLTE investigations have previously been limited to individual stars and usually only a handful of spectral lines. Contrary to the vast majority of abundance analyses available in the literatures \citep[e.g.][]{2000AJ....120.2513P, 2001A&A...370..951M, 2003MNRAS.340..304R, 2004A&A...420..183A, 2013ApJ...771...67I}, the present study will be based on NLTE line formation for \ion{Li}{1}, \ion{C}{1}, \ion{O}{1}, \ion{Na}{1}, \ion{Mg}{1}, \ion{Al}{1}, \ion{Si}{1}-\ion{Si}{2}, \ion{K}{1}, \ion{Ca}{1}, \ion{Sc}{2}, \ion{Ti}{2}, \ion{Fe}{1}-\ion{Fe}{2}, \ion{Cu}{1}, \ion{Sr}{2}, \ion{Zr}{2}, \ion{Ba}{2} and \ion{Eu}{2}. For each listed species, the original model atom was treated and tested by our previous studies (see Table\,\ref{tab:nlte} for details). The wavelength range is selected so that the lines of the NLTE elements are presented in the spectral coverage.

In this work, we aim to define a large sample, the F \& G benchmark stars, which include 51 F \& G dwarfs and subgiants in a limited range of temperatures, gravities and metallicities. These stars should be representative of the different stellar populations of the Galaxy. Most of these stars were studied under the LTE assumption in the past years. Their accurate stellar parameters have been  determined carefully by \citet[][hereafter, Paper I]{2015ApJ...808..148S}. It is important to have new abundances derived from the NLTE analysis, which will better constrain the models of the Galactic chemical evolution and the yields of the Supernovae \citep[e.g.][]{1989MNRAS.239..885M, 1995ApJS..101..181W, 1996ApJ...460..408T, 2006NuPhA.777..424N}.

In this paper, Sect.\,\ref{sect:sample} describes the stellar sample, observations, and atmospheric parameters. Details of the NLTE calculations, including the atomic models and mechanisms of the departures from LTE are given in Sect.\,\ref{sect:nlte}. Section\,\ref{sect:stars} presents the abundance results for the sample stars. In Sect.\,\ref{sect:evolution} we discuss the implications for the GCE model and nucleosynthesis, followed by a short section of conclusions.

\section{Stellar sample, observations, and atmospheric parameters}\label{sect:sample}

Our stellar sample, the observed stellar spectra, and the determination of atmospheric parameters were presented in Paper I.
Here, we remind the main points and describe briefly a reduction of the infrared (IR) spectra undertaken to remove fringes in the echelle orders, where the \ion{O}{1} 7771-5\,\AA\ triplet lines are located.

{\it The sample} includes 51 nearby stars uniformly distributed in the $-2.62 \le$ [Fe/H] $\le +0.24$ metallicity range. We selected unevolved stars, i.e. mostly dwarfs, with a few subgiants added. The Galactic thin disk stellar population is well represented in our sample by 27 stars, with [Fe/H] down to $-0.78$. We have eight thick disk stars in the $-1.47 \le$ [Fe/H] $\le -0.70$ range overlapping only a little with that of the thin disk and 16 halo stars. A membership of individual stars to the galactic stellar populations was identified based mainly on the star's kinematics.

{\it Observations.} Spectra of 48 stars were obtained for our project using the Hamilton Echelle Spectrograph mounted on the Shane 3 m telescope of the Lick observatory. Their resolving power is $R = \lambda/\Delta\lambda \simeq$ 60\,000, the spectral coverage is 3700-9300~\AA, and the signal-to-noise ratio (S/N) at 5500~\AA\ is higher than 100 for most stars. For two stars their spectra were obtained with CFHT/ESPaDOnS, as described in Paper~I. High-quality observed spectrum of HD~140283 was taken from the {\sc ESO UVESPOP} survey \citep{2003Msngr.114...10B}.
For some stars our observational material was complemented with the data from our earlier projects, namely, VLT2/UVES, 67.D-0086A (HD~74000, BD~$-04^\circ$\,3208) and 2.2 m/FOCES (HD~59374, HD~59984, HD~103095, HD~134169). We also employed the archives of CFHT/ESPaDOnS\footnote{http://www.cadc-ccda.hia-iha.nrc-cnrc.gc.ca/en/search/} for BD\,$-13^\circ$~3442, VLT2/UVES and 3.6 m/HARPS\footnote{http://archive.eso.org/wdb/wdb/adp/phase3-main/query} for HD~59374 (074.C-0364(A), $R \simeq$ 115\,000), HD~59984 (076.B-0133(A), $R \simeq$ 57\,000), HD~100563 ($R \simeq$ 115\,000), and HD~108177 ($R \simeq$ 115\,000), 1.93 m/SOPHIE\footnote{http://atlas.obs-hp.fr/sophie/} ($R$ = 76\,500) and 1.93 m/ELODIE\footnote{http://atlas.obs-hp.fr/elodie/} ($R$ = 42\,000) for HD~64090 and BD~$+66^\circ$~0268.

\begin{figure}
\epsscale{1.0}
\plotone{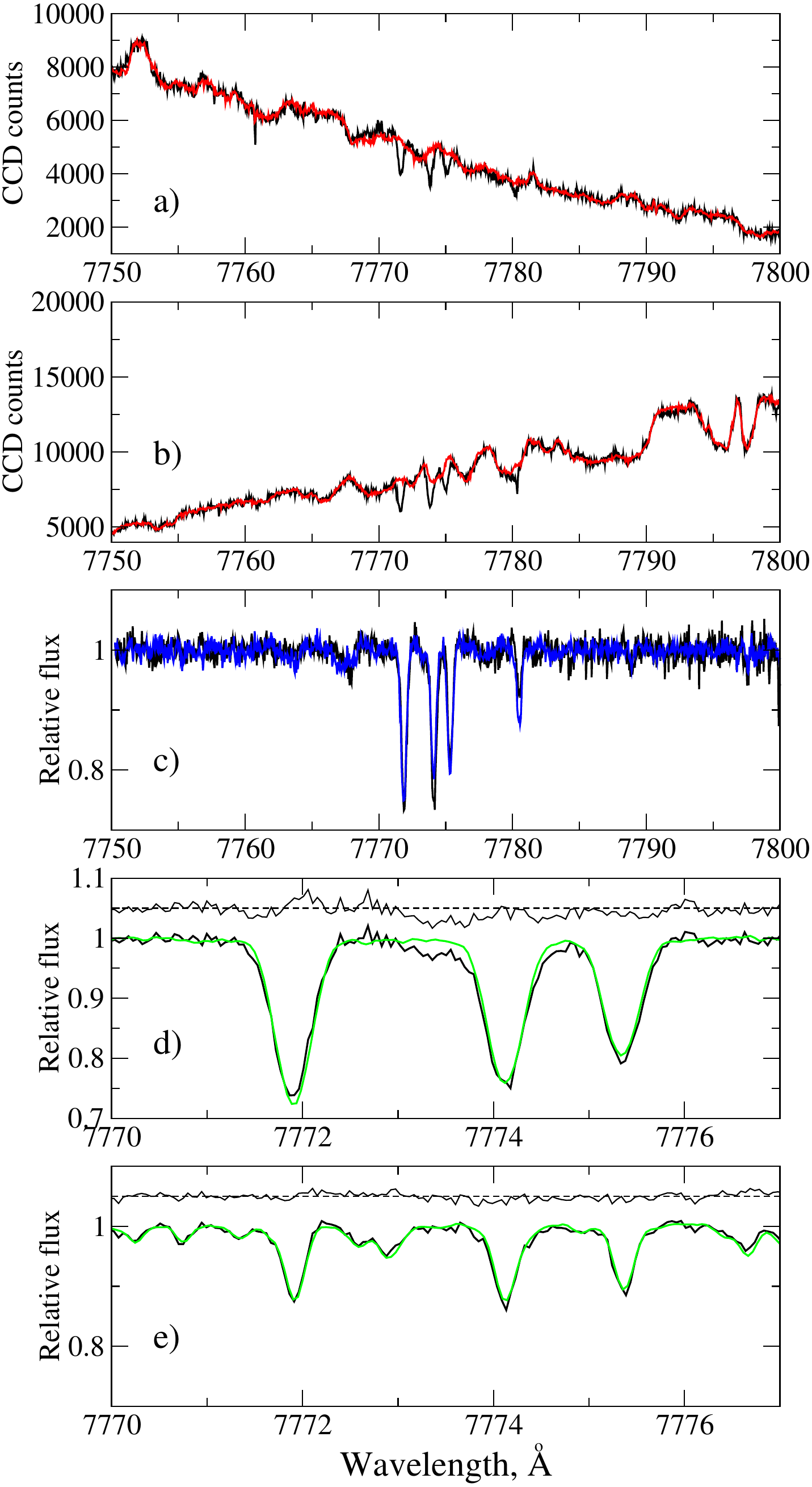}
\caption{Fringes reduction in the HD~49933 and HD~142091 spectra.
\textbf{Panel a}: fragment of the HD~49933 spectrum from the slit center of
the 97th echelle order (black curve) and the recovered fringes for the same
position on the slit (red curve). \textbf{Panel b}: the same as in panel a for
overlapping part of 98-th echelle order. \textbf{Panel c}: reduced and
normalized spectra of 97-th (black curve) and 98-th (blue curve) echelle orders.
\textbf{Panel d}: the reduced Hamilton (black curve) and the ESPaDOnS (green curve)  spectrum of HD~49933. The difference between the two spectra is shown in the upper part of the panel. \textbf{Panel e}: the same as in panel d for HD~142091.}
\label{fig:fringes}
\end{figure}

{\it Fringes reduction.}
The \ion{O}{1} 7771-5~\AA\ lines are located in the two overlapping echelle orders, 97th and 98th, of the Hamilton spectrograph (Fig.~\ref{fig:fringes}a and b). We propose the following procedure to remove the fringing effects.
The intensity of CCD fringes depends on the wavelength and the thickness of a
silicon layer of CCD. Although the standard flat fielding could be used to
remove fringes, this procedure is limited by the bright scattered light. Here we apply a statistical procedure based on a set of stellar spectra with similar exposures taken at one night. We show below that due to a comparable level of the scattered light in stellar spectra the fringes can be recovered with a reasonable precision. Moreover, different stellar radial velocities provide us with the possibility to recover the fringes even in the continuum around absorption lines. Obviously, an accuracy of this statistical procedure depends on the spectra sample volume.

The statistical approach was applied at each night to fit the fringes in the
vicinity of oxygen lines. For the observational set in March, 2011 we used seven working stars, while 14 to 19 working stars for the 2012 observations. The processing starts from raw CCD images with subtracted bias. For each star we find  positions of echelle orders and the light distribution along the slit. We then extracted 11 spectra of one pixel height along the slit and processed them independently. The 6th spectrum corresponding to the slit center is shown in Fig.~\ref{fig:fringes}a and b. All stellar and telluric lines were removed in order to use only the continuum spectrum. We select the star with the spectrum $I_0(\lambda)$ of the highest signal-to-noise ratio and reduce spectra $I_{\ast}(\lambda)$ of other working stars to the selected spectrum using the relation $I^{'}_{\ast}(\lambda)=S(\lambda)I_{\ast}(\lambda)$, where $S(\lambda)$ is smooth spline approximation of the function $I_{0}(\lambda)/I_{\ast}(\lambda)$.

The median averaging of the $I^{'}_{\ast}(\lambda)$ spectra of all working stars gives us 11 spectra of the fringes along the slit, which are then used to normalize the stellar spectra. Eleven normalized spectra for each star were averaged with weights depending on the light distribution along the slit. This procedure was applied to the second overlapping order as well. The reduced spectra are shown in Fig.~\ref{fig:fringes}c. Both spectra were then averaged with weights depending on their CCD signal level. A precise wavelength calibration is required to correctly perform the spectra averaging. We used a wavelength solution of the Ta-Th-Ar hollow cathode lamp \citep{2015ARep...59..952P} for the observations taken in 2011, March and the Ti-Ar lamp \citep{2013AJ....146...97P} for the observations of 2012.

In Fig.~\ref{fig:fringes}d and e we compare spectra of HD~49933 and HD~142091 observed with the Hamilton spectrograph ($R$ = 60\,000) and reduced in this study with the corresponding ESPaDOnS spectra, which are free of fringes. It is worth noting, the latter $R$ = 80\,000 and S/N $>200$ spectra were degraded to $R$ = 60\,000. In case of HD~49933 the root mean square (rms) of the difference between the Hamilton and ESPaDOnS spectra amounts to
0.0114 that corresponds to S/N = 90, slightly lower than S/N = 110 of the original infrared spectrum of this star. In case of HD~142091 the rms
value is 0.0059 (S/N = 170 versus the original S/N = 250). Thus, the statistical approach is efficient in removing the fringes.

{\it Stellar atmosphere parameters.} A combination of the photometric and spectroscopic methods was applied to derive a homogeneous set of the stellar atmosphere parameters: effective temperature $\Teff$, surface gravity log~g, [Fe/H], and microturbulence velocity $\vt$. Our spectroscopic analyses took advantage of employing the NLTE line formation for \ion{Fe}{1} - \ion{Fe}{2}. Paper~I estimated the systematic and statistical errors of $\Teff$ to be 50~K and 70~K, respectively, the uncertainty in log~g / $\vt$ to be 0.04~dex / 0.14~\kms, and statistical error of [Fe/H] was defined by the dispersion, $\sigma$, for lines of \ion{Fe}{2} in a given star. For most stars the latter value amounts to 0.03~dex to 0.09~dex. Hereafter, the statistical abundance error is the dispersion in the single line measurements about the mean:
$\sigma = \sqrt{\Sigma (x - x_i )^2 /(N - 1)}$.

Our sample of unevolved stars, which are uniformly distributed in a wide metallicity range, belong to the three different Galactic stellar populations and have accurate and homogeneous atmospheric parameters is suitable for Galactic chemical evolution studies.

\section{NLTE calculations}\label{sect:nlte}

\subsection{NLTE methods}

Our present investigation is based on the NLTE methods treated in our earlier studies and documented in a number of papers, where atomic data and the problems of line formation have been considered in detail. Table~\ref{tab:nlte} lists the investigated chemical species and cites the related papers. Compared with the published model atoms, collisional data were updated for several chemical species. For \ion{Ca}{2} and \ion{Sr}{2} we apply here electron-impact excitation rate coefficients from ab initio calculations of \citet{ca2_bautista} and \citet{2002MNRAS.331..875B}, respectively. For \ion{Li}{1}, \ion{Mg}{1}, \ion{Al}{1}, and \ion{Si}{1} inelastic collisions with neutral hydrogen particles are treated using accurate rate coefficients from quantum-mechanical calculations of \citet{belyaev03_Li,mg_hyd2012,Belyaev2013_Al}, and \citet {Belyaev2014_Si}, respectively. For the remaining species hydrogen collisions are computed using the formula of \citet{Steenbock1984} with a scaling factor \kH\, estimated empirically in the literature from their different influence on the different lines of a given atom in solar and stellar spectra. The references and recommended \kH\, values are indicated in Table~\ref{tab:nlte}.

In order to solve the coupled radiative transfer and statistical equilibrium equations for metals, we use a revised version of the DETAIL program \citep{detail} based on the accelerated lambda iteration, which follows the efficient method described by \citet{rh91,rh92}. The update was presented by \citet{mash_fe}. The obtained departure coefficients were then used by the codes SIU \citep{Reetz} and {\sc synthV-NLTE} \citep{Ryabchikova2015} to calculate the synthetic line profiles.

As in Paper~I, in this study we used the MARCS model structures \citep{Gustafssonetal:2008}.

\begin{deluxetable}{lcl}
\tabletypesize{\scriptsize}
\centering
\tablecaption{Atomic models used in this study and treatment of A + \ion{H}{1} inelastic collisions. \label{tab:nlte}}
\tablehead{
\colhead{Species}  & \colhead{Model atom} & \colhead{A + \ion{H}{1}}
}
\startdata
\ion{Li}{1}   & {\cite{2007A\string&A...465..587S}}  & BB03 \\
\ion{C}{1}    & \citet{2015MNRAS.453.1619A}   & \kH $^*$ = 0.3 \\
\ion{O}{1}    & \citet{sitnova_o}             & \kH\ = 1.0 \\
\ion{Na}{1}   & \citet{mg_c6}                 & \kH\ = 0.05  \\
\ion{Mg}{1}   & \citet{mash_mg13}             & BB12  \\
\ion{Al}{1}   & \citet{Baumueller_al1}        & B13  \\
\ion{Si}{1}-\ion{Si}{2} & \citet{Shi_si_sun}  & BYB14  \\
\ion{K}{1}    & {\citet{2006A\string&A...453..723Z}} & \kH\ = 0.05  \\
\ion{Ca}{1}   & \citet{mash_ca}               & \kH\ = 0.1 \\
\ion{Sc}{2}   & \citet{Zhang2008_sc}          & \kH\ = 0.1 \\
\ion{Ti}{2}   & \citet{sitnova_ti}  & \kH\ = 1.0 \\
\ion{Fe}{1}-\ion{Fe}{2} & \citet{mash_fe}     & \kH\ = 0.5 \\
\ion{Cu}{1}   & \citet{cu1_nlte_shi}          & \kH\ = 0.1 \\
\ion{Sr}{2}   & \citet{Mashonkina2001sr}      & \kH\ = 0.01 \\
\ion{Zr}{2}   & \citet{Velichko2010_zr}       & \kH\ = 1.0 \\
\ion{Ba}{2}   & \citet{Mashonkina1999}        & \kH\ = 0.01 \\
\ion{Eu}{2}   & \citet{mash_eu}               & \kH\ = 0.1 \\
\enddata
$^*$ Scaling factor to the Drawinian rates
\tablecomments{BB03 = \citet{belyaev03_Li}; BB12 = \citet{mg_hyd2012}; B13 = \citet{Belyaev2013_Al}; BYB14 = \citet{Belyaev2014_Si}.}
\end{deluxetable}

\subsection{Line list and solar abundances for a differential analysis}\label{line_list}

The lines used in the abundance analysis were selected from the lists of our NLTE papers (see Table\,\ref{tab:nlte} for references). They are listed in Table\,\ref{Tab:lines} along with the adopted atomic parameters.

The van der Waals damping was computed following the perturbation theory, where the data were available, using the van der Waals damping constants $\Gamma_6/N_{\rm H}$ at 10\,000~K as provided by \citet{BPM}. An exception was the selected lines of some elements, for which we used the $C_6$-values derived from solar line-profile fitting. If no other data were available, the $\Gamma_6/N_{\rm H}$ values from Kurucz's calculations\footnote{http://kurucz.harvard.edu/atoms.html} were employed.

Some elements considered here are represented by either a single
isotope with an odd number of nucleons (Sc), or multiple isotopes with measured wavelength differences ($\Delta\lambda \ge 0.01$\,{\AA} for \ion{Li}{1}, \ion{Cu}{1}, \ion{Sc}{2}, \ion{Ba}{2}, and \ion{Eu}{2}). Nucleon-electron spin interactions in odd-$A$ isotopes lead to hyper-fine splitting (HFS) of the energy levels, resulting in absorption lines divided into multiple components. Without accounting properly for HFS and/or isotopic splitting (IS) structure, abundances determined from the lines sensitive to these effects can be severely overestimated.

Hyperfine structure (HFS) and/or isotope structure (IS) is taken into account when necessary with the data from \citet[][\ion{Li}{1}]{1995PhRvA..52.2682S}, \citet[][\ion{Sc}{2}]{Zhang2008_sc}, \citet[][\ion{Cu}{1}]{cu1_nlte_shi}, \citet[][\ion{Sr}{2}]{1983HyInt..15..177B}, Robert Kurucz's website ({\tt http://kurucz.harvard.edu/atoms.html}, \ion{Ba}{2}), and \citet[][\ion{Eu}{2}]{Lawler_Eu}. For Li, Cu, Sr, Ba, and Eu, we use the fractional isotope abundances corresponding to the solar system matter \citep{Lodders2009}.

The Sun is used as a reference star for a subsequent stellar abundance analysis. The solar flux observations were taken from the Kitt Peak Solar Atlas \citep{Atlas}. The calculations were performed with the MARCS model atmosphere 5777/4.44/0 \citep{Gustafssonetal:2008}. A depth-independent
microturbulence of 0.9\,\kms\ was adopted. Our synthetic flux profiles were convolved with a profile that combines a rotational broadening of 1.8\,\kms\ and broadening by macroturbulence with a radial-tangential profile. The $\Vmac$ values varied mainly between 2.6 and 3.3\,\kms\ for the strong lines and between 3.4 and 4.0\,\kms\ for the weak lines. For comparison, \citet{1977ApJ...218..530G} found solar macroturbulence velocities varying between 2.9 and 3.8\,\kms\ for a small sample of the solar \ion{Fe}{1} lines. Solar LTE and NLTE abundances from the individual lines are presented in Table\,\ref{Tab:lines}.

\subsection{Departures from LTE for individual spectral lines}\label{sect:nlte3}

\begin{figure*}
\epsscale{1.0}
\plottwo{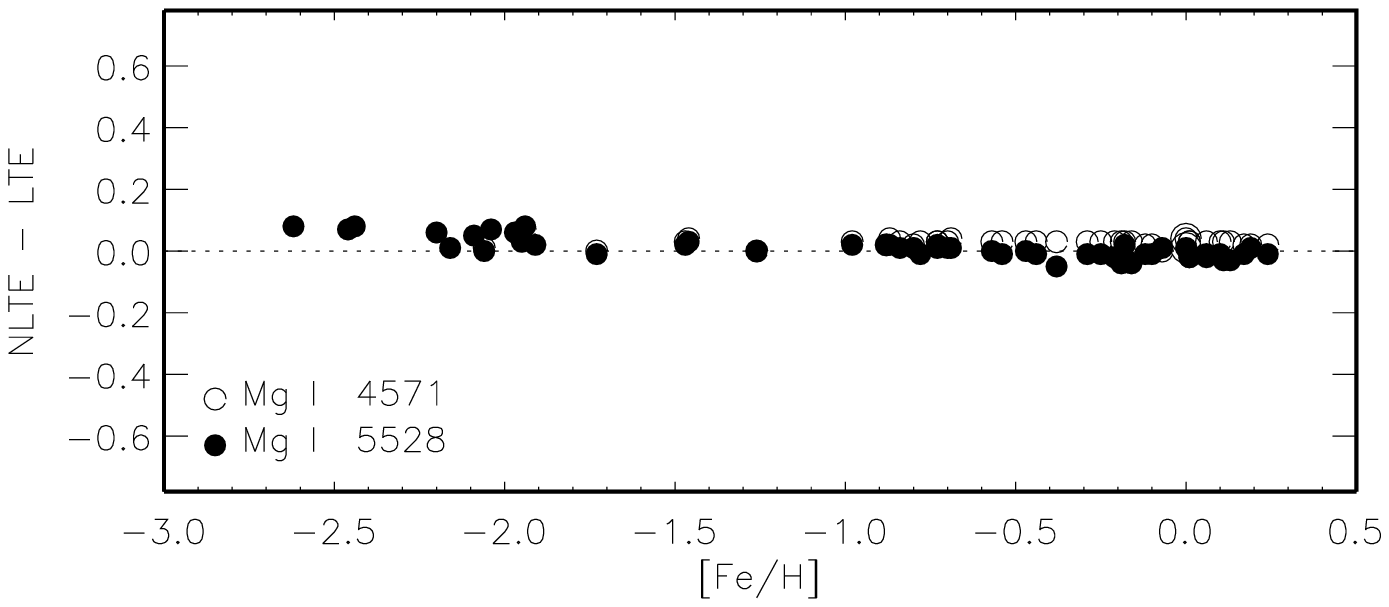}{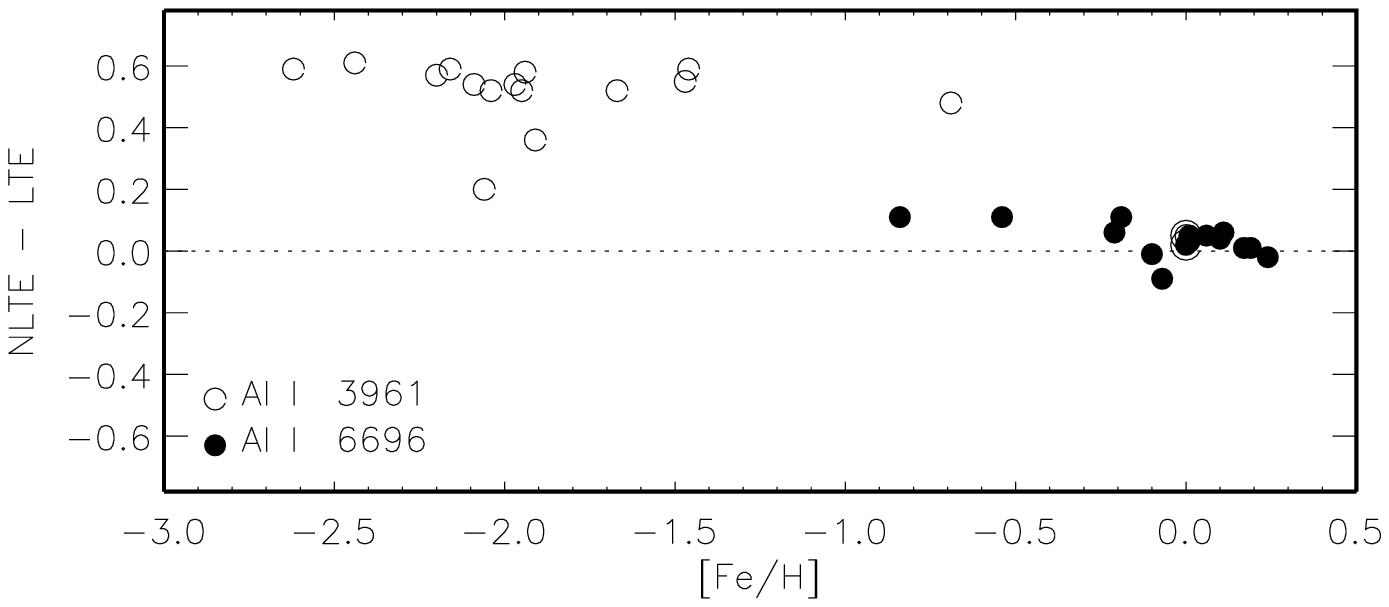}
\plottwo{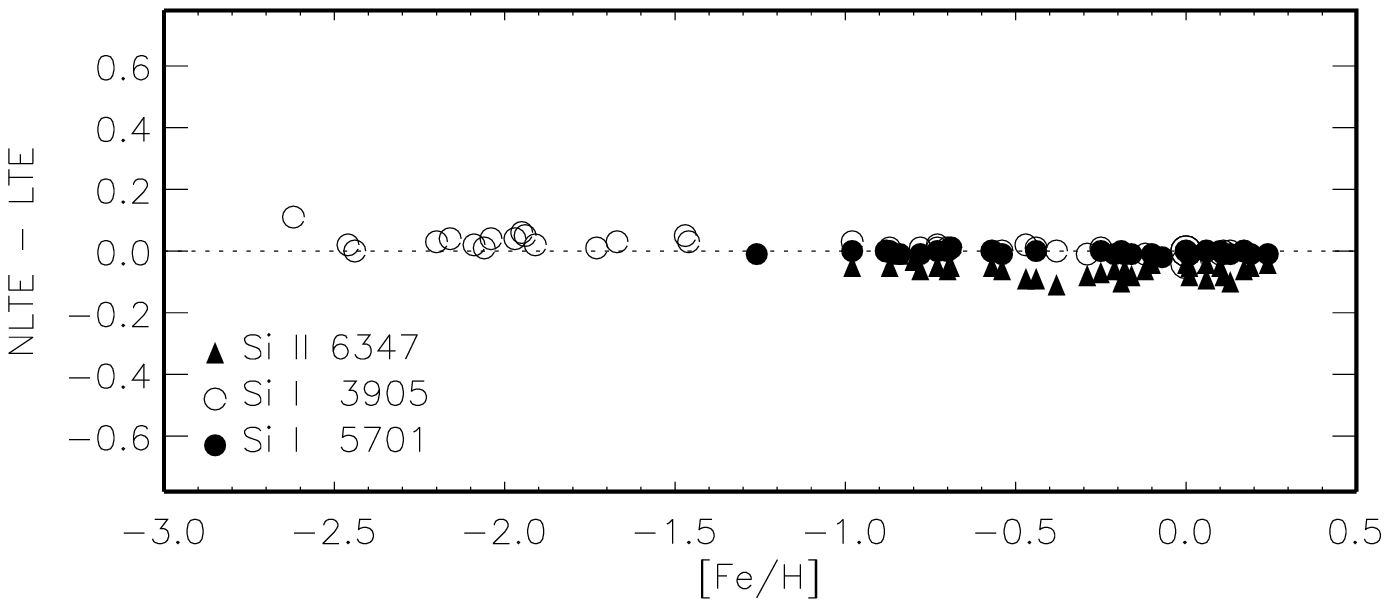}{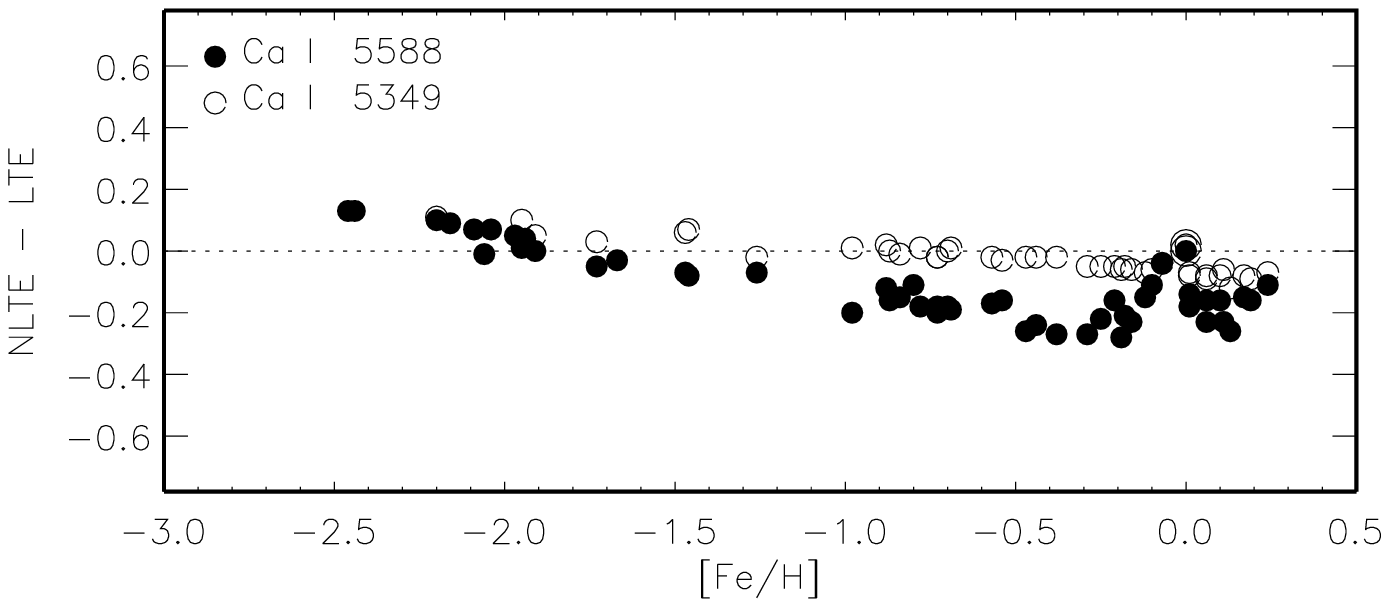}
\plottwo{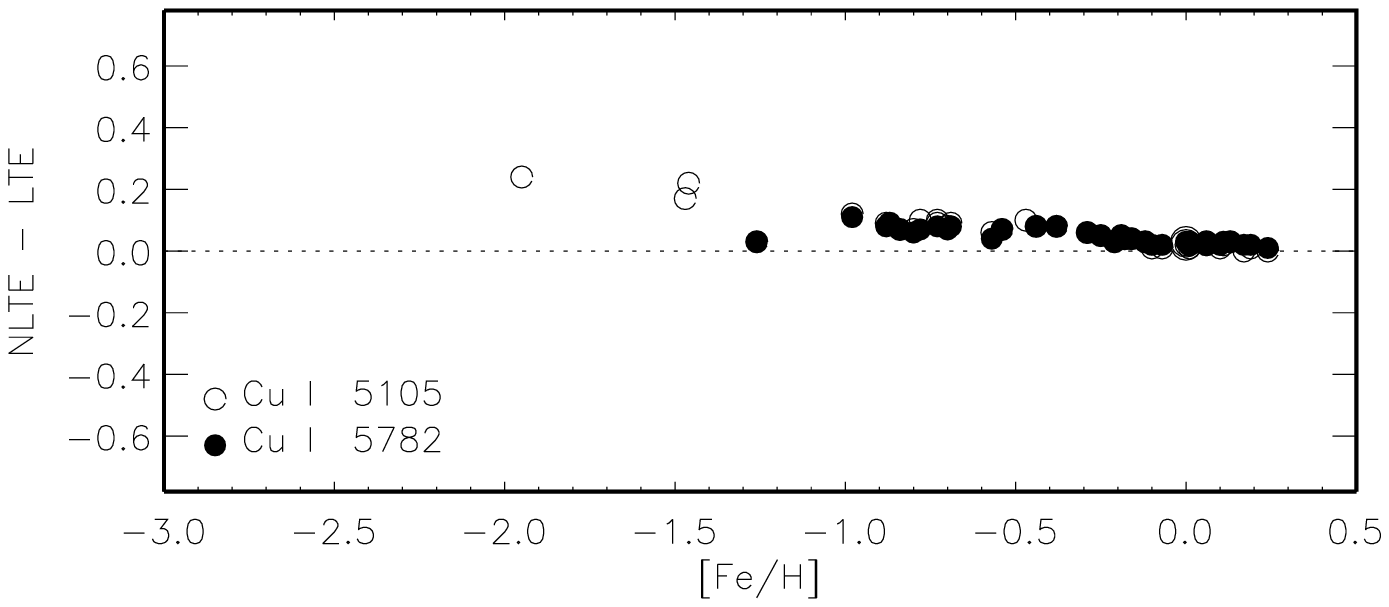}{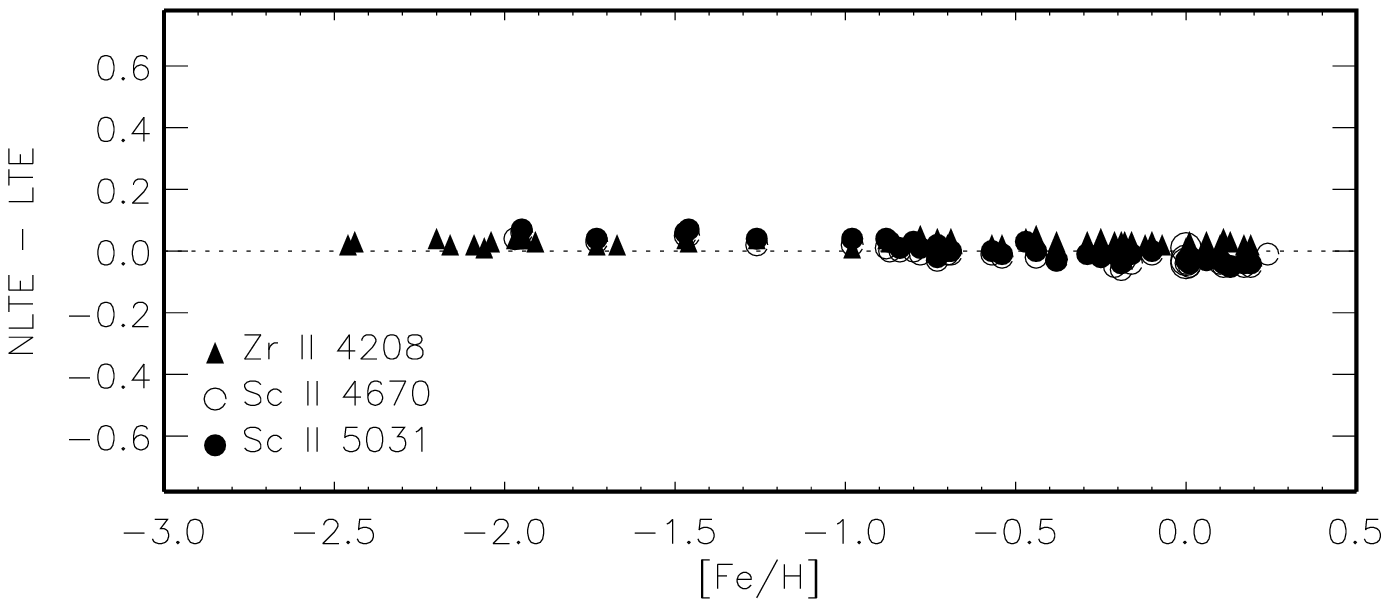}
\plottwo{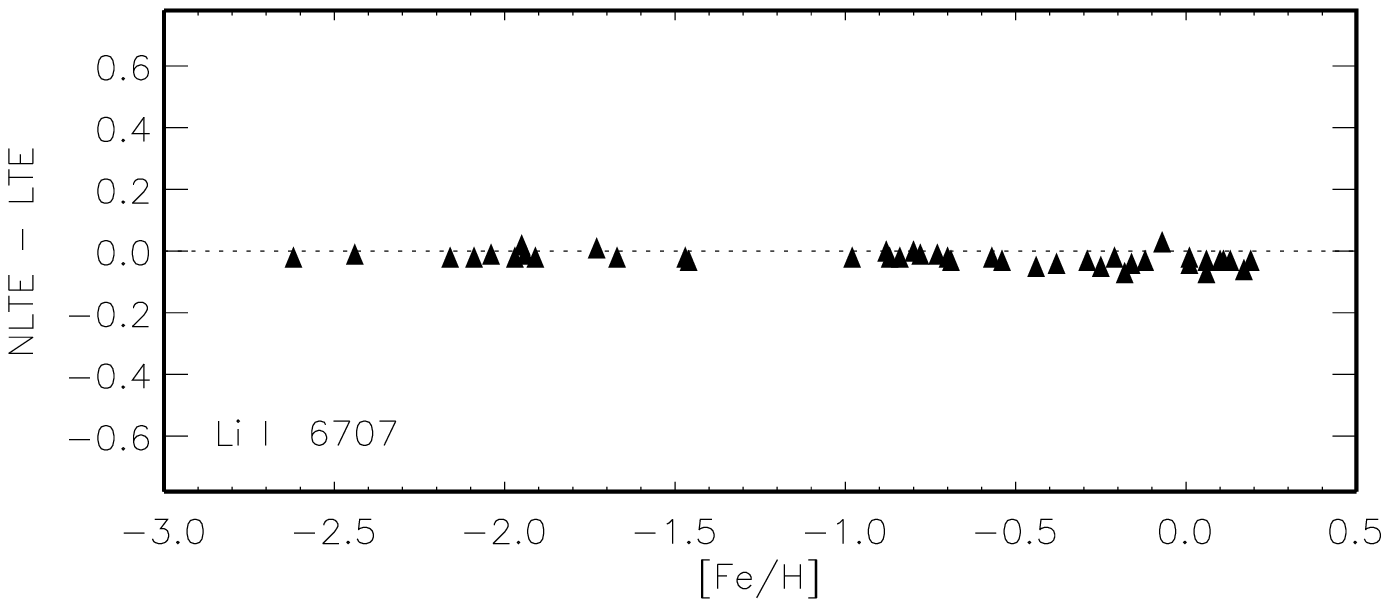}{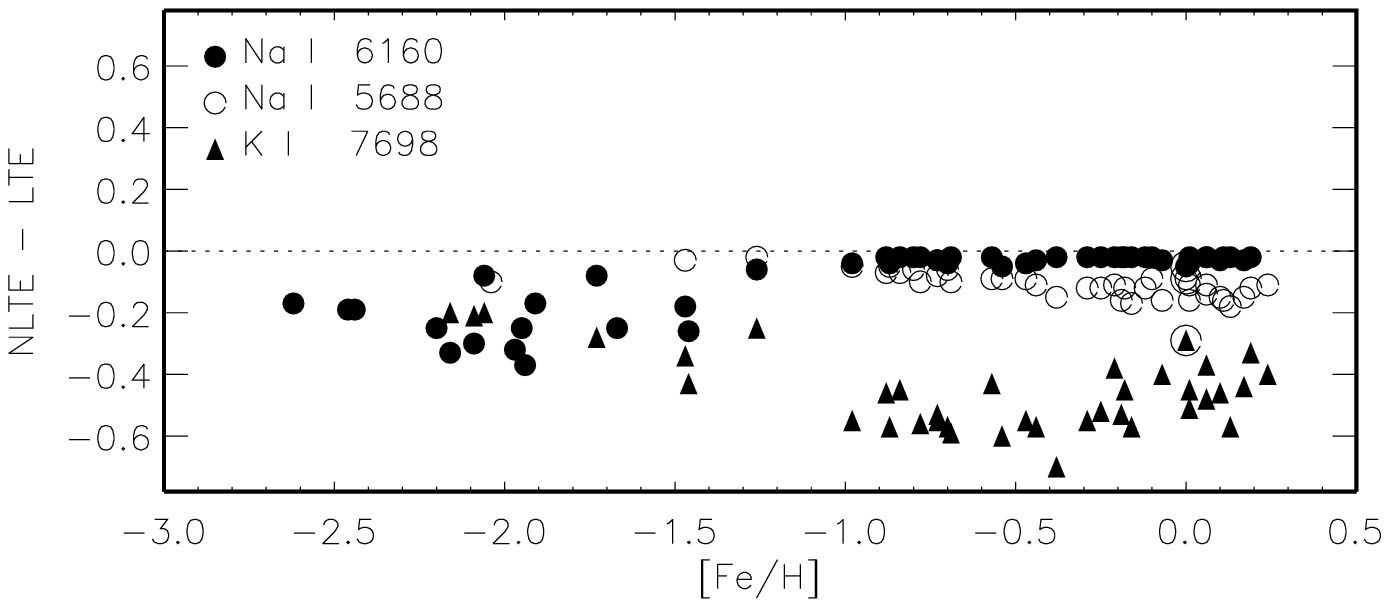}
\plottwo{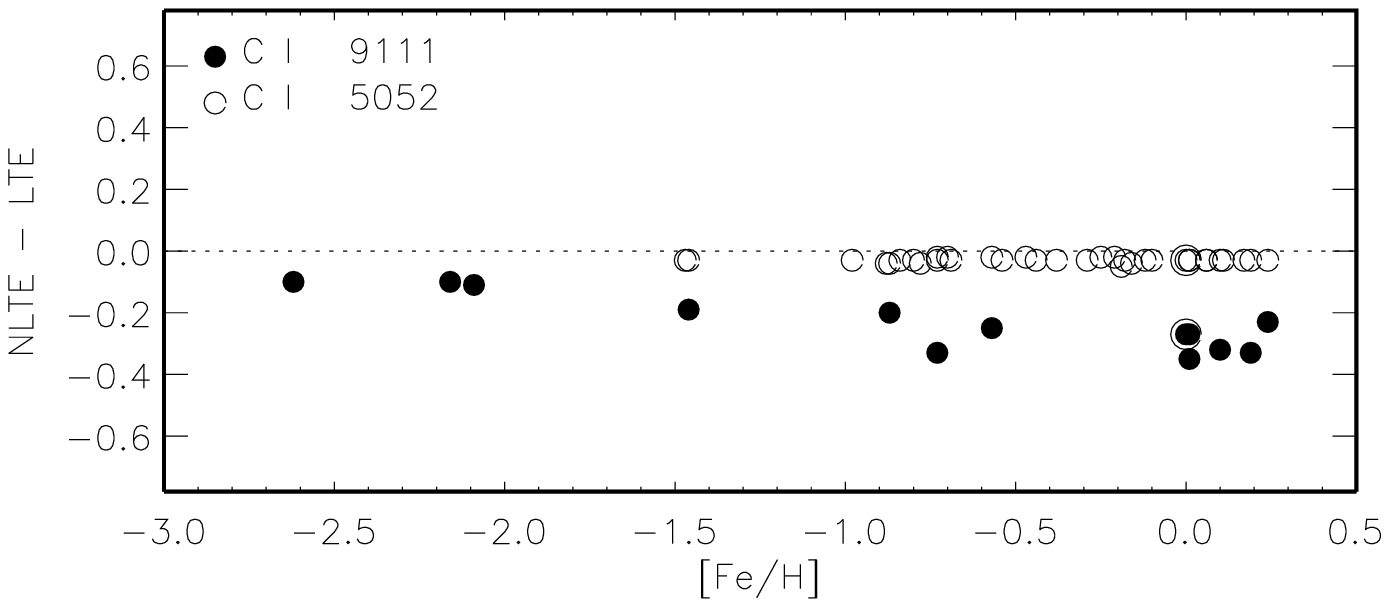}{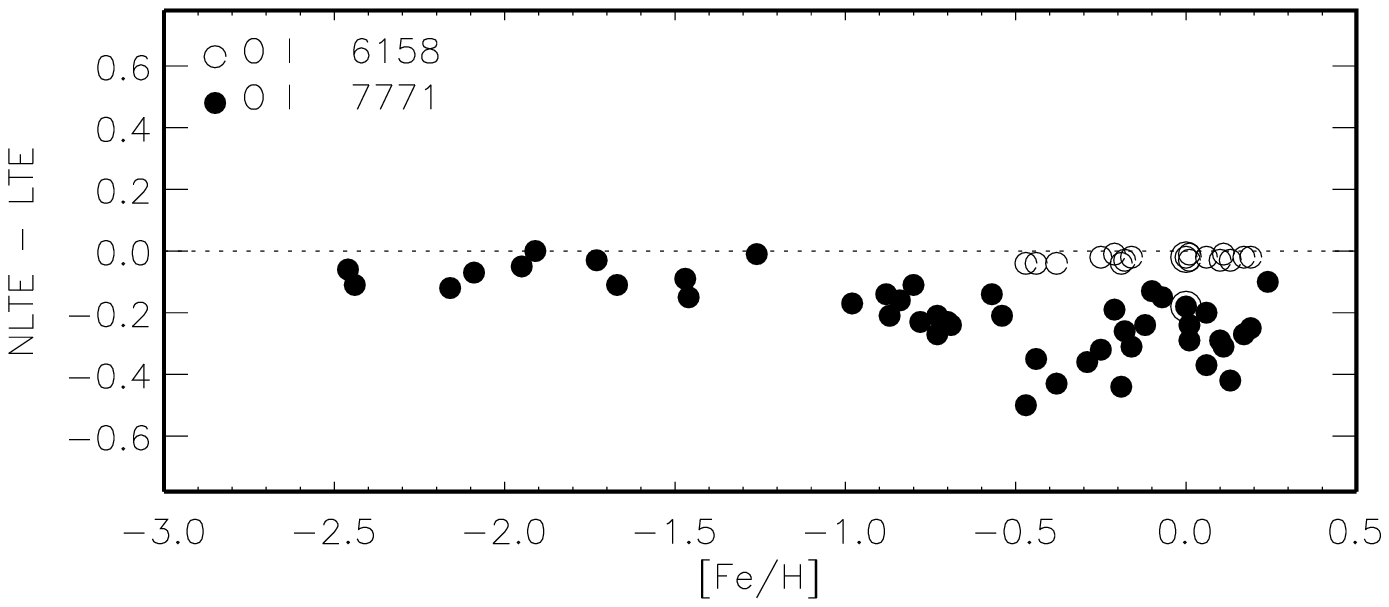}
\plottwo{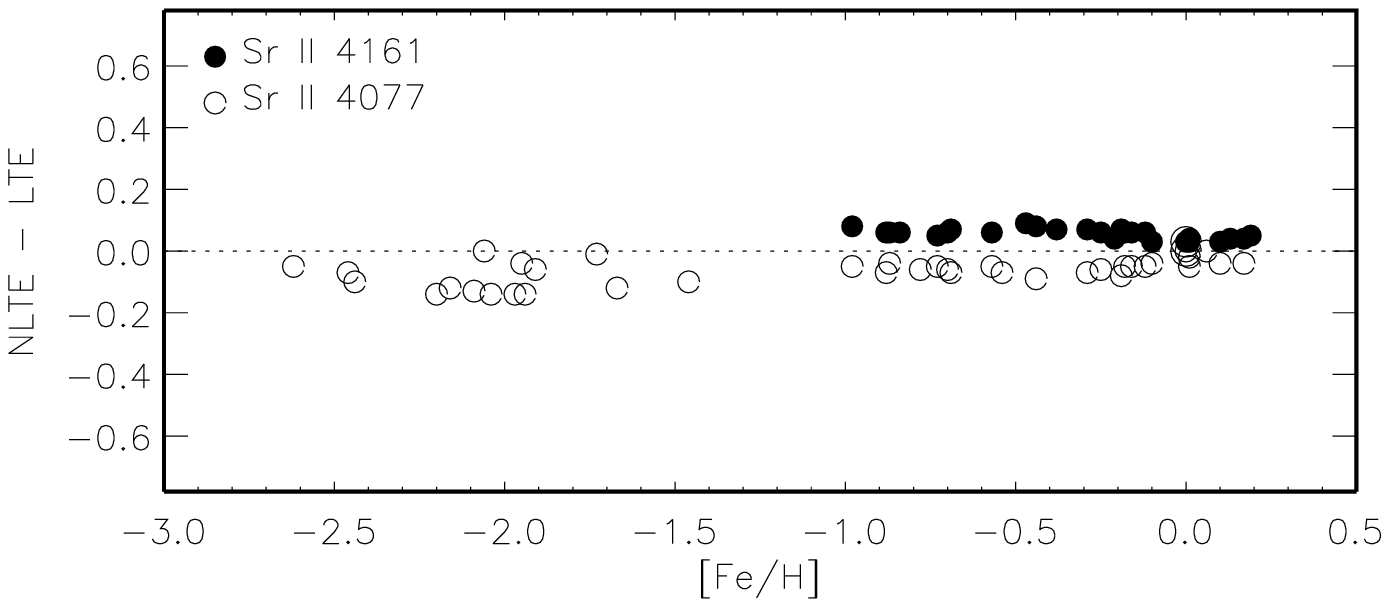}{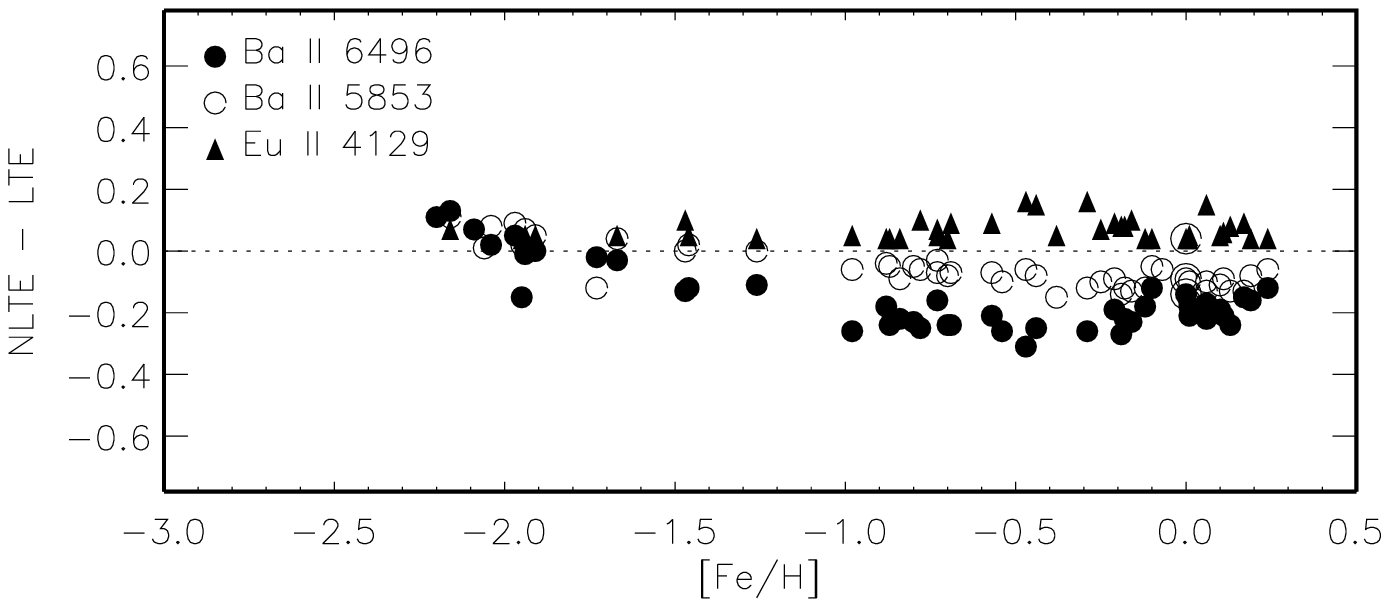}
\caption{NLTE abundance corrections for the selected lines in the investigated stars. The Sun is shown by the symbol inside the larger-size open circle. \label{Fig:dnlte}}
\end{figure*}

Our calculations show that the departures from LTE are different for lines of different chemical species and different for lines of any given species. For each individual line the NLTE effects depend on stellar parameters. Figure\,\ref{Fig:dnlte} displays the NLTE abundance corrections, $\Delta_{\rm NLTE} = \eps{NLTE} - \eps{LTE}$, for the representative lines of different species in our stellar sample, and Fig.\,\ref{Fig:fits} illustrates the departures from LTE in the line profiles. All the investigated NLTE species can be separated in five groups depending on a dominant NLTE mechanism.

1. {\it The photoionization dominated minority species \ion{Mg}{1}, \ion{Al}{1}, \ion{Si}{1}, \ion{Ca}{1}, \ion{Ti}{1}, \ion{Fe}{1}, and \ion{Cu}{1}.} Departures from LTE for these species are mainly caused by superthermal radiation of non-local origin below the thresholds of the ground state and/or low excitation levels, resulting in the overionization, i.e. depleted level populations compared with their TE values. Most investigated lines are weakened in NLTE compared with their LTE strengths, resulting in positive NLTE abundance corrections. They are overall small in the close-to-solar metallicity stars and grow toward lower metallicity due to decreasing ultraviolet (UV) opacity resulting in increasing photoionization rates. For given chemical species different lines reveal similar NLTE effects in a given model atmosphere. An exception is lines of \ion{Ca}{1}, see next paragraph. Pronounced NLTE effects for the resonance in contrast with the subordinate lines were found for \ion{Al}{1}. For example, $\Delta_{\rm NLTE}$(\ion{Al}{1} 3961\,\AA) = 0.77~dex, while $\Delta_{\rm NLTE} \le$ 0.13~dex for the subordinate lines in the model 5890/4.02/$-0.78$. This can be easy understood. The excited levels of \ion{Al}{1} though are subject to overionization, but are closely coupled to the ground state of the majority species \ion{Al}{2} via the charge-transfer reactions \ion{Al}{1}(nl) + \ion{H}{1}(1s) $\leftrightarrow$ \ion{Al}{2}(3s$^2$) + H$^-$ \citep{Belyaev2013_Al}, resulting in small $\Delta_{\rm NLTE}$ for \ion{Al}{1} 6696, 6698, 7835, 7836, 8772, 8773\,\AA. The ground state of \ion{Al}{1} is separated by 3.14~eV in energy from the excited levels, and its population is mainly decided by overionization. It is worth noting, the NLTE correction for \ion{Al}{1} 3961\,\AA\ is strongly surface gravity dependent. For example, $\Delta_{\rm NLTE}$ = 0.20~dex and 0.36~dex for the two least luminous and [Fe/H] $\simeq -2$ stars BD$+66^\circ$~0268 (5300/4.72/$-2.06$) and BD$+29^\circ$~2091 (5860/4.67/$-1.91$), while $\Delta_{\rm NLTE}$ ranges between 0.48~dex and 0.61~dex for the remaining [Fe/H] $< -0.5$ stars.

As discussed in detail by \citet{mash_ca} and \citet{mash_mg13}, lines of the photoionization dominated minority species \ion{Ca}{1} and \ion{Mg}{1} can have negative $\Delta_{\rm NLTE}$ in the close-to-solar metallicity models and positive correction in the low-metallicity models (\ion{Ca}{1} 5349, 5588\,\AA\ and \ion{Mg}{1} 5528\,\AA\ in Fig.\,\ref{Fig:dnlte}). Here, we remind briefly. The obtained NLTE abundance appears to be lower than the corresponding LTE one, if the line core forms in the layers, where the departure coefficient of the upper level drops rapidly due to photon escape from (usually) the line itself, resulting in dropping the line source function below the Planck function and enhanced absorption in the line core. In contrast, in the line wings, absorption is weaker compared with the LTE case due to overall overionization in deep atmospheric layers. Net effect is determined by a competition of the NLTE effects in the line core and the line wings. Similar NLTE mechanism leads to slightly negative $\Delta_{\rm NLTE}$  for some lines of \ion{Al}{1}, \ion{Si}{1}, and \ion{Cu}{1}.

2. {\it The collision dominated minority species \ion{Li}{1}, \ion{Na}{1}, and \ion{K}{1}.} In the stellar parameter range, with which we concern, these species are subject to the overrecombination resulting in strengthened lines of \ion{Li}{1}, \ion{Na}{1}, and \ion{K}{1} and negative NLTE abundance corrections. The origin of the overpopulation of the ground and first excited state is the photon suction process described in detail by \citet{1992A&A...265..237B}. The departures from LTE are larger for \ion{K}{1} than \ion{Na}{1} and for \ion{Na}{1} than \ion{Li}{1} because of smaller photoionization cross sections for \ion{K}{1} than \ion{Na}{1} and for \ion{Na}{1} than \ion{Li}{1}. A magnitude of $\Delta_{\rm NLTE}$ is small for the \ion{Na}{1} resonance lines in the [Fe/H] $> -1.5$ models because the lines are strong and their total absorption is mostly contributed from the line wings formed in deep atmospheric layers, where the departures from LTE are small. It is worth noting, the LTE abundances from \ion{Na}{1} 5889, 5895 \,\AA\ in the [Fe/H] $\le -1.5$ stars were derived using the measured equivalent widths because the line profiles cannot be fitted under the LTE assumption. Similarly, the LTE potassium abundances of all the stars were also derived using the measured equivalent widths.

3. {\it The majority species \ion{C}{1}, \ion{O}{1}, \ion{Si}{2}, \ion{Ti}{2}, and \ion{Fe}{2}, with negative NLTE abundance corrections for the investigated lines.} For each of these species its total number density and population of the ground state keep their TE values throughout the atmosphere. Populations of the excited levels are decided by a competition of the UV radiative pumping transitions, which produce an enhanced excitation in the line-formation layers, and photon losses in the lines, when the line optical depth drops below unity, resulting in an underpopulation of the upper levels of the corresponding transitions. For \ion{C}{1}, \ion{O}{1}, \ion{Si}{2}, \ion{Ti}{2}, and \ion{Fe}{2} the lower levels of the investigated transitions are overpopulated in line-formation layers, resulting in strengthened lines and negative NLTE abundance corrections. For different lines of \ion{Ti}{2} and \ion{Fe}{2} $\Delta_{\rm NLTE}$ is overall small. In line with the previous NLTE studies of \ion{C}{1} \citep[][and references therein]{2015MNRAS.453.1619A} and \ion{O}{1} \citep[][and references therein]{sitnova_o}, pronounced NLTE effects were computed for the infrared (IR) lines in the [Fe/H] $> -1.5$ models, with $\Delta_{\rm NLTE}$ up to $-0.5$~dex. The NLTE corrections reduce in absolute value toward lower metallicity due to shifting line-formation depth to deep atmospheric layers. For the visible lines of \ion{C}{1} and \ion{O}{1} $\Delta_{\rm NLTE}$s are overall small because the lines are weak and form in deep atmospheric  layers.

4. {\it The majority species \ion{Sc}{2}, \ion{Zr}{2}, and \ion{Eu}{2}, with positive NLTE abundance corrections for the investigated lines.} Here, for each line, NLTE leads to its weakening relative to the LTE strength, owing to the larger overpopulation of the upper than the lower level relative to the corresponding TE populations that results in the increase in the line source function above the Planck function in the line-formation layers.

5. {\it The majority species \ion{Sr}{2} and \ion{Ba}{2}, with a sign of the NLTE correction depending on the line and stellar parameters.} As found theoretically by \citet{Mashonkina1999}, NLTE may lead either to strengthening or to weakening the \ion{Ba}{2} lines depending on stellar parameters and element abundance. In our stellar sample, $\Delta_{\rm NLTE}$ is negative for \ion{Ba}{2} 5853\,\AA\ and 6496\,\AA\ in the [Fe/H] $> -1$ and [Fe/H] $> -1.8$ stars, and it becomes positive at the lower metallicity. For \ion{Sr}{2}, NLTE leads to a strengthening of the resonance lines and, in contrast, to a weakening of the subordinate line at 4161\,\AA. This can be understood as follows. In each model, the ground state keeps the TE population throughout the atmosphere and the upper level, 5p, of the resonance transition is underpopulated in the uppermost atmospheric layers due to photon losses in the resonance lines themselves resulting in an enhanced absorption of the 4077\,\AA\ and 4215\,\AA\ lines. The \ion{Sr}{2} 4161\,\AA\ line arises from the 5p-6s transition, where the upper level is overpopulated to a greater extent with regard to its LTE population than that of the lower level in the line formation layers.

\section{Determination of stellar abundances}\label{sect:stars}

To minimize the effect of the uncertainty in $gf$-values on the final
results, we applied a line-by-line differential NLTE and LTE approach, in the sense that stellar line abundances were compared with individual abundances of their solar counterparts. Throughout this study, the element abundance is determined from line profile fitting. The synthetic line profiles were computed with either the code SIU \citep{Reetz} or the codes {\sc synthV-NLTE} \citep{Ryabchikova2015} + {\sc binmag3}\footnote{http://www.astro.uu.se/$\sim$oleg/download.html}. The metal line list has been extracted from the Vienna Atomic Line Database\footnote{http://vald.astro.univie.ac.at/~vald3/php/vald.php} \citep[VALD3][]{2015PhyS...90e4005R}. Our test calculations of the \ion{C}{1} and \ion{Zr}{2} lines in a broad wavelength range from 4209~\AA\ to 9111~\AA\ in the solar model atmosphere prove that using SIU and {\sc synthV-NLTE} + {\sc binmag3} does not produce systematic shifts in derived abundances, namely the abundance difference nowhere exceeds 0.03~dex.

\begin{figure*}
\epsscale{1.0}
\plotone{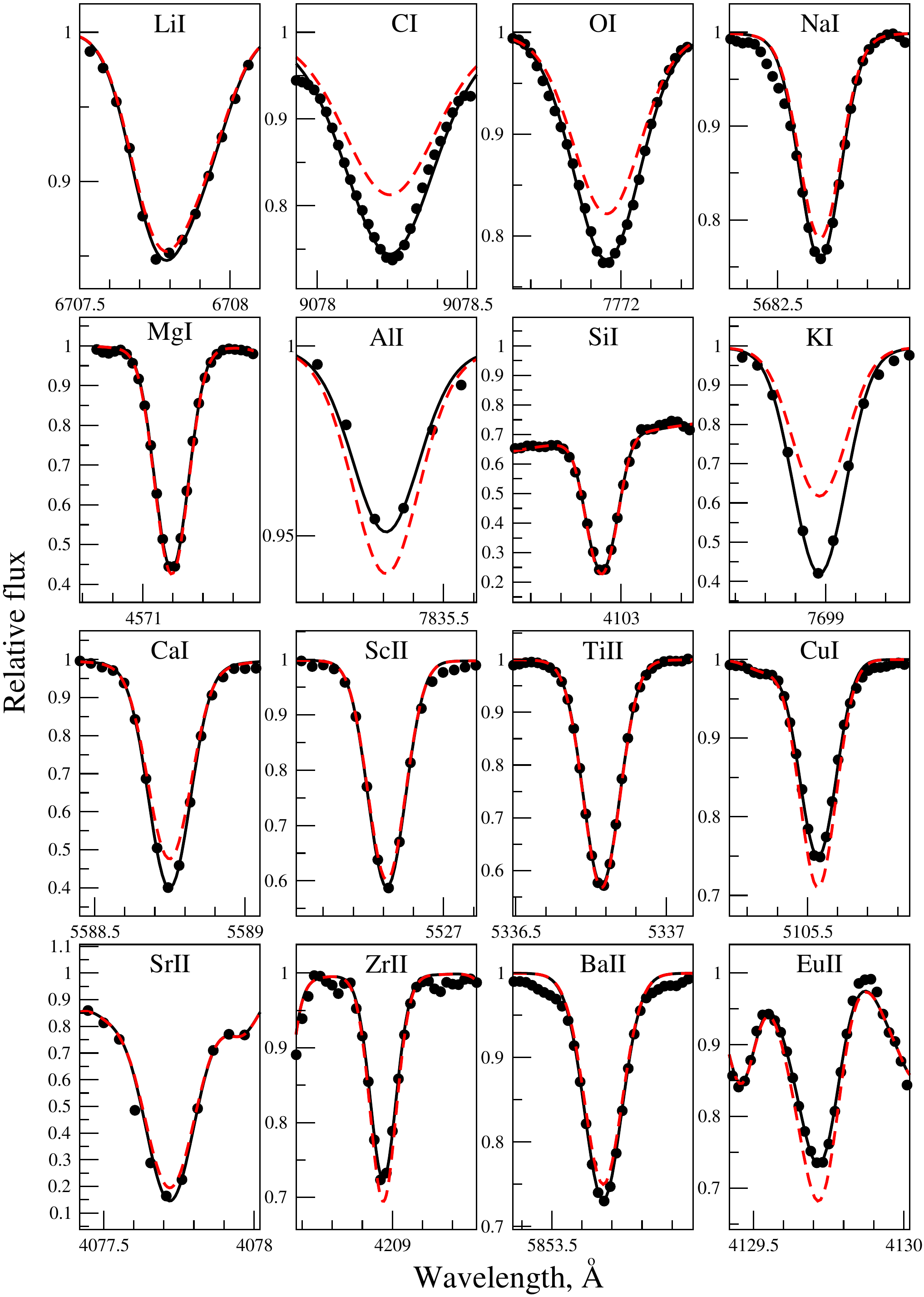}
\caption{The best NLTE fits (continuous curves) of the observed spectrum of HD~134169 (bold dots). For comparison, the LTE profiles computed with the corresponding NLTE abundances are shown by dashed curves. \label{Fig:fits}}
\end{figure*}

In order to compare the theoretical profiles with observations, they were convolved with a profile that combines instrumental broadening with a Gaussian profile, rotational broadening, and broadening by macroturbulence with a radial-tangential profile. Rotational broadening and broadening
by macroturbulence were treated separately for the six stars with $v \sin i \ge 6$~\kms, namely HD~58855 ($v \sin i$ = 10~\kms), HD~89744 (9~\kms), HD~92855 (10~\kms), HD~99984 (6~\kms), HD~100563 (10~\kms), and HD~106516 (7~\kms). We treated the overall effects of rotation and macroturbulence for the remaining stars as radial-tangential macroturbulence. The $v \sin i$ values and most probable macroturbulence velocities $\Vmac$ were determined in this study from the analysis of an extended list of lines of various chemical species. For a given star, $\Vmac$ was allowed to vary by $\pm$0.4~\kms\ (1$\sigma$). We selected a mildly MP star HD~134169 (5890/4.02/$-0.78$) to illustrate in Fig.\,\ref{Fig:fits} a quality of the line fits in a broad spectral range from 4077\,\AA\ to 9078\,\AA.

We determined abundances of 17 elements from Li to Eu and for silicon from two ionization stages. Table\,\ref{Tab:AbundanceSummary} (online material) presents the mean LTE and NLTE abundances, their error bars ($\sigma$), and the number of lines used to determine the mean abundances. For most species their abundances are based on analysis of the two to 20 lines. An exception is \ion{Li}{1}, \ion{K}{1}, and \ion{Eu}{2}, with a single line measured.

\begin{figure*}
\epsscale{1.0}
\plottwo{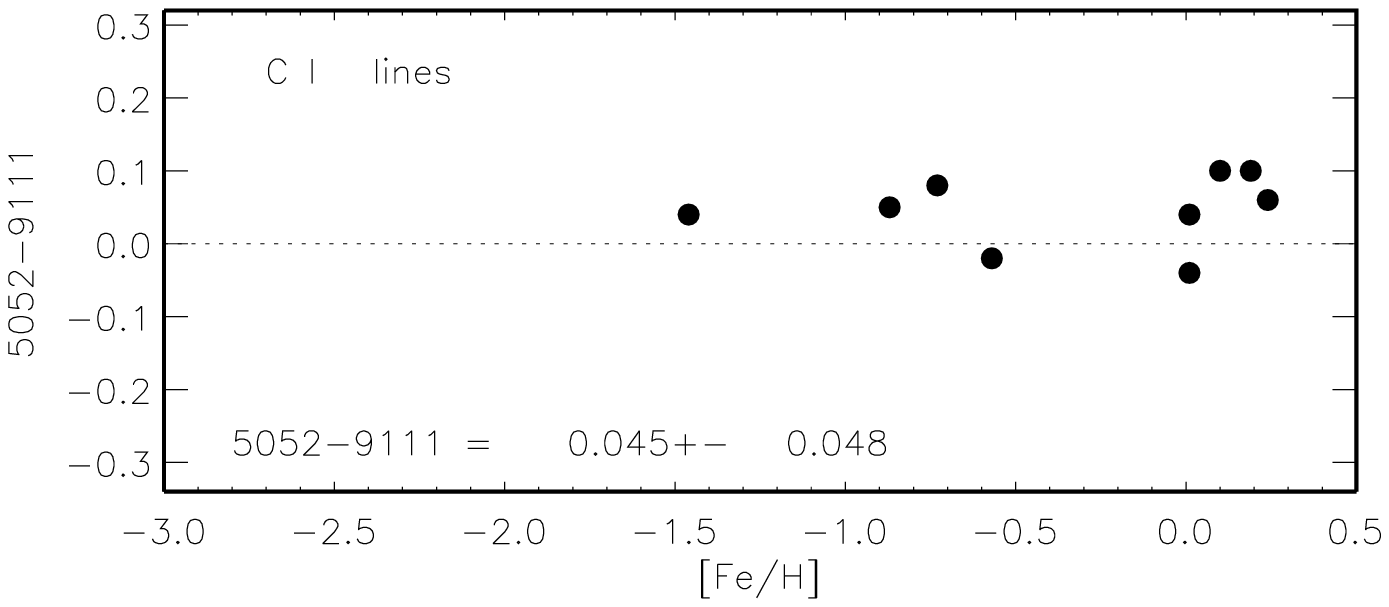}{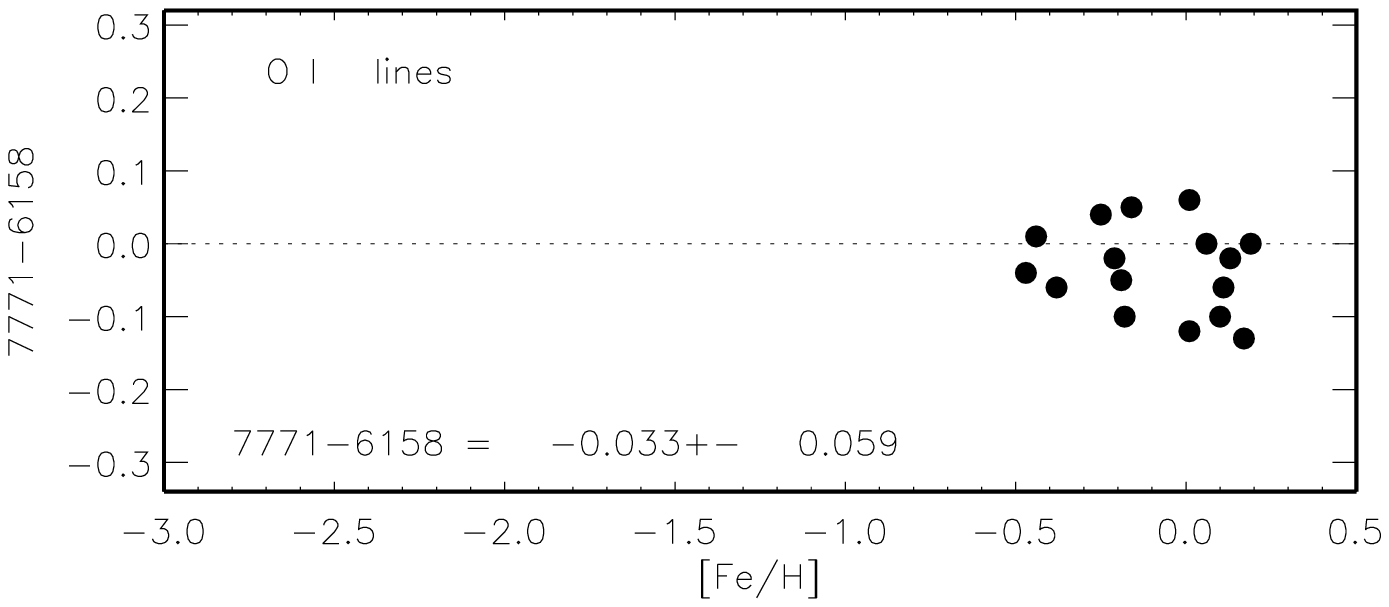}
\plottwo{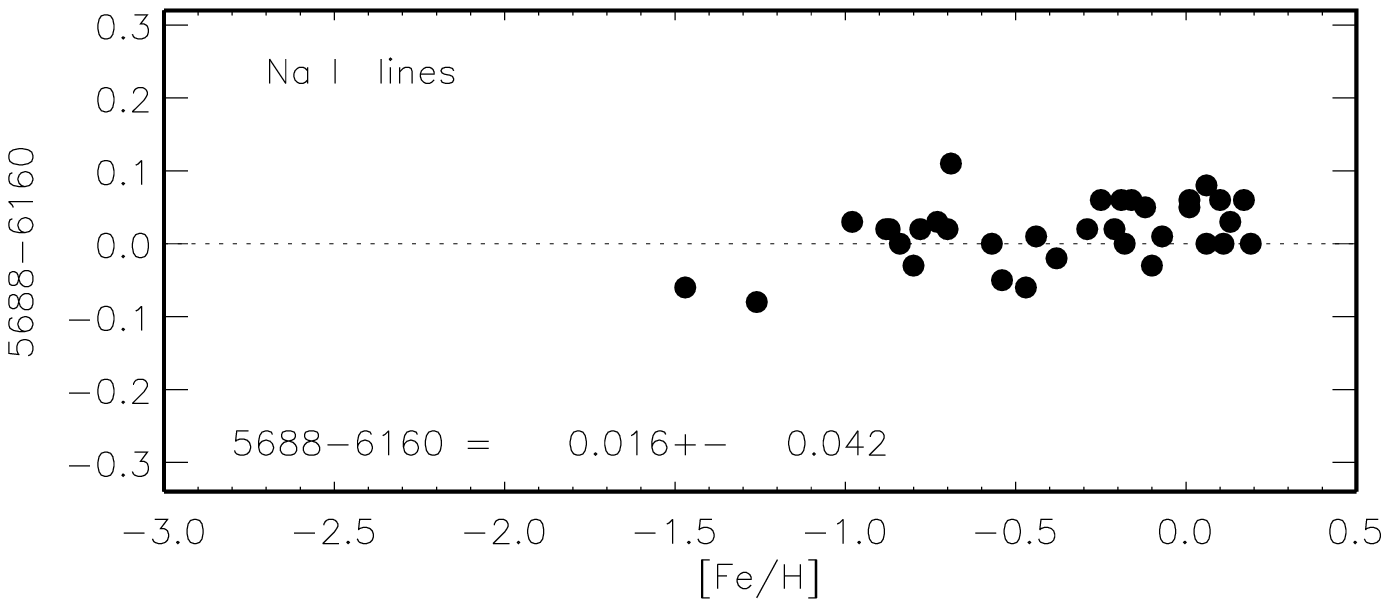}{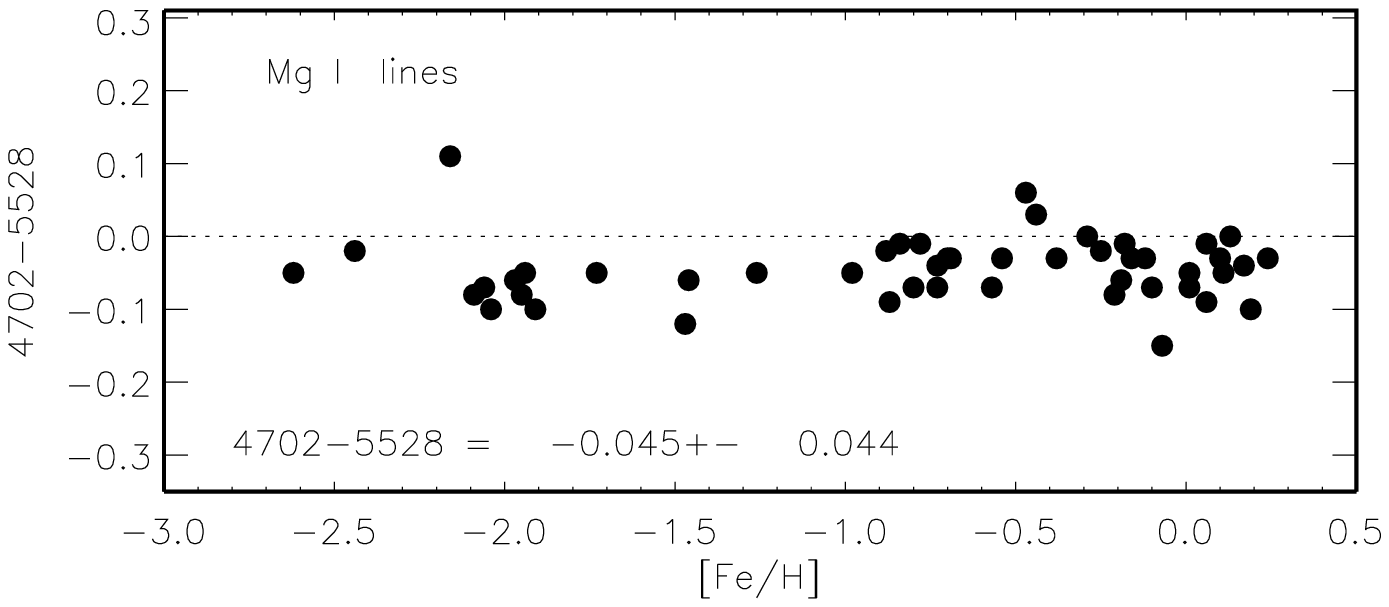}
\plottwo{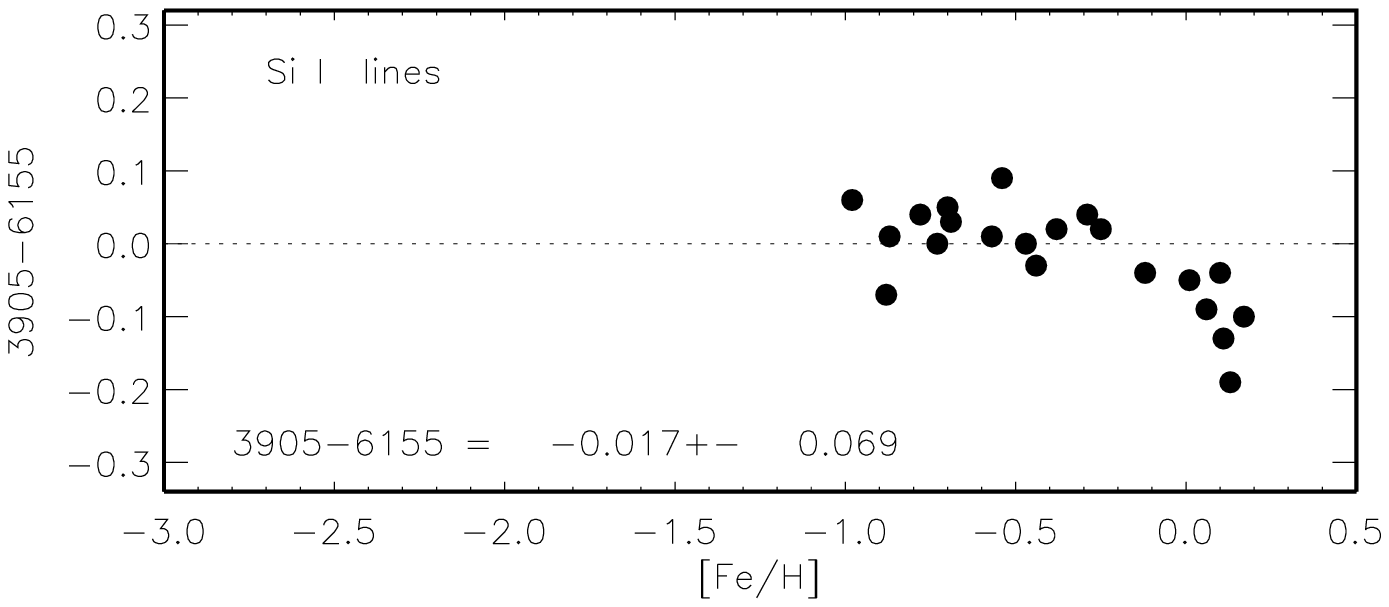}{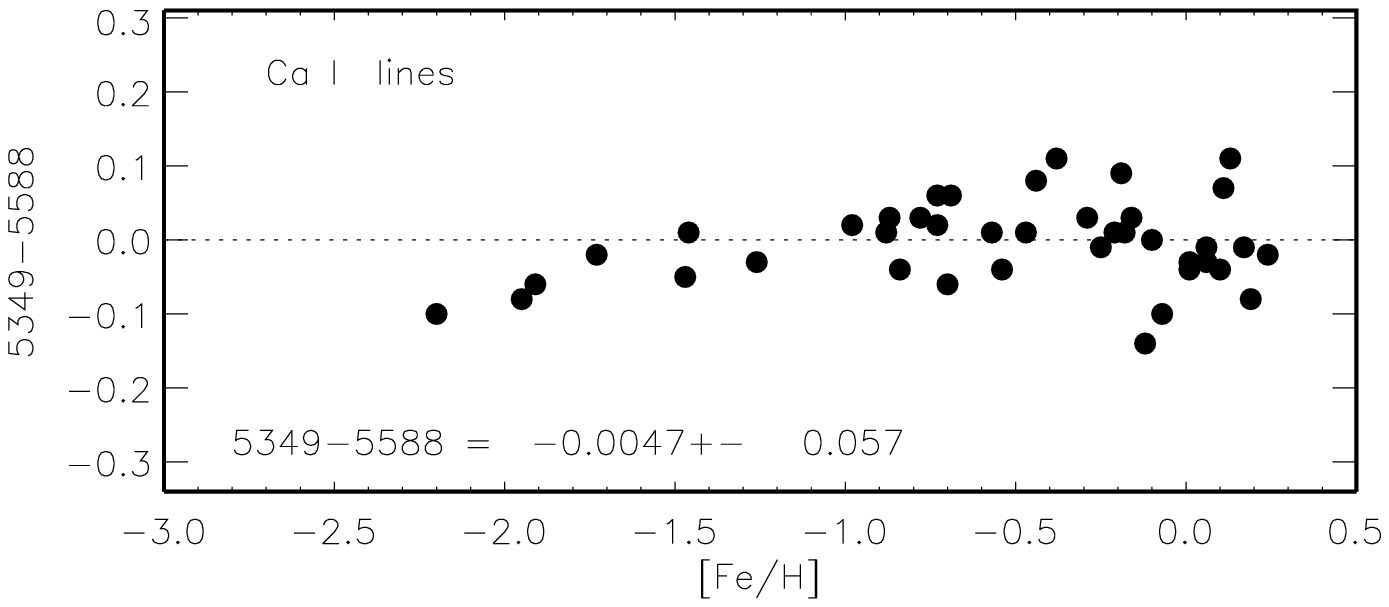}
\plottwo{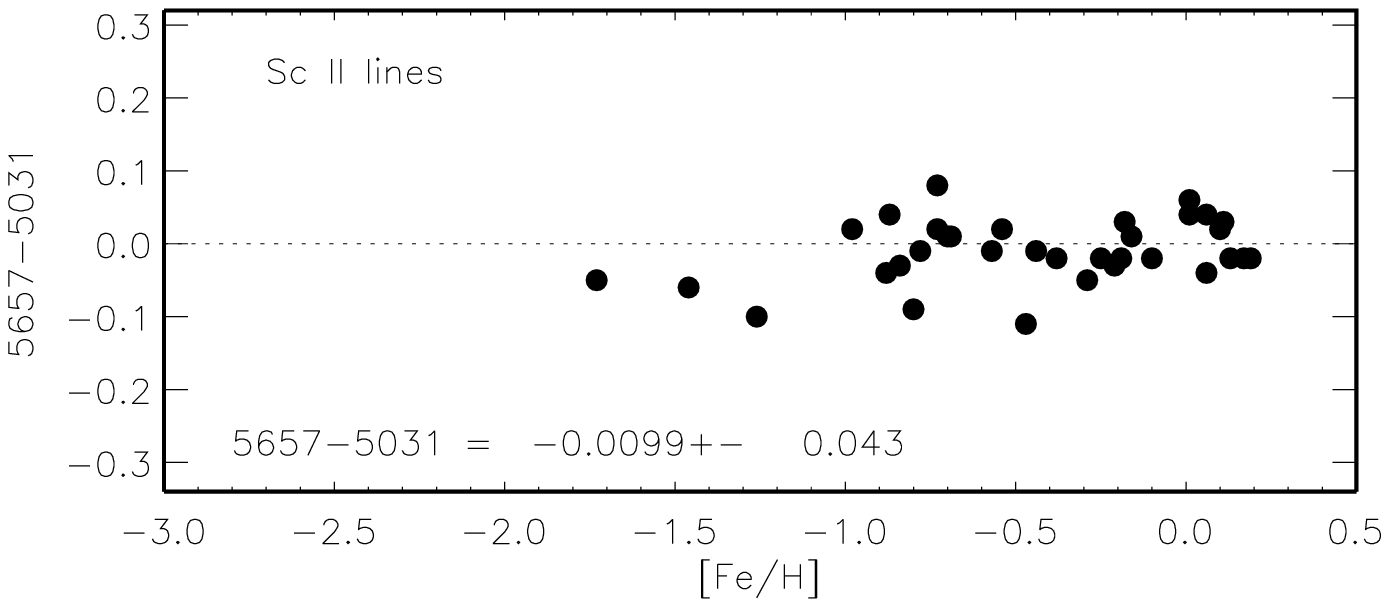}{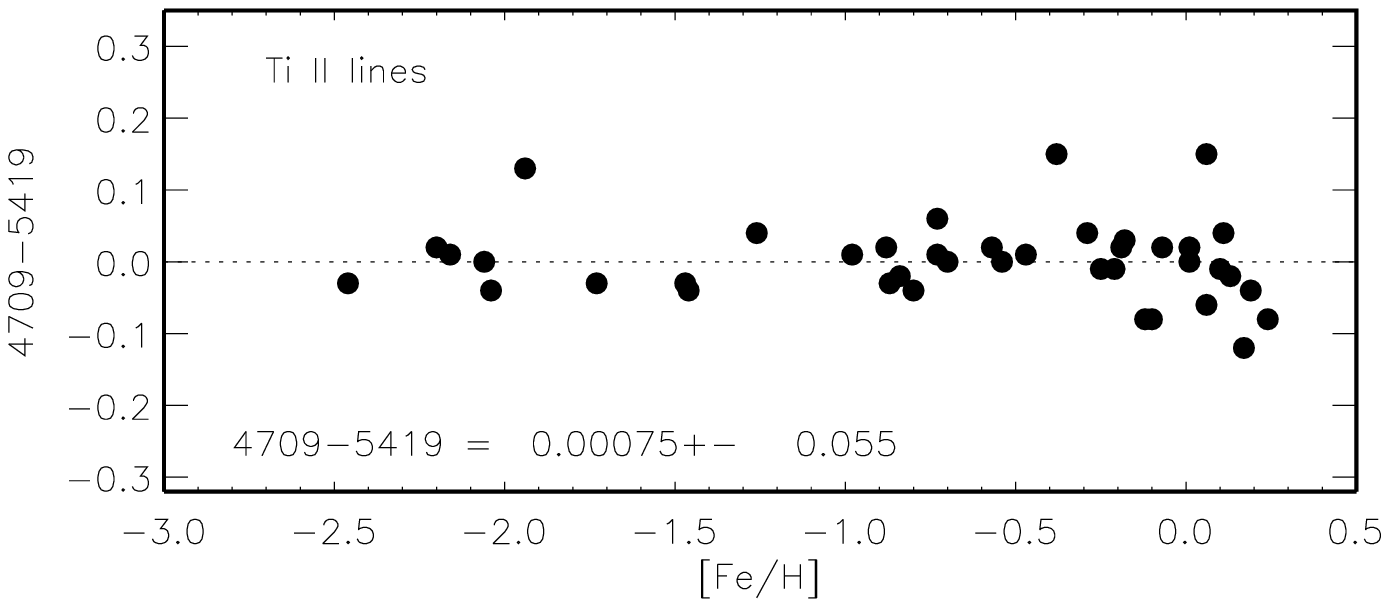}
\plottwo{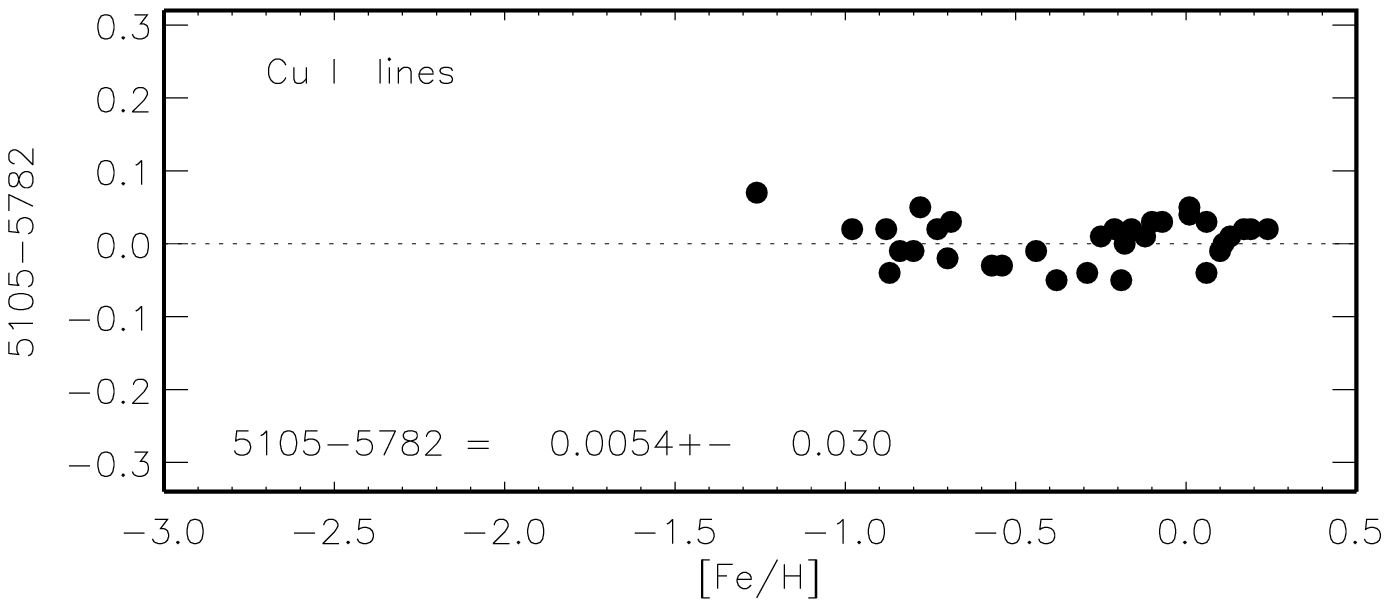}{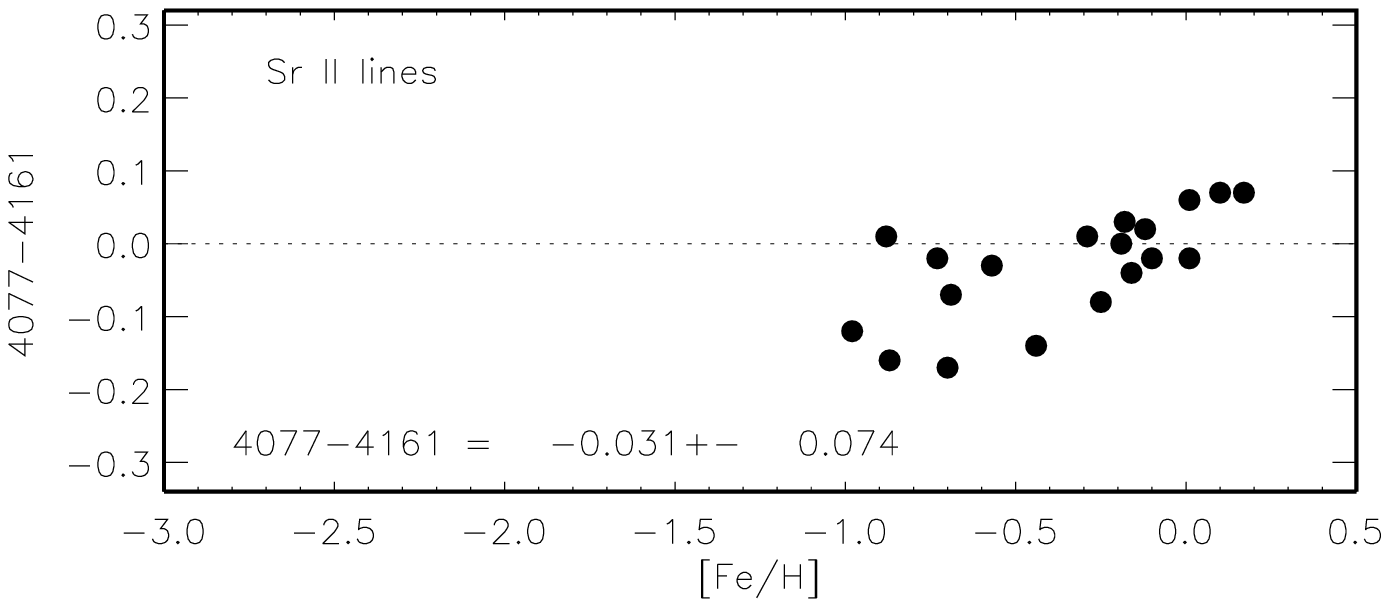}
\plottwo{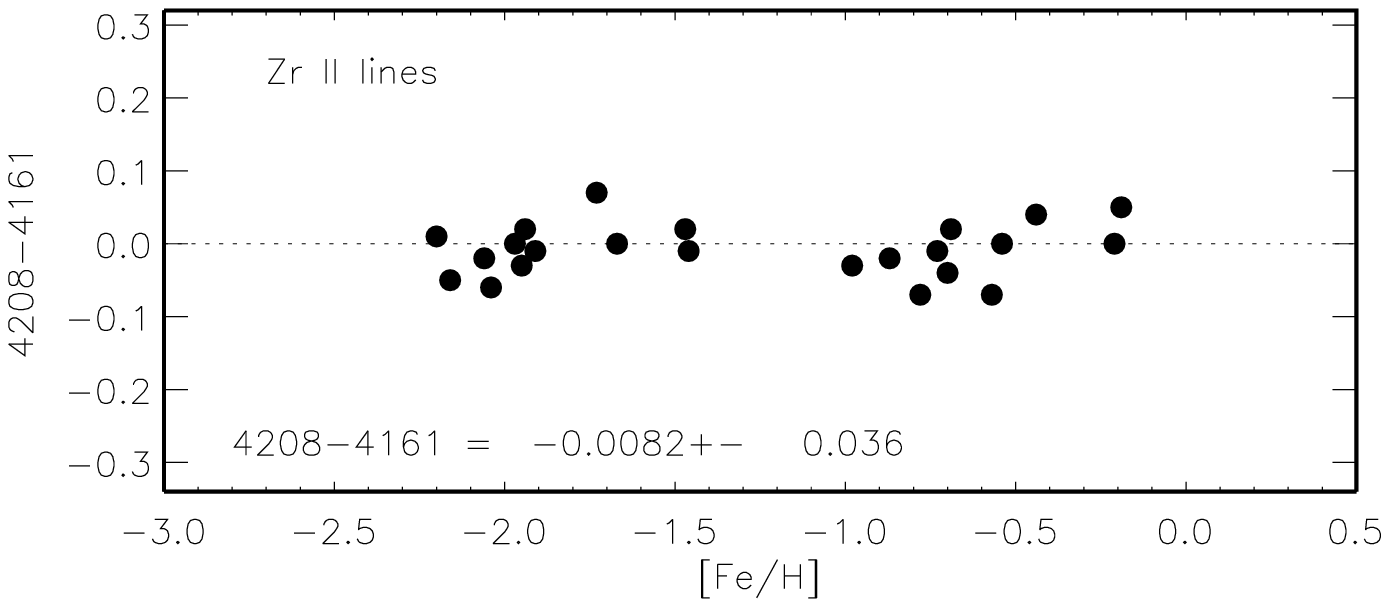}{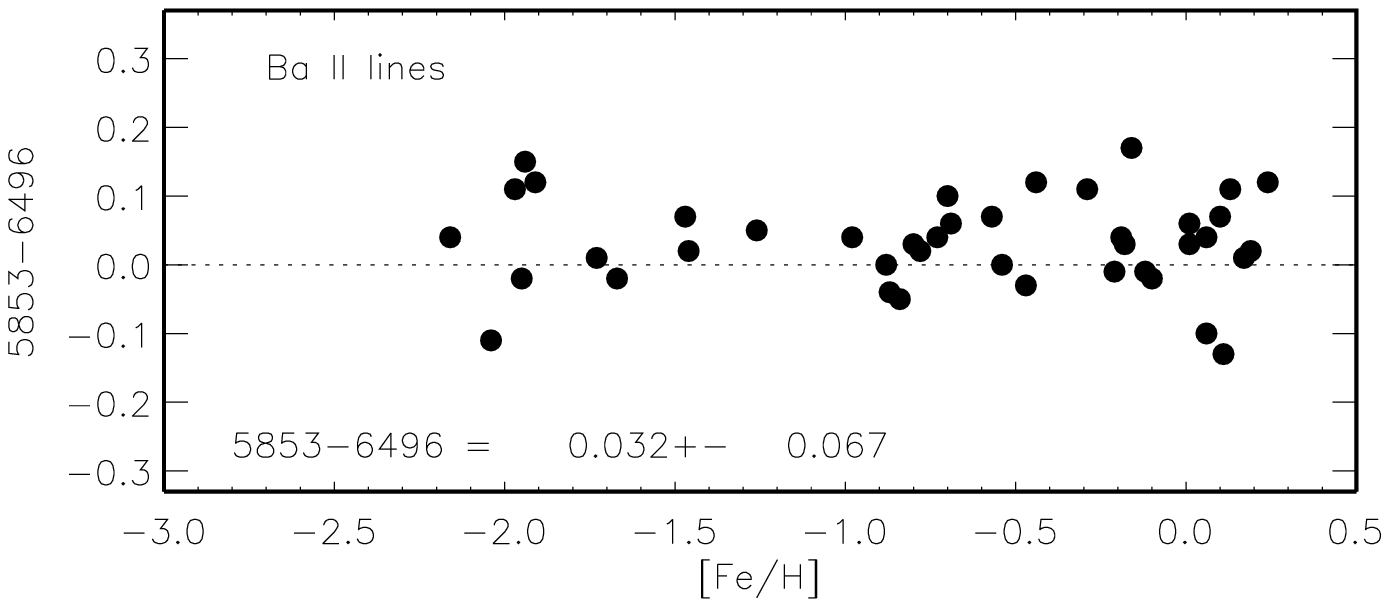}
\caption{Differences in differential NLTE abundance between individual lines. The mean difference together with the standard deviation is qouted in each panel. \label{Fig:diff2lines}}
\end{figure*}

For every species with more than one line measured the differences in differential NLTE abundance between different lines were found to be consistent within 0.05~dex, on average, for the entire stellar sample.
Figure\,\ref{Fig:diff2lines} displays the abundance differences for the selected pairs of lines. We comment below on individual chemical species.

\subsection{Notes on individual chemical species}\label{sect:species}

{\it Lithium.}
The lithium abundances are derived from the \ion{Li}{1} $6708$\,\AA\ resonance line for 42 stars. The rest nine stars do not have obvious features that are reliable enough for the lithium abundance determinations at $6708$\,\AA. The \ion{Li}{1} asymmetric profile shape (see Fig. 3 for HD\,134169) is mainly caused by the two doublet structure components, \ion{Li}{1} 6707.76\,\AA\ and \ion{Li}{1} 6707.91\,\AA. They were treated, using atomic data from Shi et al. (2007) with all the HFS components included. Taking all the blended lines in the asymmetric core region into account produces no more than 0.005\,dex change in the derived Li abundance, as showed by our test calculations for all the stars with [Fe/H] $> -0.2$ in our sample. The influence due to the presence of $^6$Li in all the halo stars are also evaluated. Assuming a meteoric isotopic ratio of 12.3 ($^7$Li/$^6$Li) gives 0.034\,dex smaller abundance for Li in average. 

{\it Carbon.}
We used three carbon abundance indicators, namely the atomic \ion{C}{1} and the molecular CH and C$_2$ lines. Suitable lines of \ion{C}{1} are located in the visible and near-IR spectral range (Table\,\ref{Tab:lines}). They all have close together an excitation energy of the lower level, $\eexc$, but different oscillator strengths, with smaller values for the visible than the near-IR lines. The \ion{C}{1} visible lines were used in the close-to-solar down to [Fe/H] = $-1.5$ stars. The IR lines are strong enough to be measured in the entire metallicity range, however, they were not used for HD~64090, BD+66$^\circ$0268, HD~24289, HD~74000, HD~108177, BD+29$^\circ$2091, and G090-003 because of strong fringes affecting the near-IR spectra. Consistent within 0.05~dex NLTE abundances from the visible and near-IR lines were found for most stars, where both groups of lines were measured, see, for example the abundance differences between \ion{C}{1} 5052\,\AA\ and 9111\,\AA\ in Fig.\,\ref{Fig:diff2lines}.

\begin{figure}
\parbox{0.5\linewidth}{\includegraphics[scale=0.35]{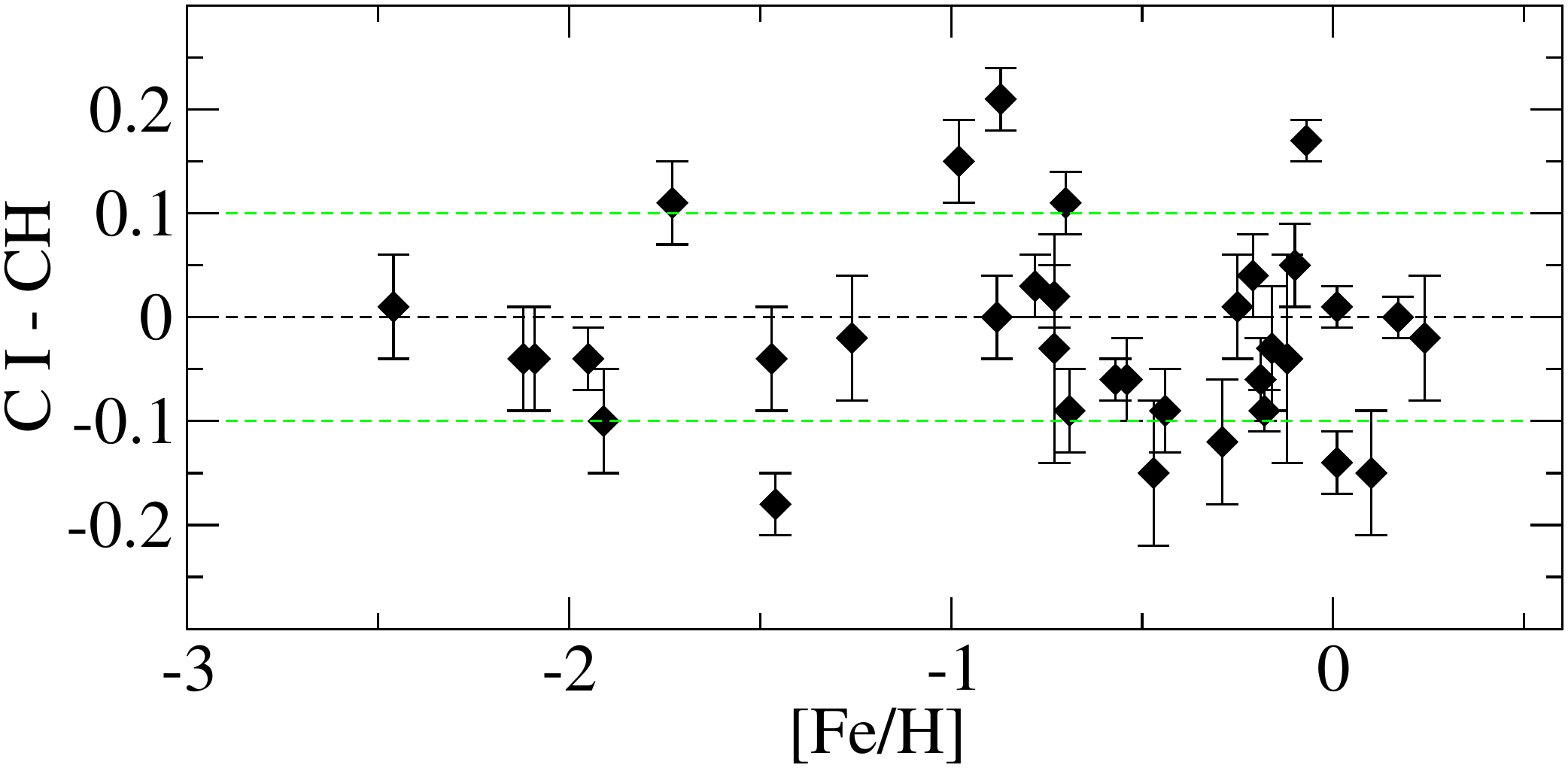}\\
\centering}
\hspace{0.1\linewidth}
\hspace{0.00\linewidth}
\hfill
\\[0ex]
\parbox{0.5\linewidth}{\includegraphics[scale=0.35]{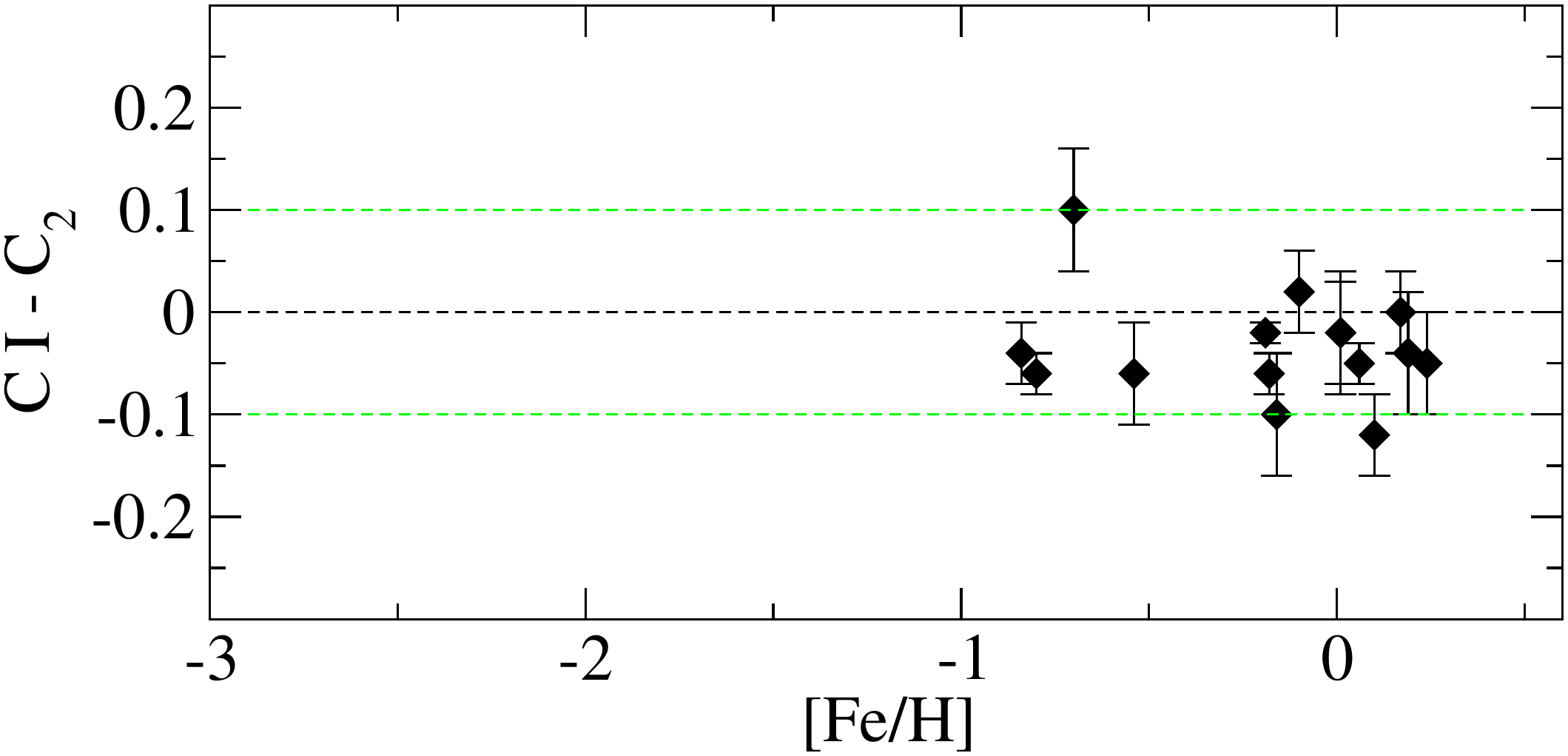}\\
\centering}
\hspace{0.1\linewidth}
\hfill
\caption{Differences in abundance derived from lines of \ion{C}{1} in NLTE and molecular lines of CH (top panel) and C$_2$ (bottom panel) in the investigated stellar sample. \label{Fig:carbon}}
\end{figure}

For the molecular CH and C$_2$ lines we use their list together with the atomic parameters from \citet{2015MNRAS.453.1619A}. The C$_2$ lines are rather weak and cannot be measured in the [Fe/H] $< -0.84$ stars, while the CH bands were detected in the entire metallicity range. An exception is the hottest stars HD~100563, BD$-13^\circ$~3442, BD$-04^\circ$~3208, and BD+24$^\circ$~1676. We obtained small abundance shift of 0.02$\pm$0.10~dex, on average,  between the molecular CH and atomic \ion{C}{1} (NLTE) lines (Fig.\,\ref{Fig:carbon}). Applying the 3D corrections from \citet{2016arXiv160507215G} to the CH G-band decreases a scatter of abundance differences only a little, resulting in CH(3D) - \ion{C}{1} = $0.01\pm0.085$. The mean abundance difference between \ion{C}{1} and C$_2$ amounts to $-0.04\pm0.05$.

Despite the fact the atomic and molecular lines give consistent results, we prefer to employ the \ion{C}{1}-based abundances for final carbon abundances. The CH-based abundances were employed for the stars with [Fe/H] $< -1$, with no \ion{C}{1} line measured.

{\it Oxygen.}
Determination of the element abundances from the \ion{O}{1} IR lines takes advantage of using the spectra with removed fringes, as described in Sect.\,\ref{sect:sample}. Employing only the visible \ion{O}{1} 6158\,\AA\ line would restricted our O abundance analysis by the [Fe/H] = $-0.5$ stars, where this line can be measured. All the \ion{O}{1} lines give consistent within the error bars abundances, as shown in Fig.\,\ref{Fig:diff2lines} for \ion{O}{1} 7771\,\AA\ and 6158\,\AA.

{\it Sodium}
abundances were determined using six \ion{Na}{1} lines including the strong \ion{Na}{1}\,{\it D} lines, because for some very metal-poor stars, only the \ion{Na}{1}\,{\it D} lines could be used for abundance determination. However, for stars with [Fe/H] $> -0.5$, the \ion{Na}{1}\,{\it D} lines ($5889$\,\AA\ and $5895$\,\AA) were not used in calculating the final average abundances.

{\it Magnesium}
abundances were determined using five neutral Mg lines as shown in Table \ref{tab:nlte}. The strong \ion{Mg}{1}\,{\it b} lines were not employed for abundance determination.

{\it Aluminum.}
Determining Al abundances is very challenging for our sample stars. We used seven lines of \ion{Al}{1}. However, for most stars their Al abundance is based on either the resonance line, \ion{Al}{1} 3961\,\AA, or the subordinate lines in the red and IR spectral region. The resonance line is strong enough to be detected in all the sample stars. However, this line falls in the wing of a strong \ion{Ca}{2} 3968\,\AA\ line, which makes the normalization very difficult for the close-to-solar metallicity stars. The other six lines in the near infrared in the Shane/Hamilton spectra suffer from the fringing effect. For HD~59374, HD~59984, and HD~134169 we used their high-quality FOCES spectra and could measure five to seven lines of \ion{Al}{1}. For each star the LTE analysis obtained a 0.25-0.45~dex lower element abundance from the resonance line compared with that from the subordinate lines. Thanks to implementing quantum-mechanical data on \ion{Al}{1}+\ion{H}{1} collisions by \citet{Belyaev2013_Al} in the SE calculations \citep{mash_al2016}, an abundance discrepancy between different lines was largely removed in NLTE.

{\it Silicon.}
We applied 11 \ion{Si}{1} and two \ion{Si}{2} lines to derive the Si abundances. Two strong ultraviolet \ion{Si}{1} lines, 3905 and 4102\,\AA, were not used for the stars with [Fe/H]$>-$0.5 because of the saturation, while for the six most MP stars, only these two lines can be used for the abundance determination.

It is worth noting, lines of \ion{Si}{1} and \ion{Si}{2} lead to consistent NLTE abundances, with the mean difference (\ion{Si}{1} - \ion{Si}{2}) = 0.00$\pm$0.05~dex for our 32 program stars. The final results are the average abundances of \ion{Si}{1} and \ion{Si}{2}.

{\it Potassium.}
Abundances of K were obtained by using the \ion{K}{1} $7699$\,\AA\ line, whereas $7664$\,\AA\ is heavily blended with the telluric O$_2$ lines for most sample stars. The potassium lines are also affected by very strong fringing effects. So, we could not determine potassium abundances for most very metal-poor stars with [Fe/H]$<-2.0$.

{\it Calcium.}
Among investigated elements in this paper, calcium covers many visible lines, and 21 \ion{Ca}{1} lines were employed in our abundance determination. The \ion{Ca}{1} resonance line at 4226\,\AA\ is very strong in all our program stars, and it was not used in the abundance analysis. For stars with [Fe/H] $>-$0.5~dex we also did not include \ion{Ca}{1} 6162 and 6439\,\AA, when calculating the average Ca abundance, because of their saturation.

{\it Scandium.} 
Because the \ion{Sc}{1} lines in metal-poor stars are extremely week, the nine lines of \ion{Sc}{2} are employed in the abundance determinations, although the number of lines are decreased to 1-3 for very metal-poor stars. Again, the NLTE corrections for all \ion{Sc}{2} lines are small.

{\it Titanium.} 
For final Ti abundances we prefer to employ lines of \ion{Ti}{2} because of small NLTE effects. Indeed, $\Delta_{\rm NLTE} \le$ 0.02~dex, in absolute value, everywhere for the \ion{Ti}{2} lines in our calculations.

{\it Copper.}
We applied three \ion{Cu}{1} lines for determinations of copper abundances, namely $5105$\,\AA, $5218$\,\AA, and $5782$\,\AA, which are the same as that in our very recent study \citep{2016A&A...585A.102Y}. Among them, $5105$\,\AA\ is the strongest and least blended (the only weak \ion{Fe}{2} line at the very blue wing) line, and thus is a good indicator of the copper abundance. The $5218$\,\AA\ line is weak, and its blue wing is blended by a \ion{Fe}{1} line, which usually has a comparable equivalent width with the \ion{Cu}{1} line. We thus took the two lines together into the consideration during the line profile fitting. The $5782$\,\AA\ line is also blended by several weak lines (i.e. \ion{Cr}{1}, \ion{Cr}{2}, \ion{Fe}{1}, \ion{Fe}{2}). Taking all the lines into account gives a consistent copper abundance with the other two lines. The Cu abundance difference between including and ignoring the blended lines near $5782$\,\AA\ is $\sim 0.02$~dex, on average. The \ion{Cu}{1} lines are weakened towards lower metallicity, and no copper abundance can be derived from these three lines for stars with [Fe/H] $< -1.5$.

{\it Strontium.}
Three lines of \ion{Sr}{2} were employed in the abundance determinations. The subordinate line at 4161~\AA\ was measured in the [Fe/H] $\ge -0.98$ stars, and it gives the Sr abundance in line with that from \ion{Sr}{2} 4077~\AA, with a mean difference of 0.03$\pm$0.07~dex for 19 common stars (Fig.\,\ref{Fig:diff2lines}). The \ion{Sr}{2} 4215.5~\AA\ line is heavily blended by \ion{Fe}{1} 4215.426~\AA\ and by a few CN molecular lines in the far blue and red line wings, and it was not used for the [Fe/H] $> -1$ stars. In the more MP stars, the two resonance lines give consitent abundances, with \ion{Sr}{2} 4077 -- \ion{Sr}{2} 4215 = $-0.01\pm0.08$ for 17 common stars.

{\it Zirconium.}
Only three lines of \ion{Zr}{2} are suitable for stellar abundance determinations. The \ion{Zr}{2} 4208.98\,\AA\ line is strong enough to be measured in the entire range of metallicity. An exception is our most MP star BD~$-13^\circ$~3442 ([Fe/H] = $-2.62$), where no \ion{Zr}{2} line was detected. To account for the blending Cr~I 4208.95\,\AA\ line ($\eexc$ = 3.85~eV, log~$gf = -0.528$ according to VALD) correctly, we controled the chromium LTE abundance using a nearby line of Cr~I  4209.365\,\AA, with $\eexc$ = 3.85~eV and log~$gf = -0.263$ (VALD). In contrast, \ion{Zr}{2} 5112\,\AA\ is unblended, but weak and can only be measured in the [Fe/H] $> -0.88$ stars. Another line, \ion{Zr}{2} 4161.21\,\AA, is located in red wing of \ion{Fe}{1} 4161.08\,\AA. The blending effect reduces toward lower metallicity, and \ion{Zr}{2} 4161\,\AA\ provides a reliable abundance at [Fe/H] $< -0.19$. We obtained consistent abundances from all the lines, with a mean abundance difference of $-0.01\pm0.04$~dex between \ion{Zr}{2} 4208\,\AA\ and 4161\,\AA\ and of 0.00$\pm$0.06~dex between \ion{Zr}{2} 4208\,\AA\ and 5112\,\AA.

{\it Barium.}
For the majority of stars their barium abundance was determined from the \ion{Ba}{2} subordinate lines, which are almost free of HFS effects. According to our estimate for \ion{Ba}{2} 6497\,{\AA}, neglecting HFS makes a difference in the solar abundance of no more than 0.01\,dex. We avoided employing the \ion{Ba}{2} 4554~\AA\ and 4934~\AA\ resonance lines for the [Fe/H] $> -2$ stars, where they are saturated and the derived element abundance depends on the Ba isotope mixture adopted in the calculations because the lines are strongly HFS affected. In the three most MP stars, HD~140283, BD~+24$^\circ$~1676 and BD~$-13^\circ$~3442, the subordinate lines of \ion{Ba}{2} cannot be extracted from noise, and the barium abundance given in Table~\ref{Tab:AbundanceSummary} was determined from the resonance lines. It is worth noting, \ion{Ba}{2} 4554~\AA\ and 4934~\AA\ are rather weak in each of these stars, and a change in the Ba abundance derived from these lines does not exceed few hundredth when moving from the solar mixture \iso{134}{Ba} : \iso{135}{Ba} : \iso{136}{Ba} : \iso{137}{Ba} : \iso{138}{Ba} = 2.4 : 6.6 : 7.9 : 11.2 : 71.7 \citep{Lodders2009} to the r-process one \iso{135}{Ba} : \iso{137}{Ba} : \iso{138}{Ba} = 24 : 22 : 54 \citep{Travaglio1999}. For example, \ion{Ba}{2} 4554~\AA\ in HD~140283 has an equivalent width ($EW$) of 20\,m\AA, and the abundance shift between using the solar and the r-process Ba isotope mixture amounts to 0.02~dex.

{\it Europium.}
Three lines of \ion{Eu}{2} were employed in the abundance determinations. The subordinate line at 6645~\AA\ was measured in the [Fe/H] $\ge -0.78$ stars, and it appears to give systematically higher abundance compared with that from \ion{Eu}{2} 4129~\AA, with a mean difference of 0.14$\pm$0.13~dex for 17 common stars. This line was nowhere used to obtain the final Eu abundance. The resonance line of \ion{Eu}{2} at 4204.878-4205.117~\AA\ is blended by numerous metal lines, which cannot be taken into account correctly even using the synthetic spectrum approach. As a result, the abundance difference between \ion{Eu}{2} 4129~\AA\ and 4205~\AA\ was obtained to be $-0.06\pm0.06$~dex for 37 common stars. We avoided using \ion{Eu}{2} 4205~\AA\ in a determination of the final Eu abundance. An exception is the six stars with [Fe/H] between $-1.73$ and $-2.20$, where \ion{Eu}{2} 4129~\AA\ could not be measured due to either strong blending by the SiH 4129.609, 4129.666, 4129.774~\AA\ lines in the cool dwarfs HD~64090 and BD~+66$^\circ$~0268 or a bad quality of the observed spectra. No line of \ion{Eu}{2} can be extracted from noise in our more MP stars.

\subsection{Uncertainties in derived abundances}

We choose a mildly metal-deficient star HD~134169 ([Fe/H] = $-0.78$) to perform a detailed error analysis and to estimate the uncertainties in the abundance measurements for all the investigated species. Stochastic errors ($\sigma_{obs}$) caused by random uncertainties in the continuum placement, line profile fitting, and $gf$-values, are represented by a dispersion in the measurements of multiple lines around the mean, as given in Table~\ref{Tab:AbundanceSummary} when $N \ge 2$ lines of an element are observed. Systematic uncertainties include those that exist in the adopted stellar parameters. Table~\ref{tab:uncertainty} summarizes the various sources of
uncertainties. For each species we choose a representative line indicated in Col.~3 to calculate an abundance shift due to a change of $-70$~K in $\Teff$, +0.07~dex in log~g, and $-0.1$~\kms\ in $\vt$. The quantity $\Delta(T,g,\xi)$ listed in Col.~7 is the total impact of varying each of the three parameters, computed as the quadratic sum of Cols.~4, 5, and 6.

\begin{deluxetable}{lccrrcc}
\tabletypesize{\scriptsize}
\tablecaption{Error budget for elements in HD~134169. \label{tab:uncertainty}}
\tablehead{
\colhead{Atom} & \colhead{$\sigma_{obs}$} & \colhead{$\lambda$} & \colhead{$\Delta T$} & \colhead{$\Delta \log g$} & \colhead{$\Delta \xi$} & \colhead{$\Delta$} \\
               &                          & \colhead{(\AA)}     & \colhead{$-80$\,K} & \colhead{0.07} & \colhead{$-0.1${\scriptsize \kms}} & \colhead{($T,g,\xi$)} \\
\colhead{(1)} & \colhead{(2)} &\colhead{(3)} &\colhead{(4)} &\colhead{(5)} &\colhead{(6)} & \colhead{(7)}
}
\startdata
\ion{Li}{1} &      & 6707 & $-0.08$ & $-0.01$ &  0.01 &  0.08 \\
\ion{C}{1}  & 0.01 & 5380 &  0.06 &  0.04 &  0.00 &  0.07 \\
\ion{O}{1}  & 0.01 & 7771 &  0.06 &  0.02 &  0.00 &  0.06 \\
\ion{Na}{1} & 0.04 & 5688 & $-0.02$ &  0.02 &  0.00 &  0.03 \\
\ion{Mg}{1} & 0.04 & 5528 & $-0.07$ & $-0.04$ &  0.01 &  0.08 \\
\ion{Al}{1} & 0.10 & 8772 & $-0.05$ & $-0.02$ &  0.00 &  0.05 \\
\ion{Si}{1} & 0.05 & 6145 &  0.00 &  0.03 &  0.00 &  0.03 \\
\ion{K}{1}  &      & 7698 &  0.02 &  0.02 &  0.03 &  0.04 \\
\ion{Ca}{1} & 0.05 & 5588 & $-0.02$ &  0.01 &  0.03 &  0.04 \\
\ion{Sc}{2} & 0.03 & 5526 & $-0.11$ & $-0.07$ &  0.02 &  0.13 \\
\ion{Ti}{2} & 0.03 & 5336 & $-0.01$ &  0.04 &  0.04 &  0.06 \\
\ion{Cu}{1} & 0.03 & 5218 &  0.01 &  0.07 &  0.00 &  0.07 \\
\ion{Sr}{2} & 0.06 & 4077 & $-0.04$ &  0.01 &  0.01 &  0.04 \\
\ion{Zr}{2} & 0.05 & 4208 & $-0.01$ &  0.04 &  0.01 &  0.04 \\
\ion{Ba}{2} & 0.01 & 6496 &  0.01 &  0.04 &  0.06 &  0.07 \\
\ion{Eu}{2} &      & 4129 & $-0.02$ &  0.04 &  0.00 &  0.04 \\
\enddata
\end{deluxetable}

\subsection{Notes on individual stars}\label{sect:notes}

{\it Planet-host stars.}
In our sample, five stars have been reported to harbor one or more planets according the catalog listing of The Extrasolar Planets Encyclopaedia\footnote{http://exoplanet.eu/catalog/}. Three of them, HD~30562, HD~82943, and HD~89744, are metal rich stars with [Fe/H] $\simeq$ 0.1-0.2~dex, while HD~115617 and HD~142091 have solar or slightly subsolar metallicities. We did not detect any special characteristics in element abundance ratios for these planet host stars.

{\it Stars with the thin-disc kinematics, but the thick-disk chemistry.} 
Our two most MP stars with a thin-disc kinematics, HD~105755 ([Fe/H] = $-0.73$) and HD~134169 ([Fe/H] = $-0.78$), reveal $\alpha$- and r-process enhancements typical of the thick disk stars, with [Mg/Fe] = 0.29 and 0.34 and [Eu/Ba] = 0.50 and 0.51.

\begin{figure}
\epsscale{1.0}
\plotone{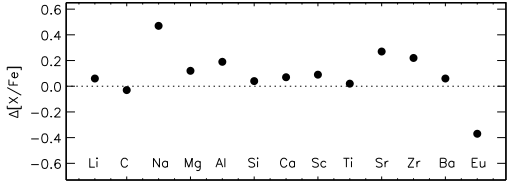}
\caption{Differences in element abundance ratios, [X/Fe], between HD~74000 and HD~24289. All the results are from the NLTE calculations.  \label{Fig:hd74000}}
\end{figure}

{\it Halo star HD~74000 ([Fe/H] = $-1.97$)}
reveals typical abundances of the $\alpha$-process elements, but overabundance of sodium and underabundance of europium
compared with the stars of close metallicity (Fig.\,\ref{Fig:alpha}, \ref{Fig:odd}, \ref{Fig:heavy}, \ref{Fig:ratios2}, and \ref{Fig:ratios3}). We selected HD~24289 ([Fe/H] = $-1.94$) to show a difference in element abundance pattern between HD~74000 and the [Fe/H] $\simeq -2$ stars (Fig.\,\ref{Fig:hd74000}). Its peculiar [Na/Fe], [Na/Mg], [Eu/Fe], and [Eu/Ba] ratios were reported earlier by \citet{mg_c6} and \citet{2003A&A...397..275M}. This star is also known for its extreme nitrogen overabundance, with [N/Fe] = 0.9 \citep{1987PASP...99..335C}, and probably not representative of a standard evolutionary scenario for our Galaxy.

{\it Halo star G090-003 ([Fe/H] = $-2.04$).}
We draw an attention to high abundances of Na and Al in this star. The \ion{Na}{1} resonance lines in its observed spectrum, both are affected by the emissions of, probably, the telluric origin. Since a quality of the spectrum of G090-003 is very good, we could measure the Na abundance from \ion{Na}{1} 5688\,\AA. It gives a nearly 0.2~dex lower abundance compared with that from the resonance lines, however, still large. Although the Na and Al abundances in G090-003 are higher than that in stars of close metallicity, their ratio is close to solar one, like in other stars (Fig.\,\ref{Fig:ratios2}).

\section{The Galactic chemical evolution}\label{sect:evolution}

\subsection{Stellar abundance trends}

\begin{figure}
\epsscale{1.0}
\plotone{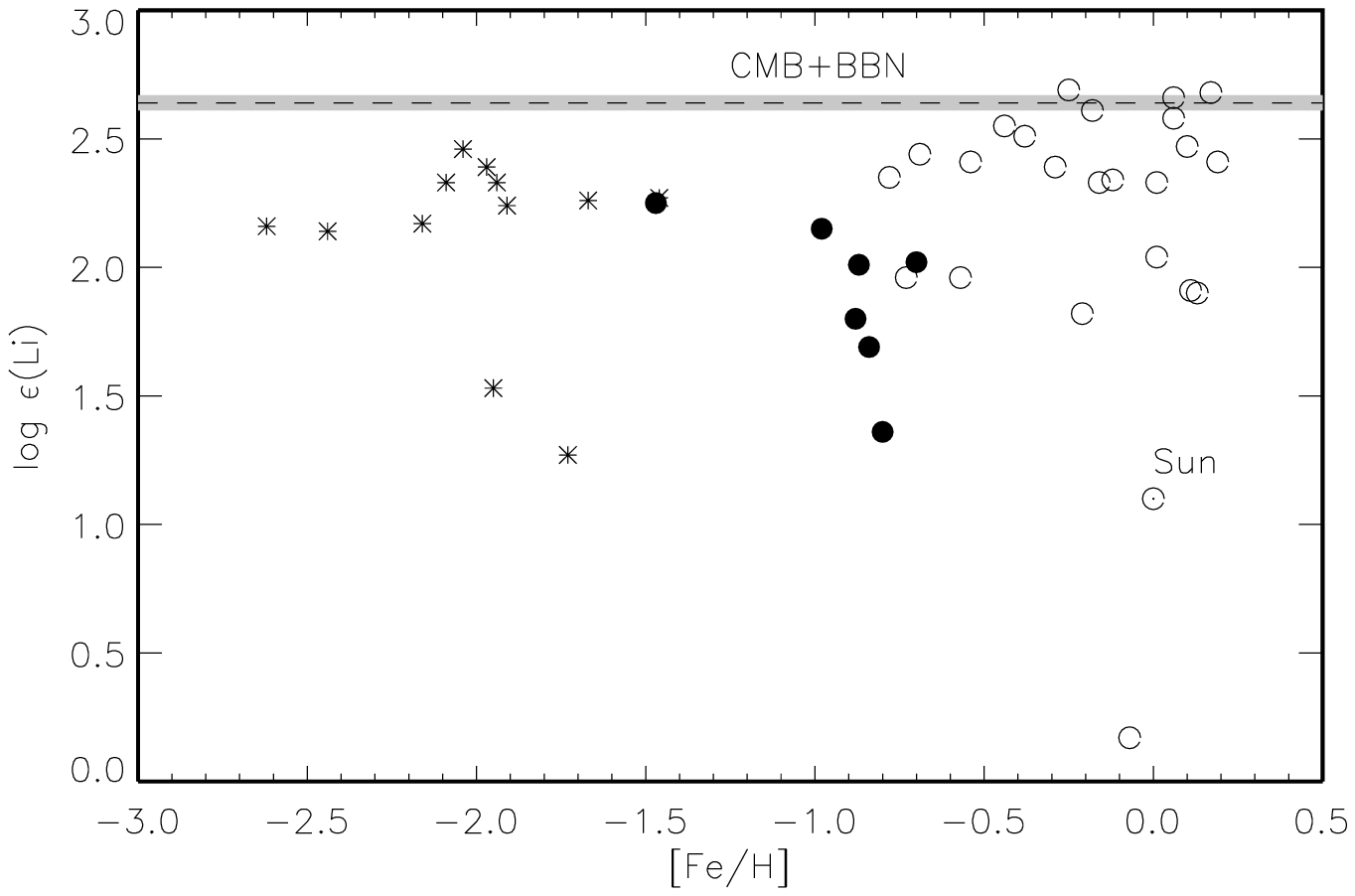}
\plotone{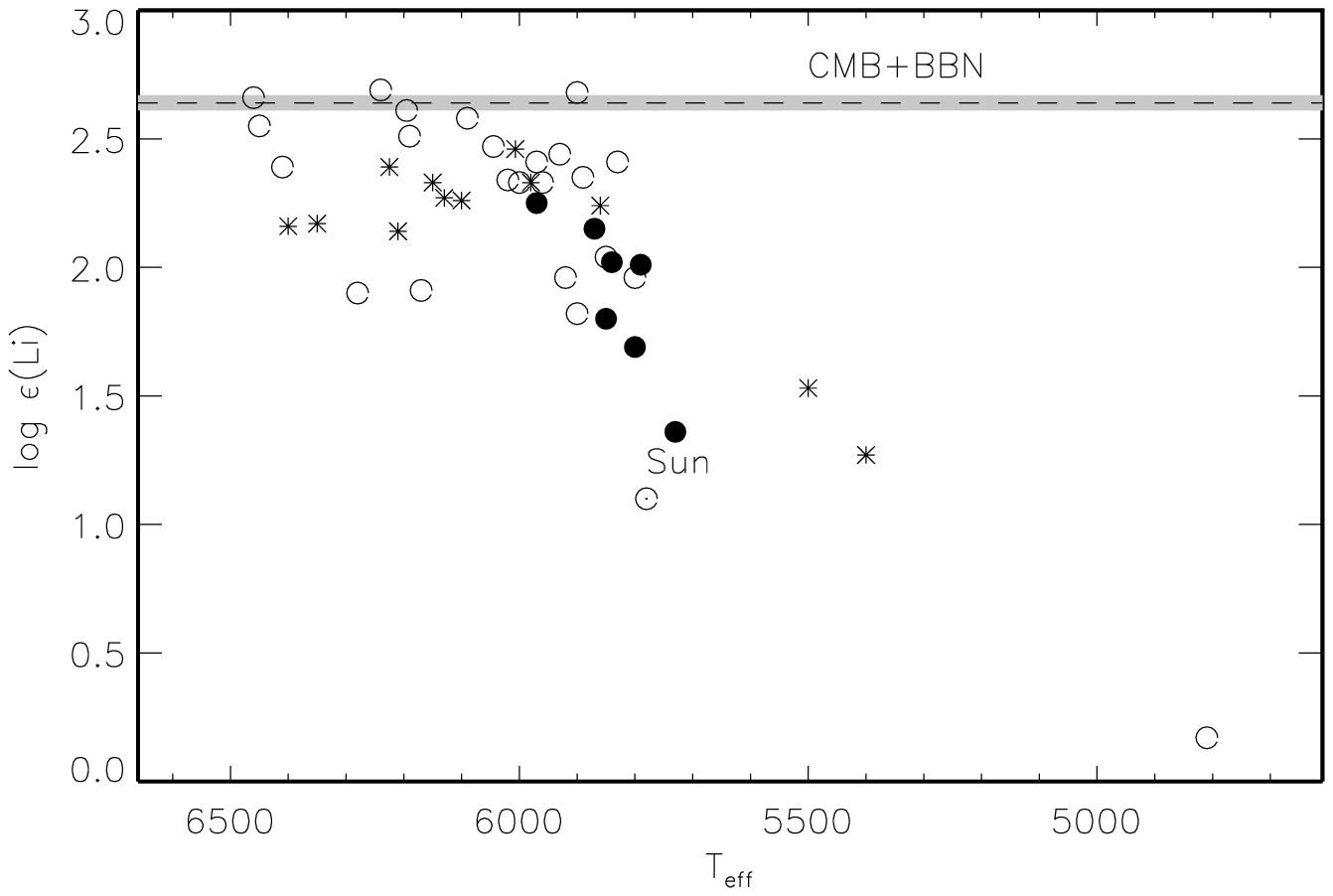}
\caption{Stellar NLTE abundances of Li as a function of metallicity (top panel) and effective temperature (bottom panel). Different symbols correspond to different stellar populations, namely the thin disc (open circles), the thick disc (filled circles), and the halo (asterisks). The dashed line and shaded area show the predicted primordial lithium abundance, $\eps{Li}$(CMB+BBN) = 2.64$\pm$0.03 \citep{2007ApJS..170..377S}.  \label{Fig:li}}
\end{figure}

Stellar abundances for different elements, classified from their nucleosynthesis histories, for a large sample of stars with different metallicities play a key role in the study of the chemical evolution of these elements themselves, their origins and the chemical evolution of the Galaxy.
This study is of particular importance because it presents, for the first time, abundances of many elements in a broad metallicity range that were homogeneously derived from the NLTE analysis. Among all the investigated species lithium holds a specific position, because it is of primordial origin and considered a key diagnostic to test and constrain our description of the early Galaxy, of stellar interiors and evolution, and of spallation physics. Elements beyond carbon are of stellar origin. Their abundances suffer from the so-called even-odd effect, which gives rise to different yields for different elements despite of their same nucleosynthesis path. Therefore, in the C to Ti range we group the even-nuclear charge ($Z$) elements and the odd-$Z$ elements. Elements beyond the iron group are believed to be produced in the neutron-capture nuclear reactions. We discuss separately Sr to Eu and copper, because for the latter its production mechanisms are still debated.

{\it Lithium.}
We found that Li abundances of the warm ($\Teff \ge$ 5800~K) halo stars are surrounding the well defined plateau at $\eps{Li}$ = 2.2 (Fig.\,\ref{Fig:li}). This is in line with the earlier discovery of a remarkably flat and constant Li abundance among Galactic halo dwarf stars spanning a wide range of effective temperatures and metallicities --- the so-called Spite plateau \citep{1982Natur.297..483S}. Careful re-analysis of the literature data led \citet{2005A&A...442..961C} to deduce $\eps{Li}$ = 2.177$\pm$0.071 for the [Fe/H] $\le -1.5$ stars with $\Teff \ge$ 5700~K. With the baryon-to-photon ratio defined accurately by the Wilkinson Microwave Anisotropy Probe (WMAP), standard Big Bang Nucleosynthesis (BBN) predicts a
primordial lithium abundance of $\eps{Li}$ = 2.64$\pm$0.03 \citep{2007ApJS..170..377S} to 2.72$\pm$0.06 \citep{2012ApJ...744..158C}. Several physical mechanisms were proposed to reduce the Li abundance at the surface of halo stars compared with the pristine one \citep[see, e.g.][]{2005A&A...442..961C,2006Natur.442..657K,2015MNRAS.452.3256F}, however, the theoretical models see considerable difficulties to reconcile a non negligible depletion of lithium with both the flatness and the small dispersion along the Spite plateau. It is worth noting, the NLTE corrections for the only line, \ion{Li}{1} 6707\,\AA, used in the abundance determinations are mostly negative and small in absolute value (Fig.\,\ref{Fig:dnlte}). Our two coolest ($\Teff \le$ 5500~K) halo stars have a more than 0.7~dex lower Li abundance compared with the Spite plateau. Similar temperature dependence was already noticed by \citet{2005A&A...442..961C}, and they suggested that ``the most massive of the halo stars have had a slightly different Li history than their less massive contemporaries''.

The Li abundances of the thick disk stars are very similar to that of the halo stars of close temperature. This puts strong constraints on the possible Li depletion mechanism(s).
In the thin disk stars $\eps{Li}$ varies between 1.8 and 2.7, and a temperature dependence is not evident. An outlier is the cool giant HD~142091 ($\Teff$ = 4810~K, log~g = 3.12), where $\eps{Li}$ = 0.17, in line with the star's evolutionary status. We confirm an existence of the Li-desert at $\Teff \simeq 6000$~K and $\eps{Li} \simeq 1.8$, as found by \citet{2012ApJ...756...46R}.

\begin{figure*}
\epsscale{1.0}
\plottwo{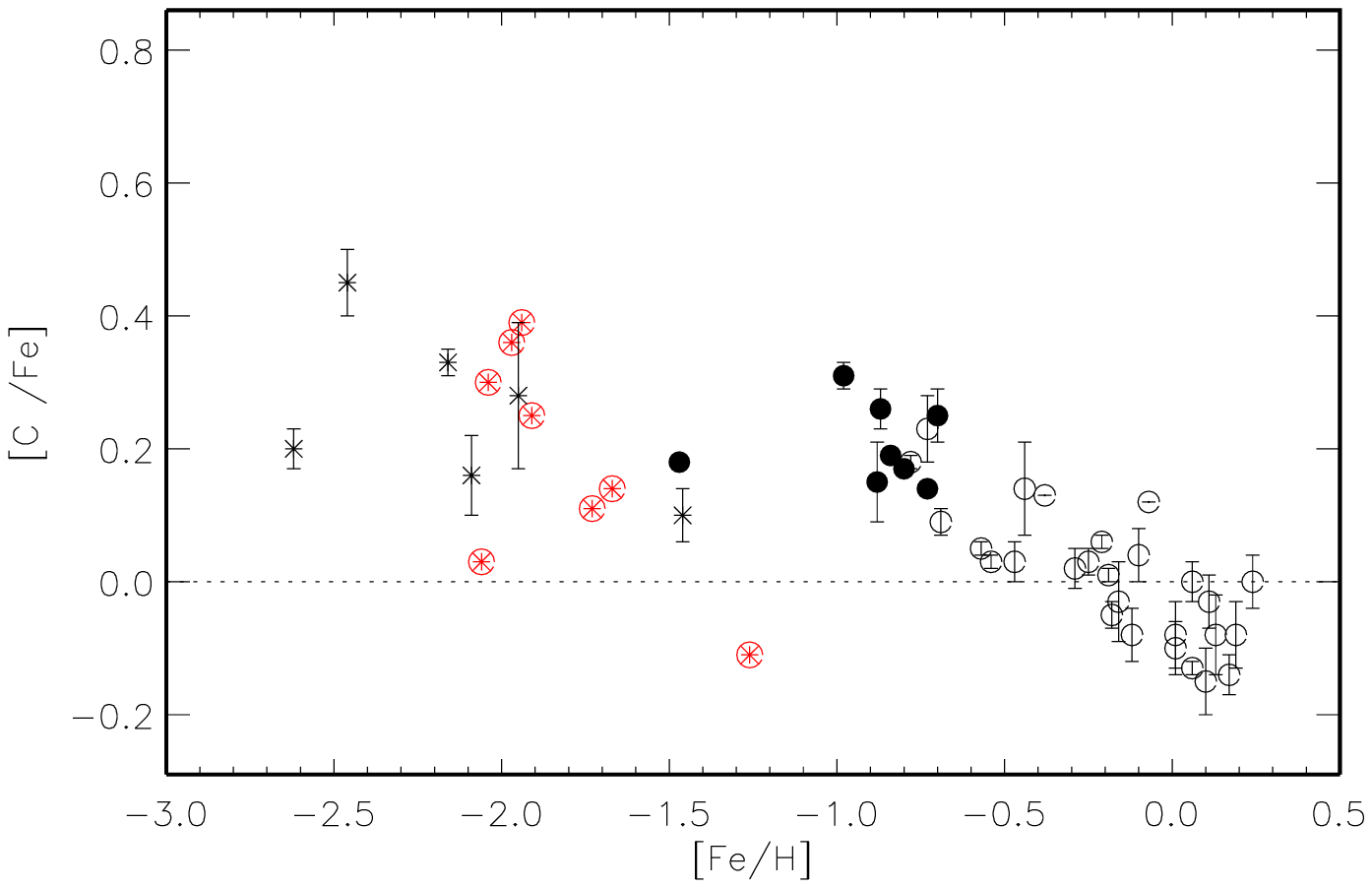}{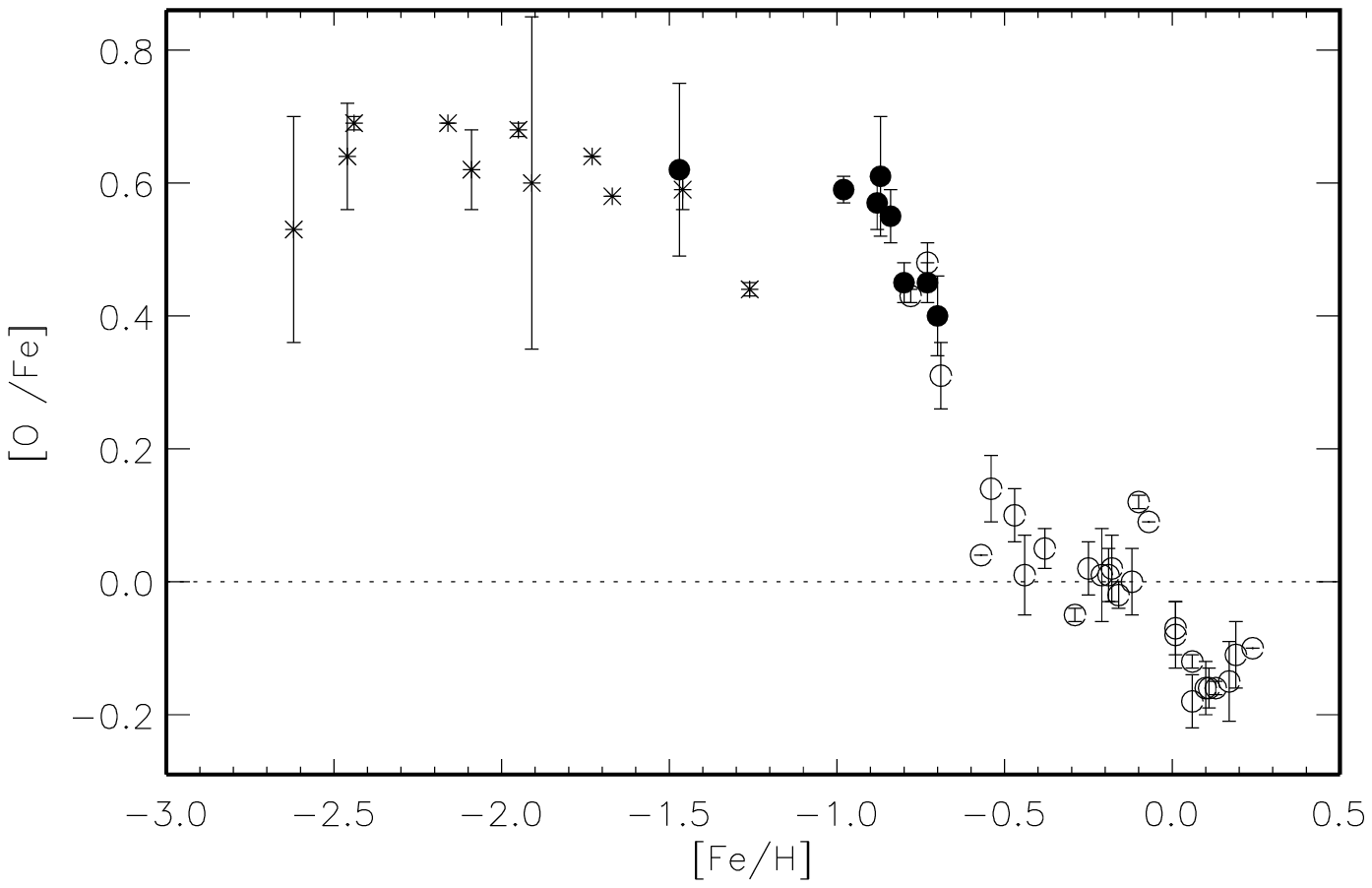}
\plottwo{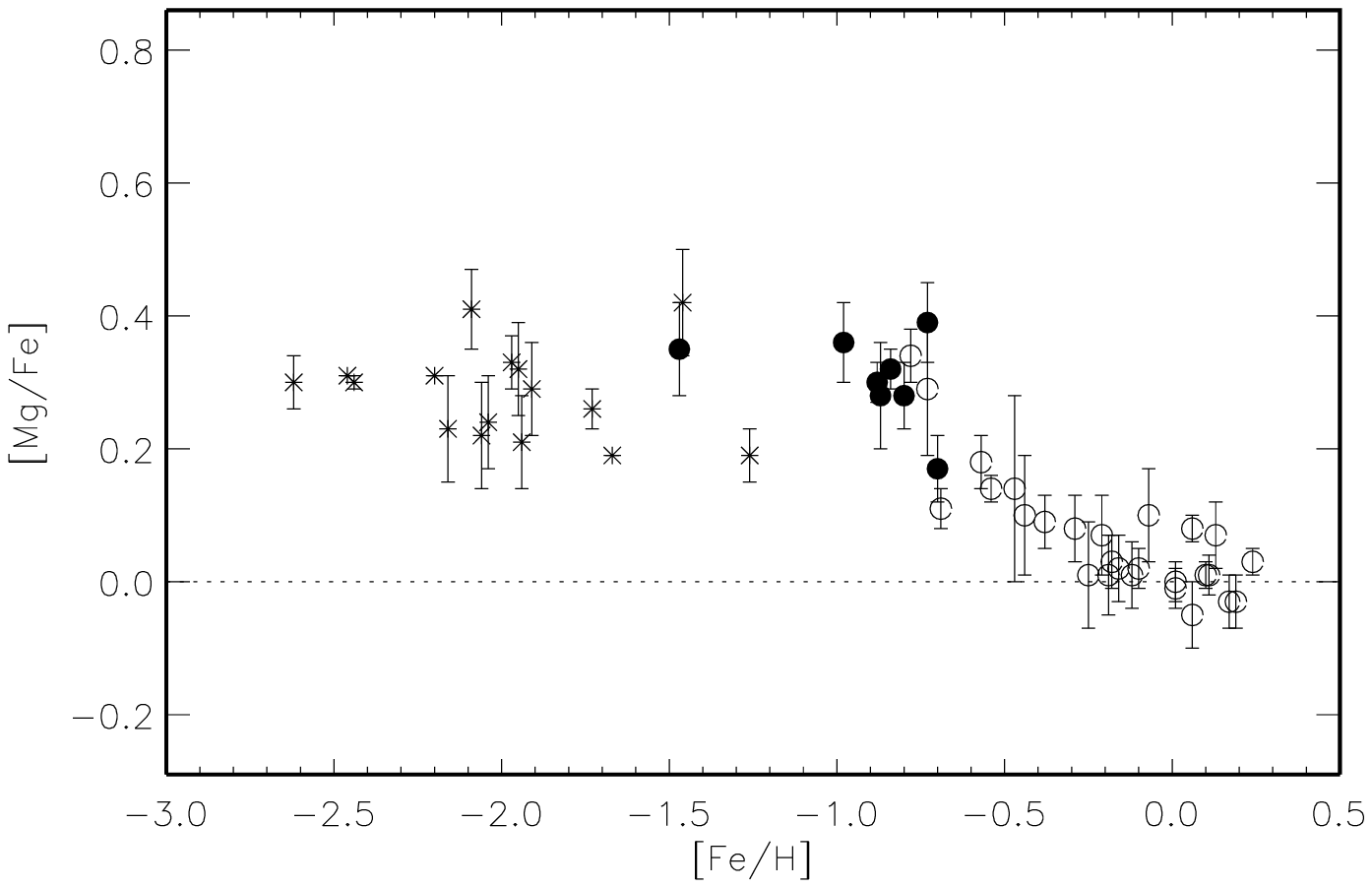}{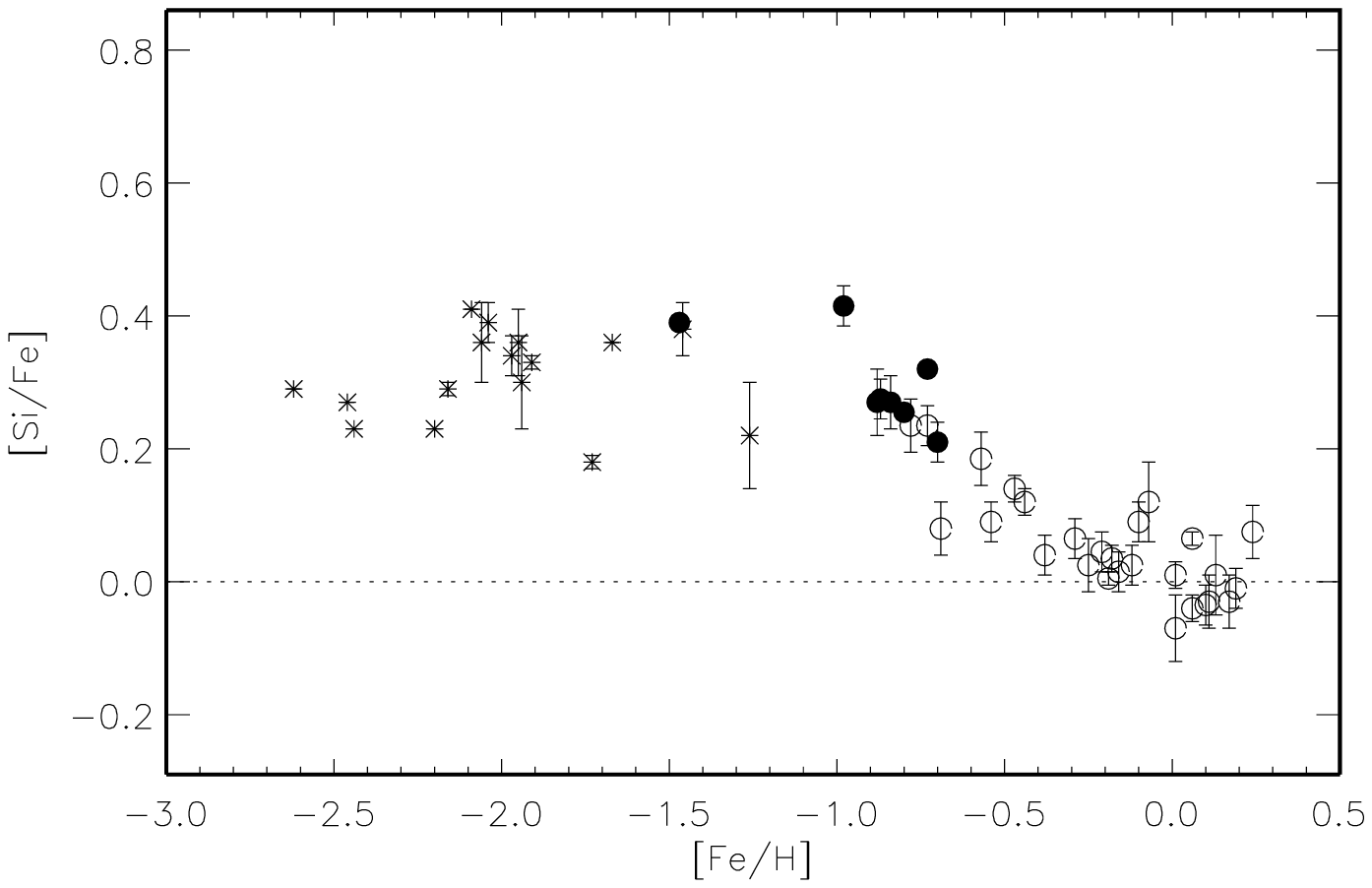}
\plottwo{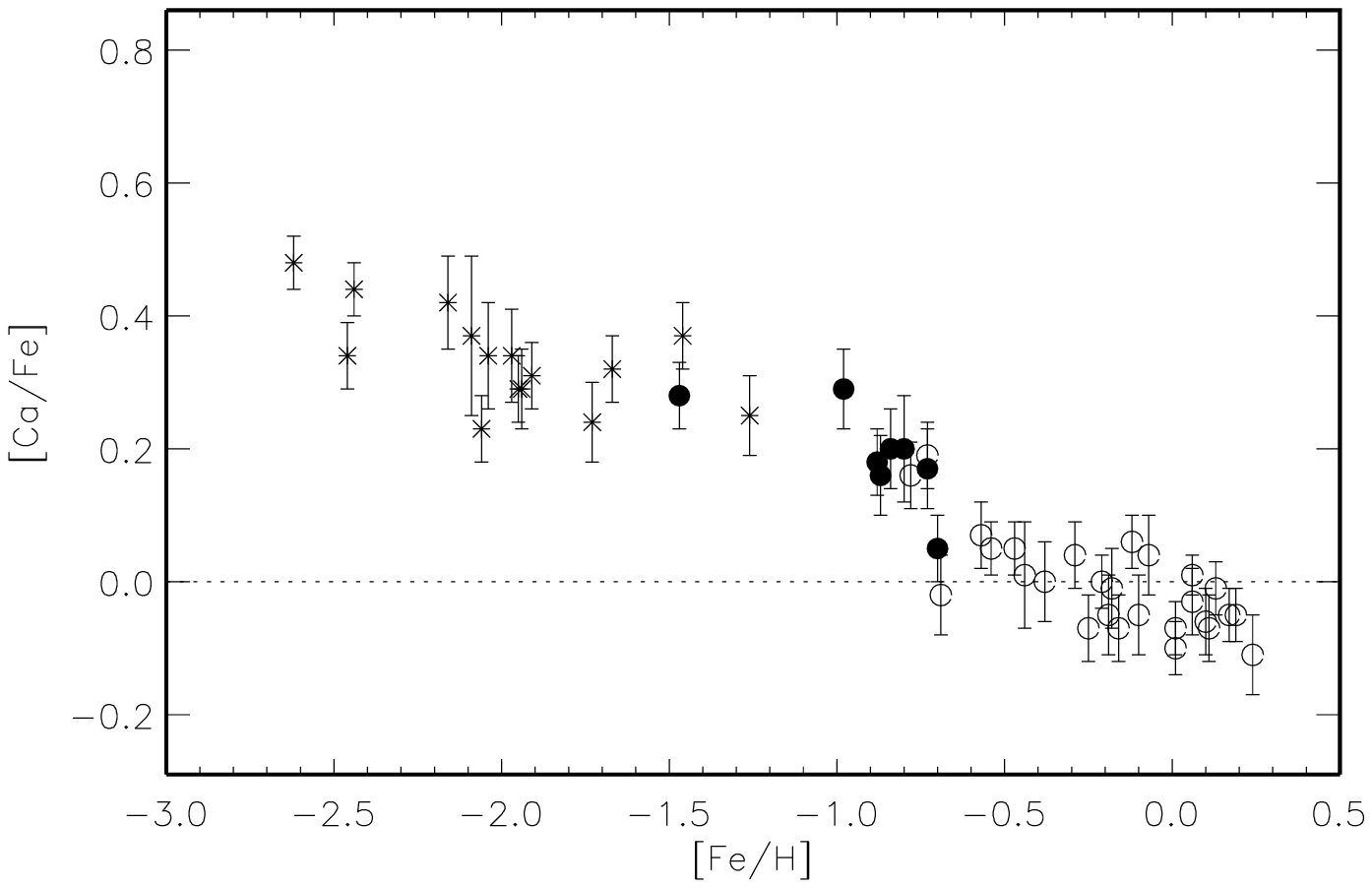}{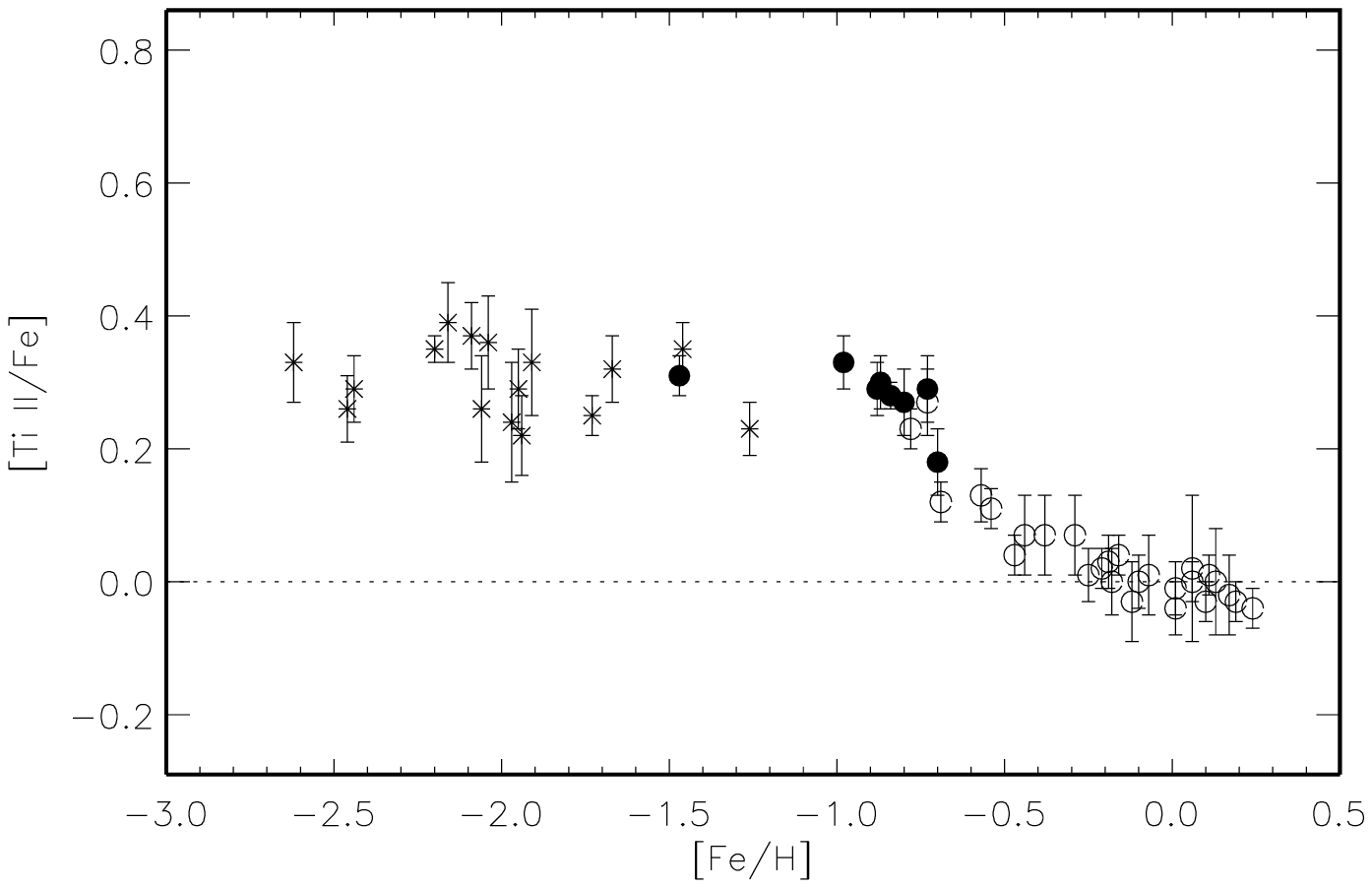}
\caption{Stellar element-to-iron NLTE abundance ratios: even-Z elements C, O, Mg, Si, Ca, and Ti. The same symbols are used as in Fig.\,\ref{Fig:li}. Asterisks inside the circles show the halo stars with only the molecular CH lines available. \label{Fig:alpha}}
\end{figure*}

\begin{figure*}
\epsscale{1.0}
\plottwo{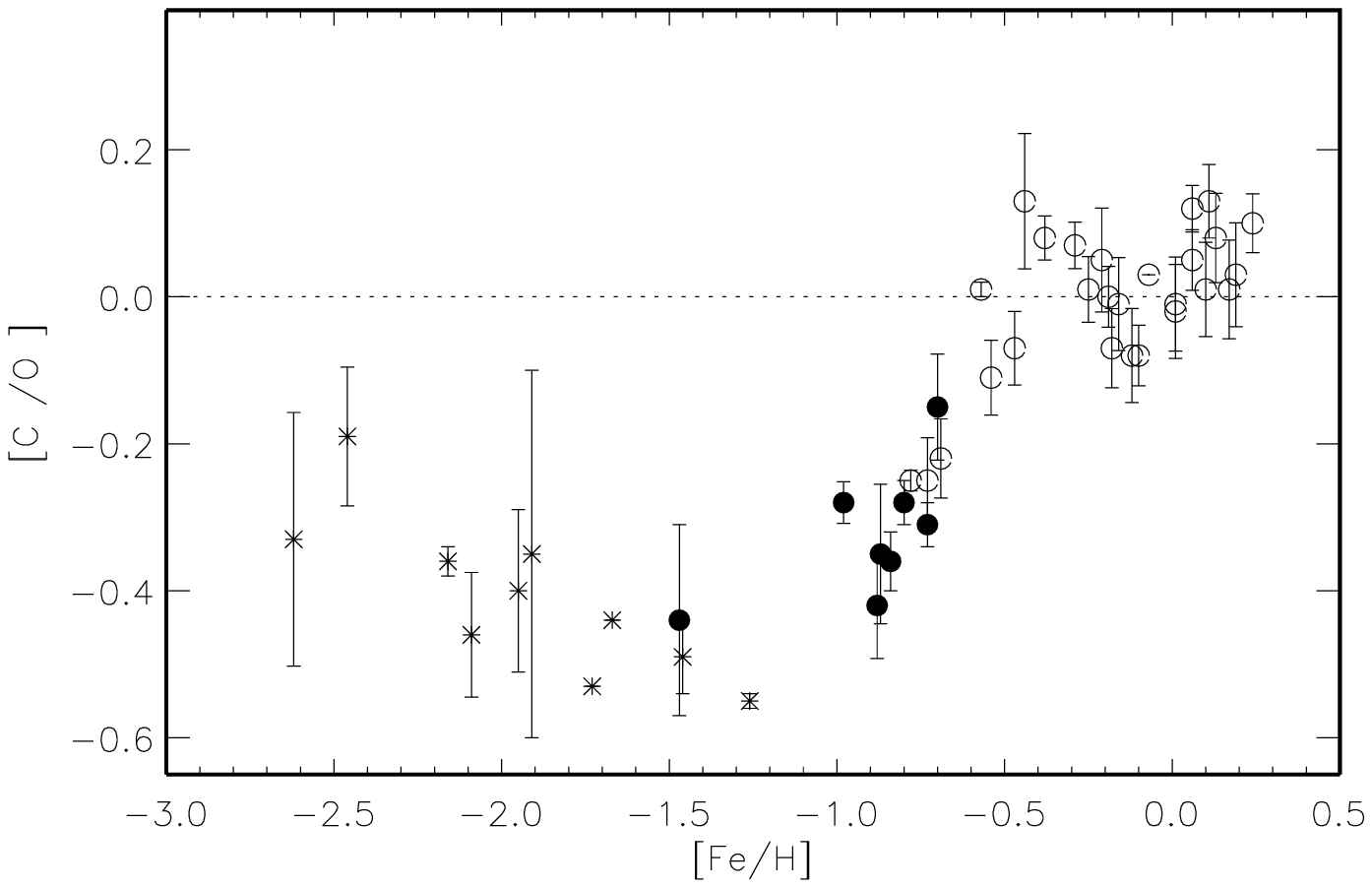}{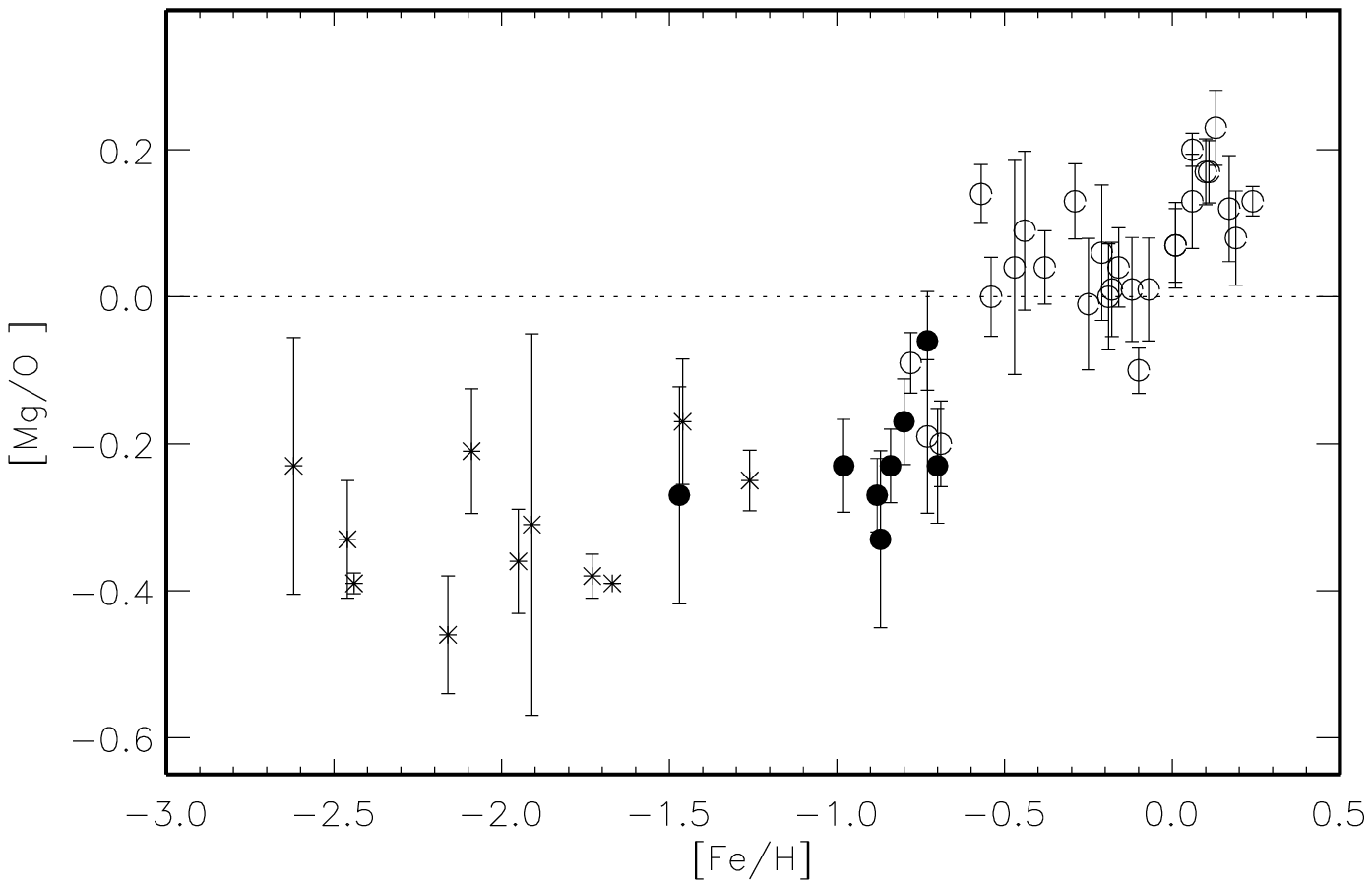}
\plottwo{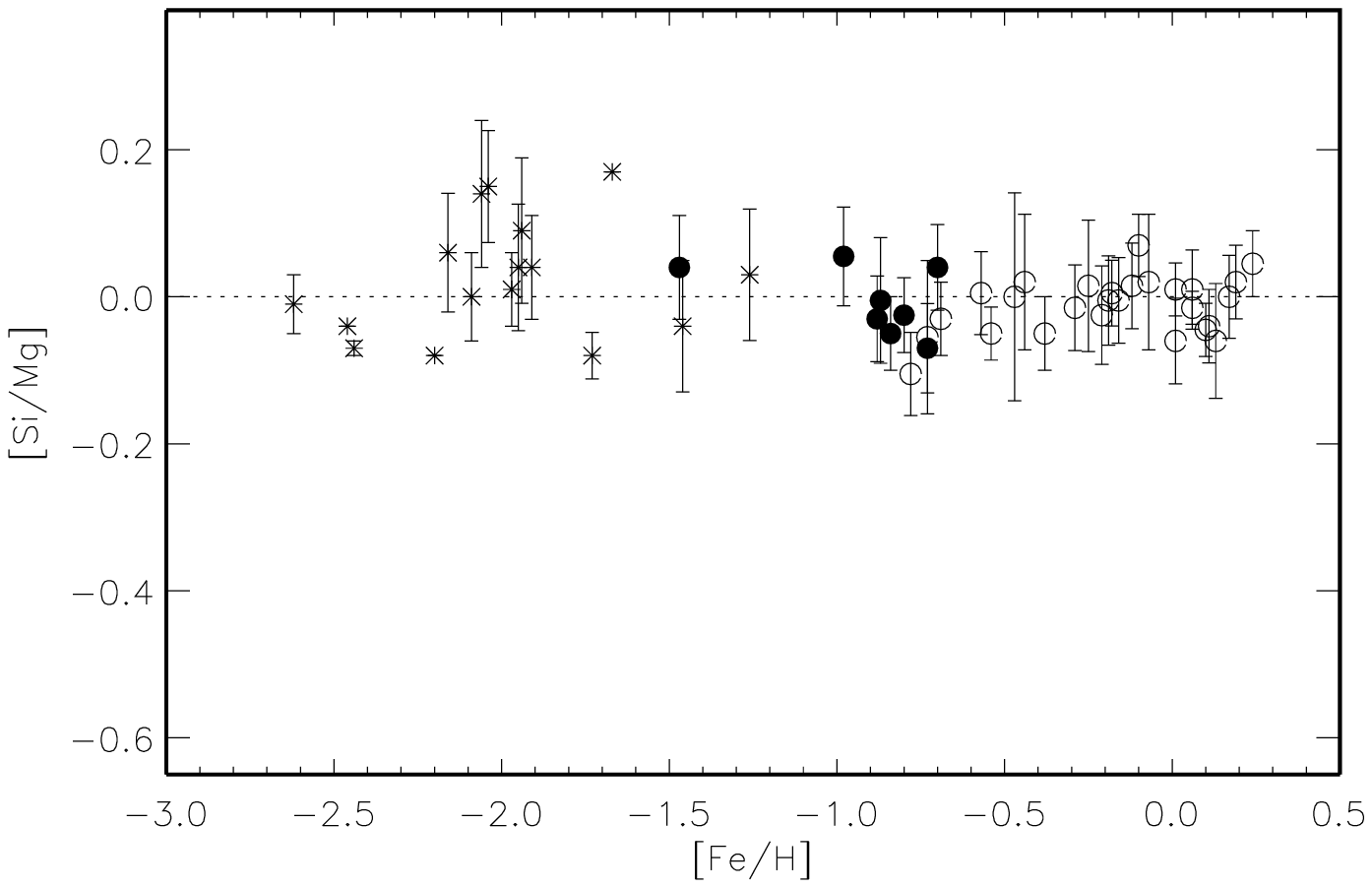}{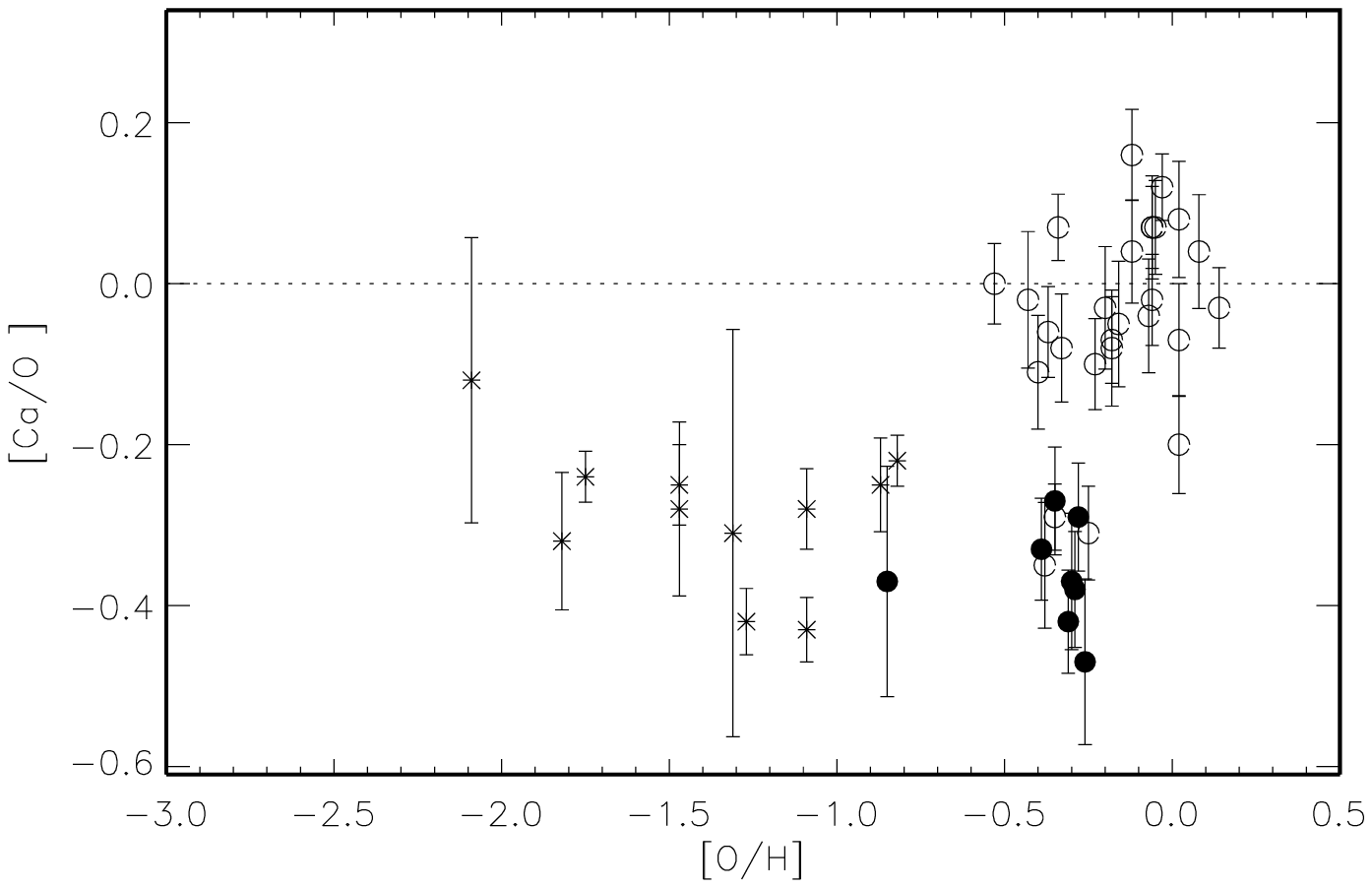}
\caption{Stellar NLTE abundance ratios between the even-Z elements. The same symbols are used as in Fig.\,\ref{Fig:li}. \label{Fig:ratios1}}
\end{figure*}

{\it Even-Z elements.}
Carbon abundance increases relative to the Fe one, when metallicity decreases from super-solar values down to [Fe/H] $\simeq -0.8$ (Fig.\,\ref{Fig:alpha}), and reaches [C/Fe] = 0.21$\pm$0.06, on average, in the thick disk stars. A substantial scatter of [C/Fe] in the halo stars seems to be due to including stars with the C abundance derived from the molecular CH lines. For example, the lowest [C/Fe] values of $-0.11$ and 0.03 were obtained for HD\,103095 and BD\,$+66^{\circ}$\,0268, respectively. However, this is not supported by analysis of the [C/O] ratios (Fig.\,\ref{Fig:ratios1}). The stars of close metallicity show very similar [C/O] ratios, independent of whether the C abundance is based on the CH or \ion{C}{1} lines.

Carbon NLTE abundances were calculated by \citet[][hereafter, F06]{2006A&A...458..899F} for a sample of $-3.2 < {\rm [Fe/H]} < -0.7$ dwarfs, using observations of \citet{2004A&A...414..931A}. The [C/Fe] NLTE ratios were obtained to be close to solar one, in contrast to our results. We consider two sources of this discrepancy. The first one is a different treatment of inelastic collisions with \ion{H}{1} atoms. The final carbon NLTE abundances were obtained by F06 assuming negligible collisions with hydrogen. We did include collisions with \ion{H}{1} and used \kH\ = 0.3. The abundance difference between applying \kH\ = 0 and 0.3 is non-negligible. Calculations of F06 with \kH\ = 0 resulted in 0.1-0.15~dex lower abundances compared with those for \kH\ = 1 (Fig.~9 in F06). Our calculations with \kH\ = 0, 0.3, and 1 show that the abundance difference between applying \kH\ = 0 and 1 is larger than that between \kH\ = 0.3 and 1. For example, in HD~59374 (5850/4.38/$-0.88$) $\Delta\eps{}$(\kH\ = 1 - \kH\ = 0) = 0.08~dex, while it amounts 0.03~dex, when comparing the \kH\ = 1 and 0.3 based abundances. The second source concerns, probably, with a different treatment of background opacity. As shown by \citet{2015MNRAS.453.1619A}, their NLTE abundance corrections agree well with those of F06 in the [Fe/H] $\ge -1$ model atmospheres, when applying common \kH\ = 1, and they are less negative at lower metallicities, by 0.08~dex in the 6000/4/$-2$ model and by 0.16~dex at [Fe/H] = $-3$. Our test calculations show that a variation in background opacity, for example excluding H$_2^+$, metal lines, quasi H$_2$ molecular absorption, can lead to stronger departures from LTE and to 0.2~dex more negative NLTE corrections for lines of \ion{C}{1} in the 5777/3.70/$-2.38$ 

Oxygen-to-iron NLTE abundance ratios in stars more metal-poor than [Fe/H] $\simeq -0.9$ form a plateau at [O/Fe] = 0.61, with a rather small scatter at a given metallicity (Fig.\,\ref{Fig:alpha}). An exception is the halo star HD~103095 that has a 0.16~dex lower O/Fe ratio.  For higher metallicity, [O/Fe] shows a downward trend that continues up to super-solar metallicities. It can be seen that the thick disk and the thin disk stars reveal a common behavior in the overlapping metallicity range, although it is rather narrow.
Our results agree well with those of \citet{2014A&A...562A..71B} who applied the empirical formula to take the NLTE effects into account. However, a scatter of our data for stars of close metallicity is certainly smaller. \citet{2015MNRAS.454L..11A} inferred a similar value of [O/Fe] $\simeq 0.5$ in the $-2.5 \le$ [Fe/H] $< -1$ range using stellar parameters and observed equivalent widths of the \ion{O}{1} lines from the literature and performing detailed 3D NLTE radiative transfer calculations.

From LTE analysis of the $-3.2 \le$ [Fe/H] $\le -0.7$ dwarf and subgiant stars, \citet{2004A&A...414..931A} deduced that ``C/O drops by a factor of 3-4 as O/H decreases from solar to about 1/10 solar", in line with the earlier findings \citep[see references in][]{2004A&A...414..931A}. Their new result was a discovery of the upturn in C/O at [O/H] = $-1$. Having applied the NLTE corrections to the LTE abundances of \citet{2004A&A...414..931A}, F06 recovered a similar behaviour of C/O, with the upturn at [Fe/H] $\simeq -1.2$, where [C/O] $\simeq -0.6$ to $-0.7$ depending on \kH\ value and grows at lower metallicities. Our data are qualitatively similar, namely C/O is, on average, solar in the thin disk stars with [Fe/H] $> -0.6$ and decreases steeply at lower metallicities, down to [C/O] = $-0.55$ at [Fe/H] = $-1.26$ (Fig.\,\ref{Fig:ratios1}). We confirm the upturn in [C/O] at [Fe/H] $\simeq -1.2$. The eleven more metal-poor stars form a linear regression of [C/O] = $-0.78 - 0.19$ [Fe/H], with $\sigma$ = 0.06. The observed C/O trend is important for better understanding nucleosynthesis in the early Galaxy.

Multiple abundance determinations can be found in the literature for Mg, Si, Ca, and Ti. However, homogeneous NLTE abundances of all the four species and also oxygen in the stellar sample covering a broad metallicity range were obtained in this study for the first time. Magnesium, silicon, calcium, and titanium reveal a common behavior that is typical of the $\alpha$-process elements. They are enhanced relative to Fe in the halo and the thick disk stars, with nearly constant [X/Fe] ratios at [Fe/H] $< -0.8$ and similar for different elements (Fig.\,\ref{Fig:alpha}). For example, 16 halo stars have, on average, [Mg/Fe] = 0.28$\pm$0.07, [Si/Fe] = 0.31$\pm$0.07, and [Ti/Fe] = 0.30$\pm$0.05. For [Ca/Fe] there is a hint of its increasing towards lower metallicity. The mean amounts to [Ca/Fe] = 0.32$\pm$0.08. For each element [X/Fe] decreases at [Fe/H] $> -0.8$ and reaches the solar value at the solar metallicity. In the overlapping metallicity range the thin and thick disk stars have similar [$\alpha$/Fe] ratios.

We obtained that abundance ratios among Mg, Si, Ca, and Ti are close to solar value, independent of metallicity (see Si/Mg in Fig.\,\ref{Fig:ratios1}), while each of these elements is deficient relative to oxygen in the halo and the thick disk stars (Mg/O and Ca/O in Fig.\,\ref{Fig:ratios1}). Most thin disk stars have, on average, solar [$\alpha$/O] ratios. Outliers are the three stars, HD~59984, HD~105755, and HD~134169, with a thin-disc kinematics, but a low Fe abundance of [Fe/H] = $-0.69$, $-0.73$, and $-0.78$, respectively. Their [$\alpha$/Fe] and [$\alpha$/O] ratios suggest a thick-disk origin. A step-like increase of [$\alpha$/O] in the thick disk-to-thin disk transition is, in particular, clearly seen, when plotting these elemental ratios as a function of [O/H].

Obtained [Mg/Fe] ratios of our halo and thick disk stars are 0.1~dex lower compared with [Mg/Fe] $\simeq$ 0.4 derived for the nearby [Fe/H] $< -1$ stars by \citet{2008MNRAS.384..173F, 2011MNRAS.414.2893F} from the LTE analysis. \citet{2012A&A...545A..32A} derived the LTE abundances of large sample of nearby stars, and using Mg, Si, and Ti as representatives of the $\alpha$-process elements, they concluded that an $\alpha$-enhancement for the thick disk and the stars with [Fe/H] $< -0.5$ is close to 0.3~dex.

\begin{figure*}
\epsscale{1.0}
\plottwo{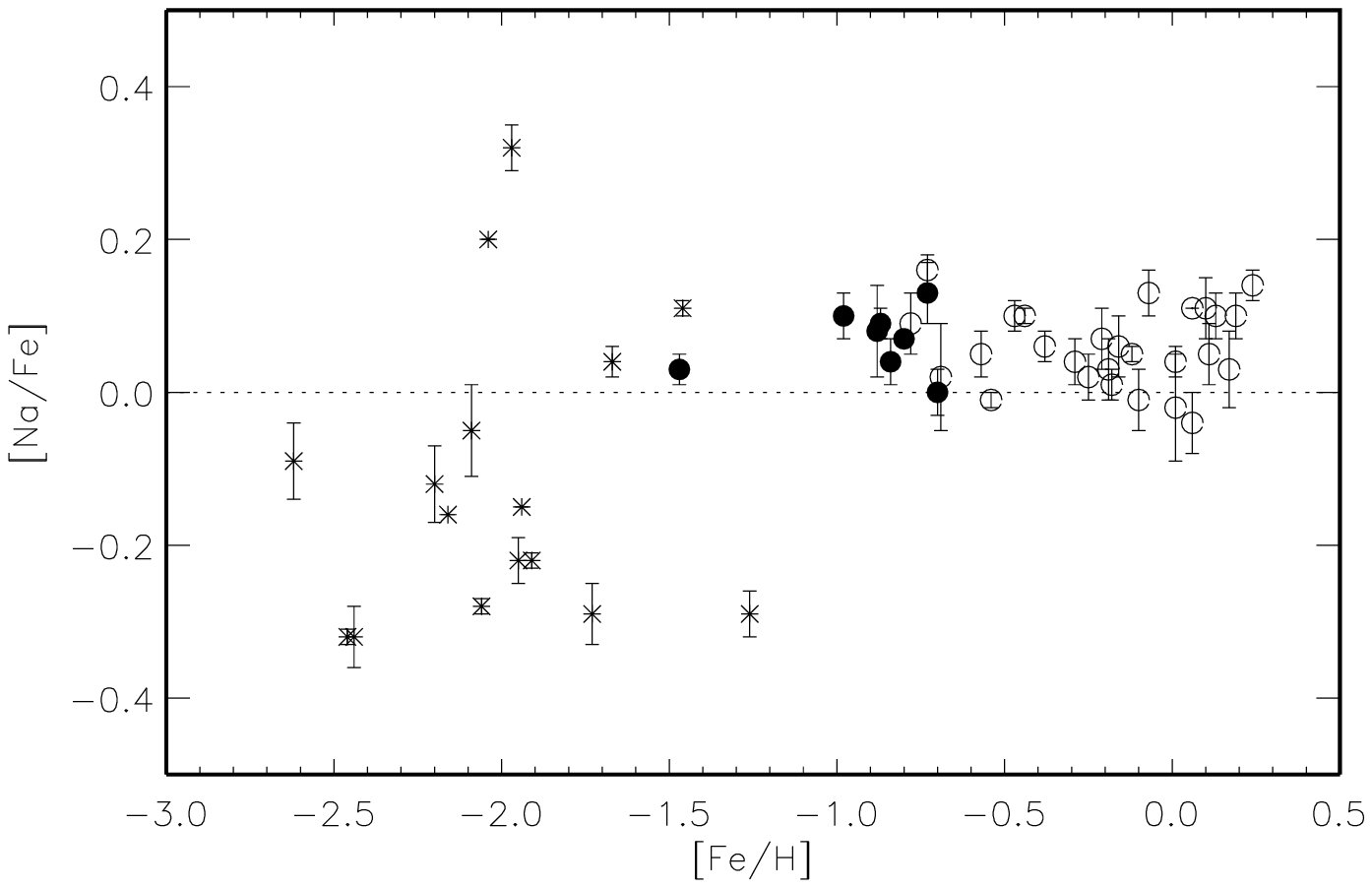}{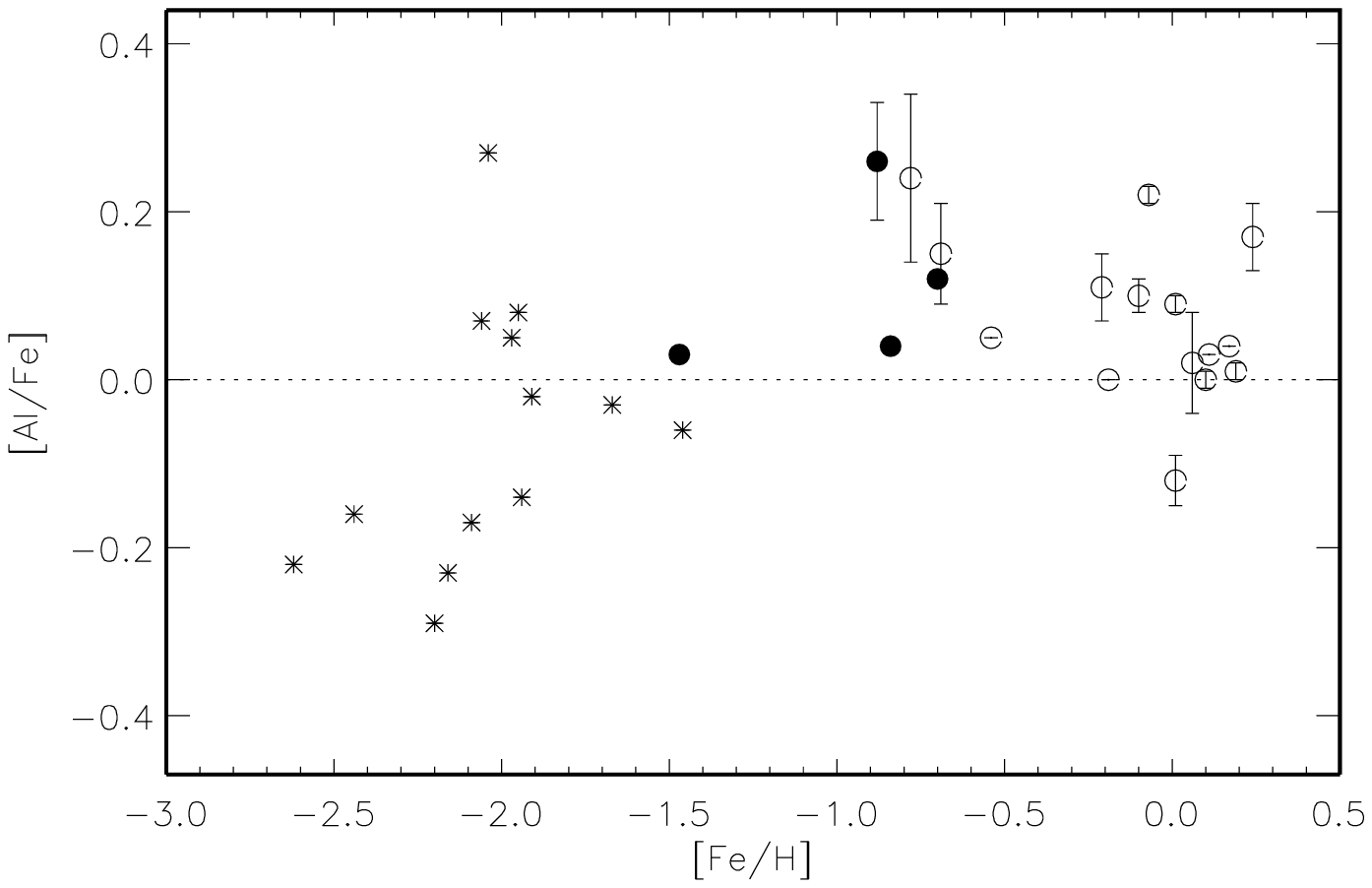}
\plottwo{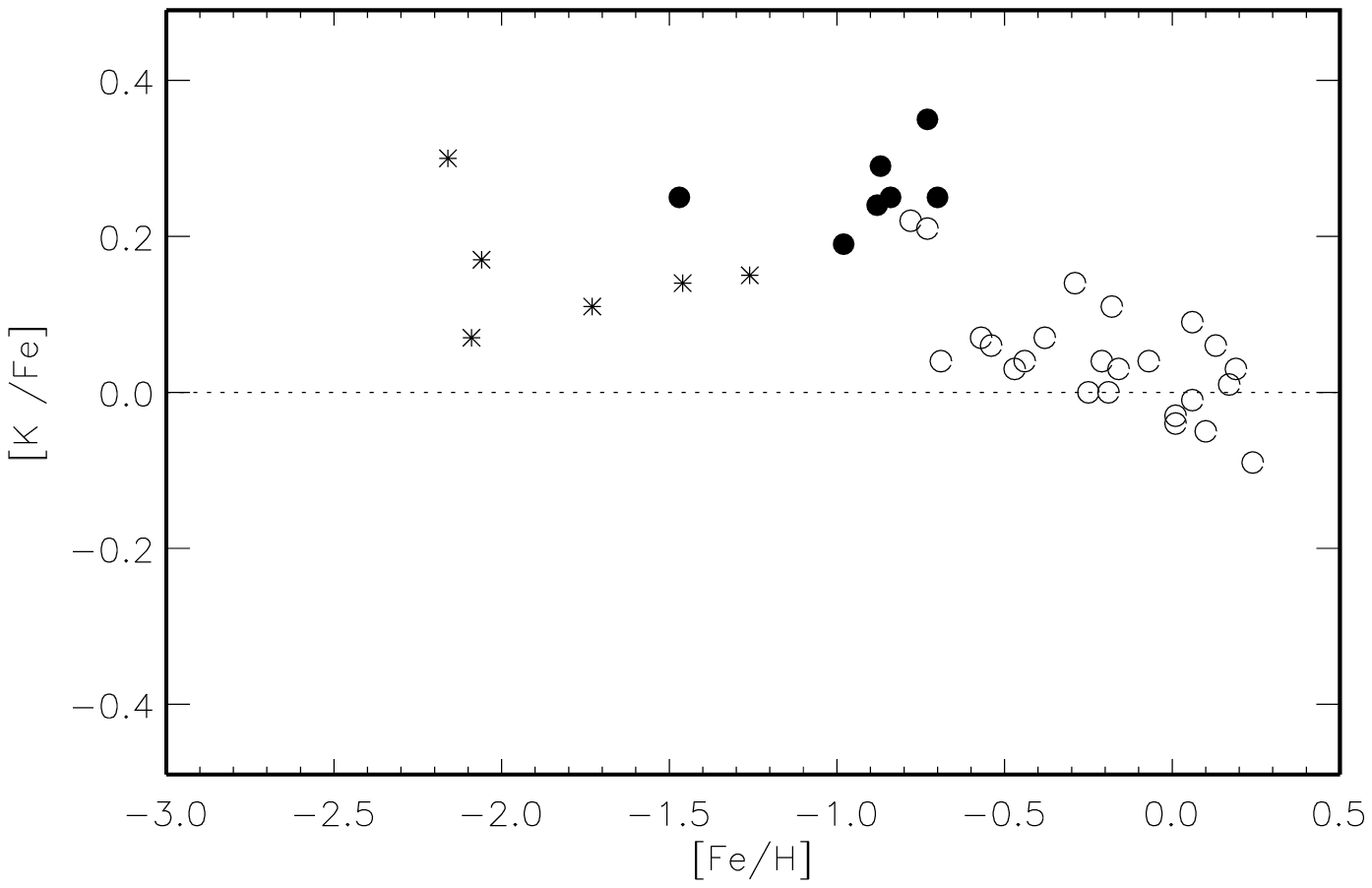}{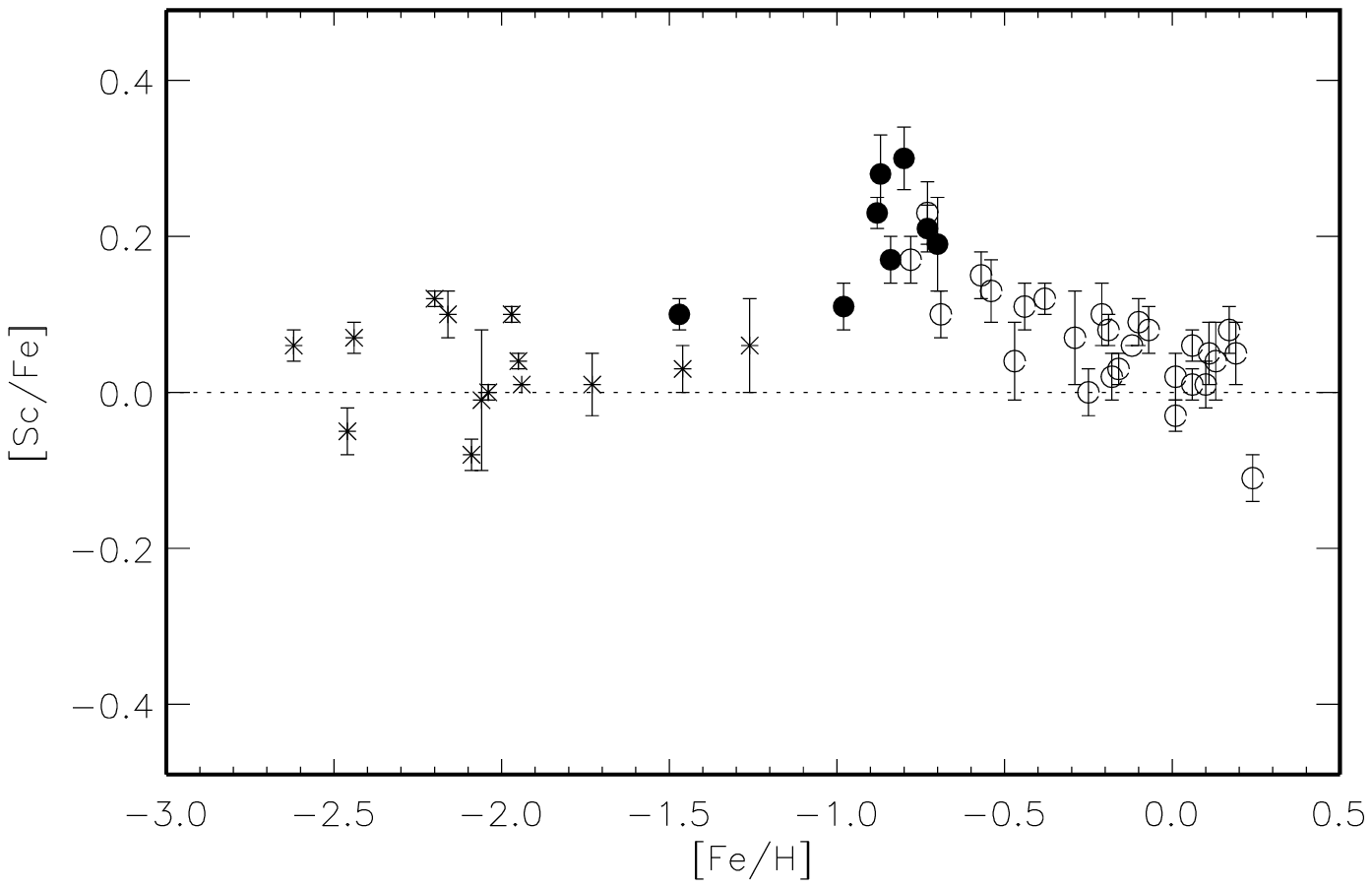}
\caption{The same as in Fig.\,\ref{Fig:alpha} for the odd-$Z$ elements Na, Al, K, and Sc. \label{Fig:odd}}
\end{figure*}

\begin{figure*}
\epsscale{1.0}
\plottwo{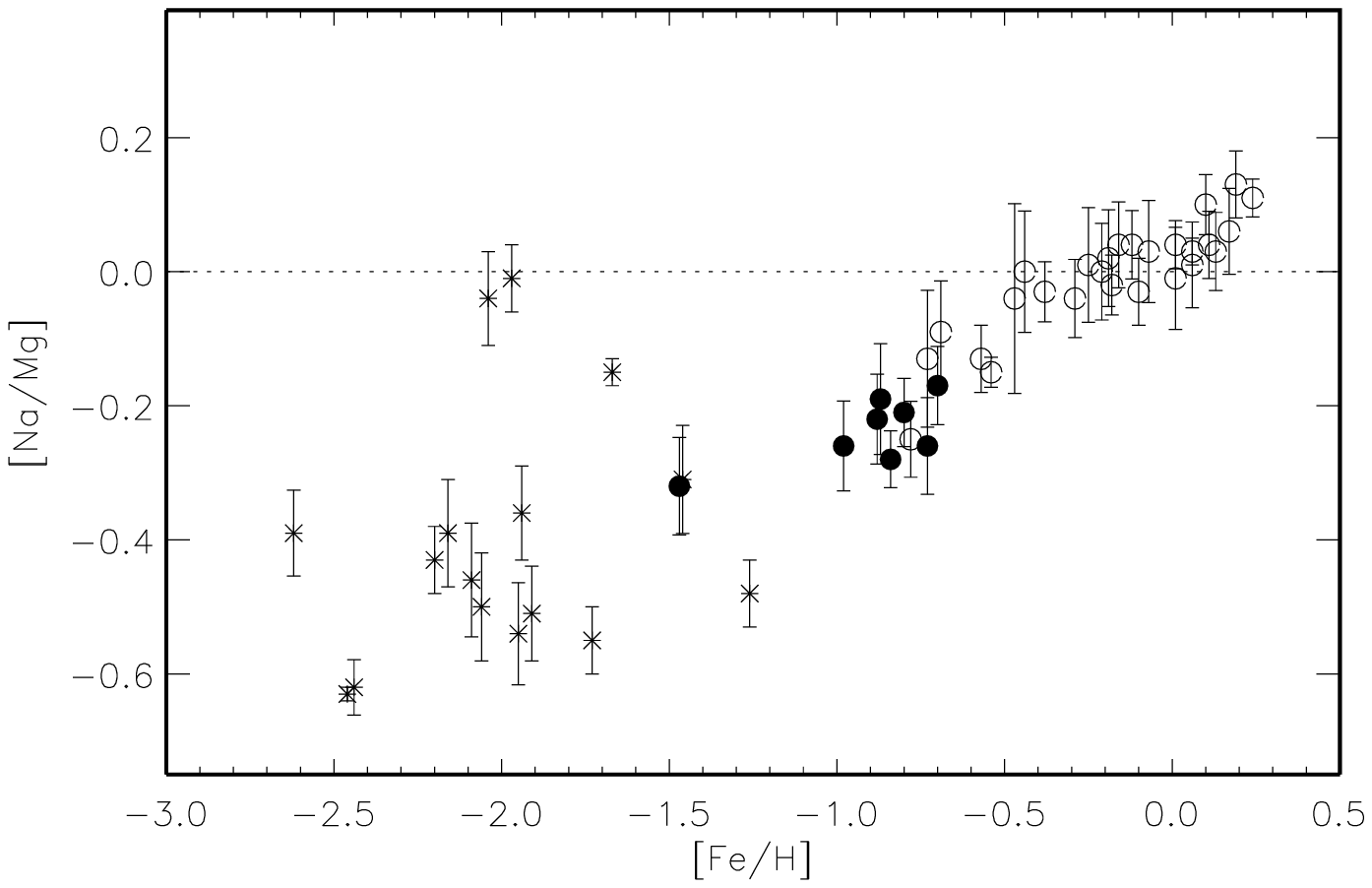}{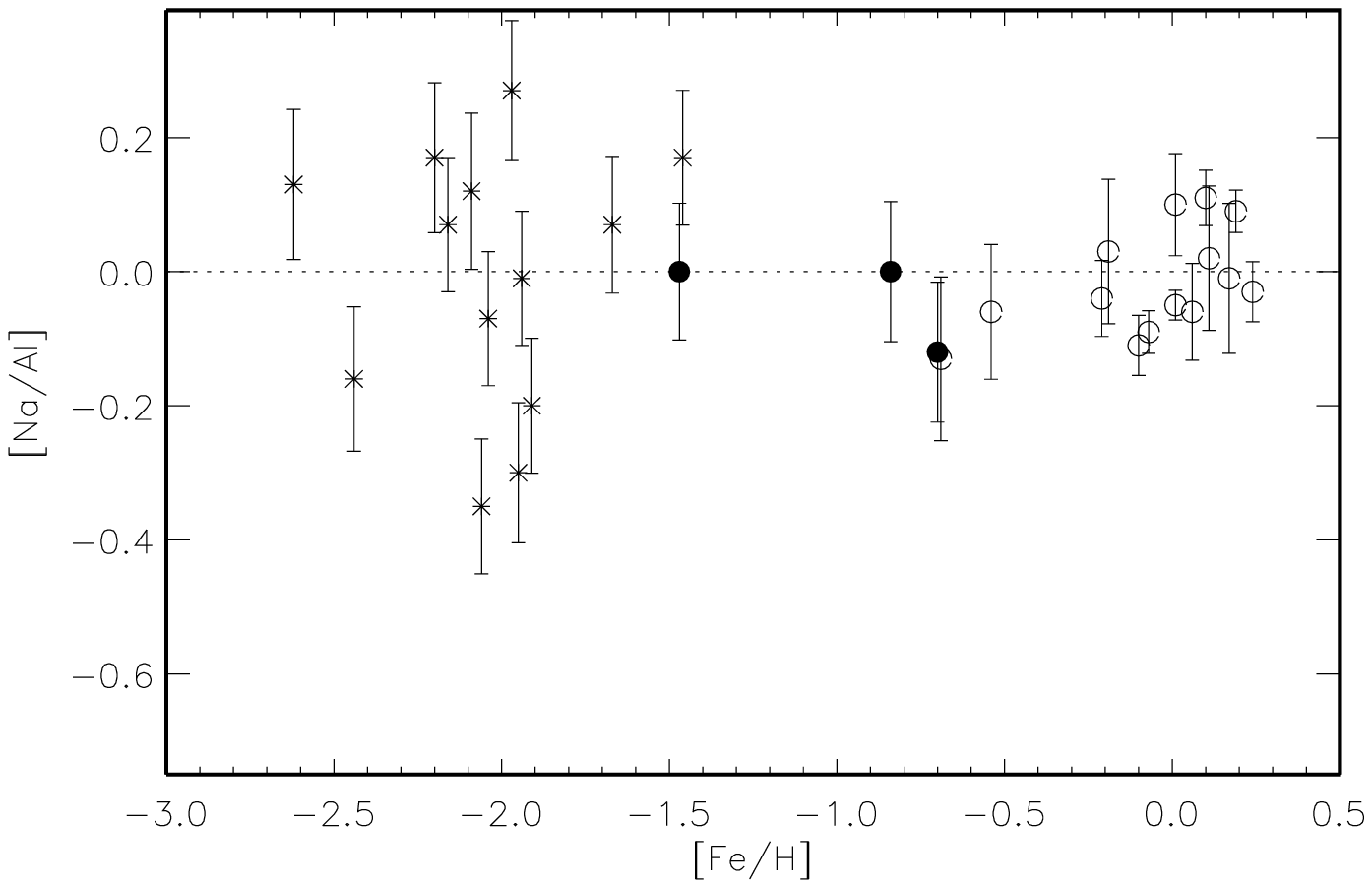}
\plottwo{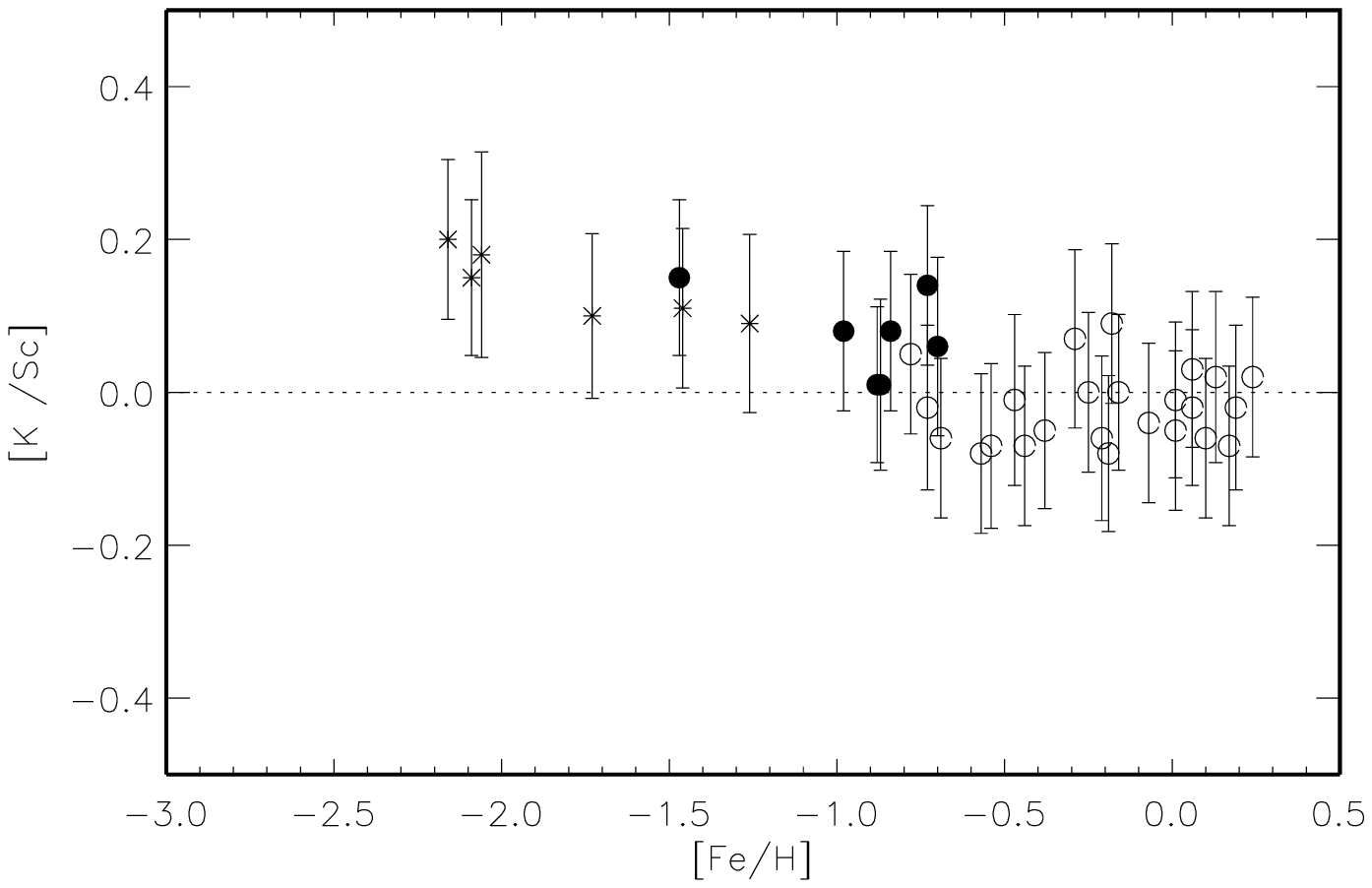}{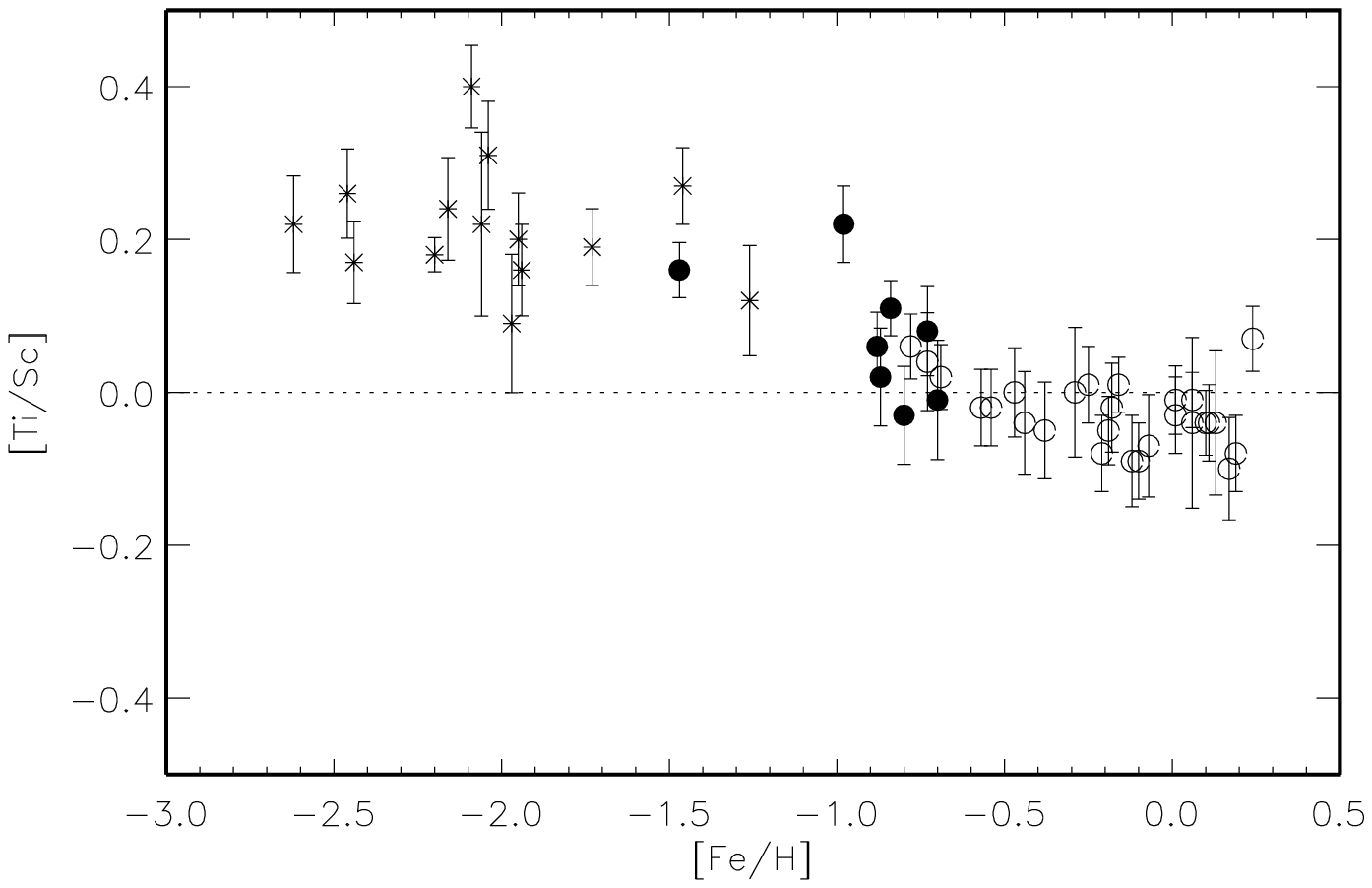}
\caption{Stellar NLTE abundance ratios involving the odd-$Z$ elements. The same symbols are used as in Fig.\,\ref{Fig:li}. \label{Fig:ratios2}}
\end{figure*}

{\it Odd-Z elements.}
We consider Na and Al together, although conclusions related to Al are less firm due to lower accuracy of the derived Al abundances as discussed in Sect.\,\ref{sect:species}. Both Na and Al follow the Fe abundance in the thin and thick disc stars (Fig.\,\ref{Fig:odd}). In most halo stars the Na/Fe and Al/Fe ratios are subsolar, with a rather large scatter of data. In contrast, a well-defined downward trend is observed for Na/Mg, when [Fe/H] decreases from super-solar values to $-1$, and the more MP stars form a plateau at [Na/Mg] $\simeq -0.5$ (Fig.\,\ref{Fig:ratios2}). The Na/Al ratios seem to be solar, independent of metallicity. The two halo stars, HD\,74000 and G090-003, are clear outliers, with [Na/Fe] $\ge$ 0.2 and [Na/Mg] $\simeq$ 0. The latter (G090-003) has also high [Al/Fe] = 0.27, but normal Na/Al. See notes on these two stars in Sect.\,\ref{sect:notes}. One more star, HD~108177, has higher [Na/Fe] = 0.04 and [Na/Mg] = $-0.15$ compared with the halo stars of similar metallicity.

A metal-poor plateau for Na/Mg was reported in the earlier NLTE studies by \citet{2006A&A...451.1065G}, with [Na/Mg] = $-0.7$ for the $-3.1 \le$ [Fe/H] $< -1.8$ dwarfs, and \citet{2010A&A...509A..88A}, with [Na/Mg] $\simeq -0.8$ for the $-4.2 \le$ [Fe/H] $< -2$ giants. The difference between our value, [Na/Mg] $\simeq -0.5$, and the literature data is, most probably, due to overestimated magnesium NLTE abundances in \citet{2006A&A...451.1065G} and \citet{2010A&A...509A..88A}. For example, the latter paper reported the mean [Mg/Fe] = 0.61 for their stellar sample, while, in this study, a MP plateau was obtained at [Mg/Fe] = 0.28. The difference in Mg abundances is, in turn, probably due to different treatment of inelastic collisions with \ion{H}{1} atoms. Our study takes advantage of employing the \ion{Mg}{1} + \ion{H}{1} collision rates from quantum-mechanical calculations of \citet{mg_hyd2012}, while \citet{2006A&A...451.1065G} and \citet{2010A&A...509A..88A} used the formula of \citet{Steenbock1984} with \kH\ = 0.05 and 1/3, respectively.

The heavier odd-$Z$ elements, K and Sc, behave like the $\alpha$-elements in the thin and thick disk stars, at [Fe/H] $> -1$ (Fig.\,\ref{Fig:odd}). Indeed, K/Fe and Sc/Fe grow towards lower metallicity from the solar value to [K/Fe] $\simeq 0.25$ and [Sc/Fe] $\simeq 0.2$.  In the halo stars, potassium remains to be enhanced relative to Fe, however, with the lower magnitude, [K/Fe] $< 0.2$, while Sc/Fe is close to the solar value. As a result, the trends are non-monotoneous, and a group of the thick disk stars at [Fe/H] around $-0.8$ looks like a local peak. In contrast, the K/Sc, Ca/Sc, and Ti/Sc ratios reveal a remarkably monotoneous behavior, with a rather small scatter of data for stars of close metallicity (see [K/Sc] and [Ti/Sc] in Fig.\,\ref{Fig:ratios2}). This suggests a common site of the K-Ti production.

\begin{figure}
\epsscale{1.0}
\plotone{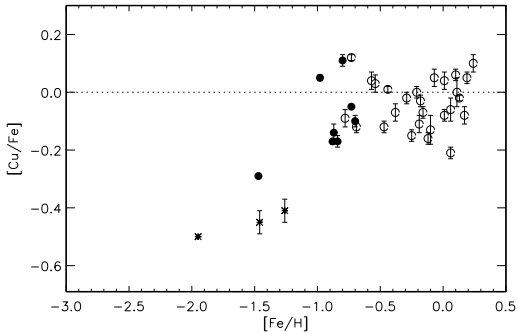}
\caption{The same as in Fig.\,\ref{Fig:alpha} for copper. \label{Fig:cu}}
\end{figure}

{\it Copper.}
The thin and thick disk stars with [Fe/H] $\ge -1$ reveal very similar and close-to-solar Cu/Fe ratios (Fig.~\ref{Fig:cu}). The statistics is very poor at the lower metallicity, where we could measure the Cu abundances for three halo stars and the thick disk star HD~94028. Copper is underabundant relative to Fe in all the four stars, with [Cu/Fe] between $-0.41$ and $-0.50$ in the halo stars and a slightly higher value of $-0.29$ in HD~94028. Our data combined with the three $-1.3 <$ [Fe/H] $< -1$ stars from \citet{2015ApJ...802...36Y} suggest the upward trend, where [Cu/Fe] increases from $-0.5$ at [Fe/H] $\simeq -2$ to about $-0.1$ at [Fe/H] $\simeq -0.8$.

\begin{figure*}
\epsscale{1.0}
\plottwo{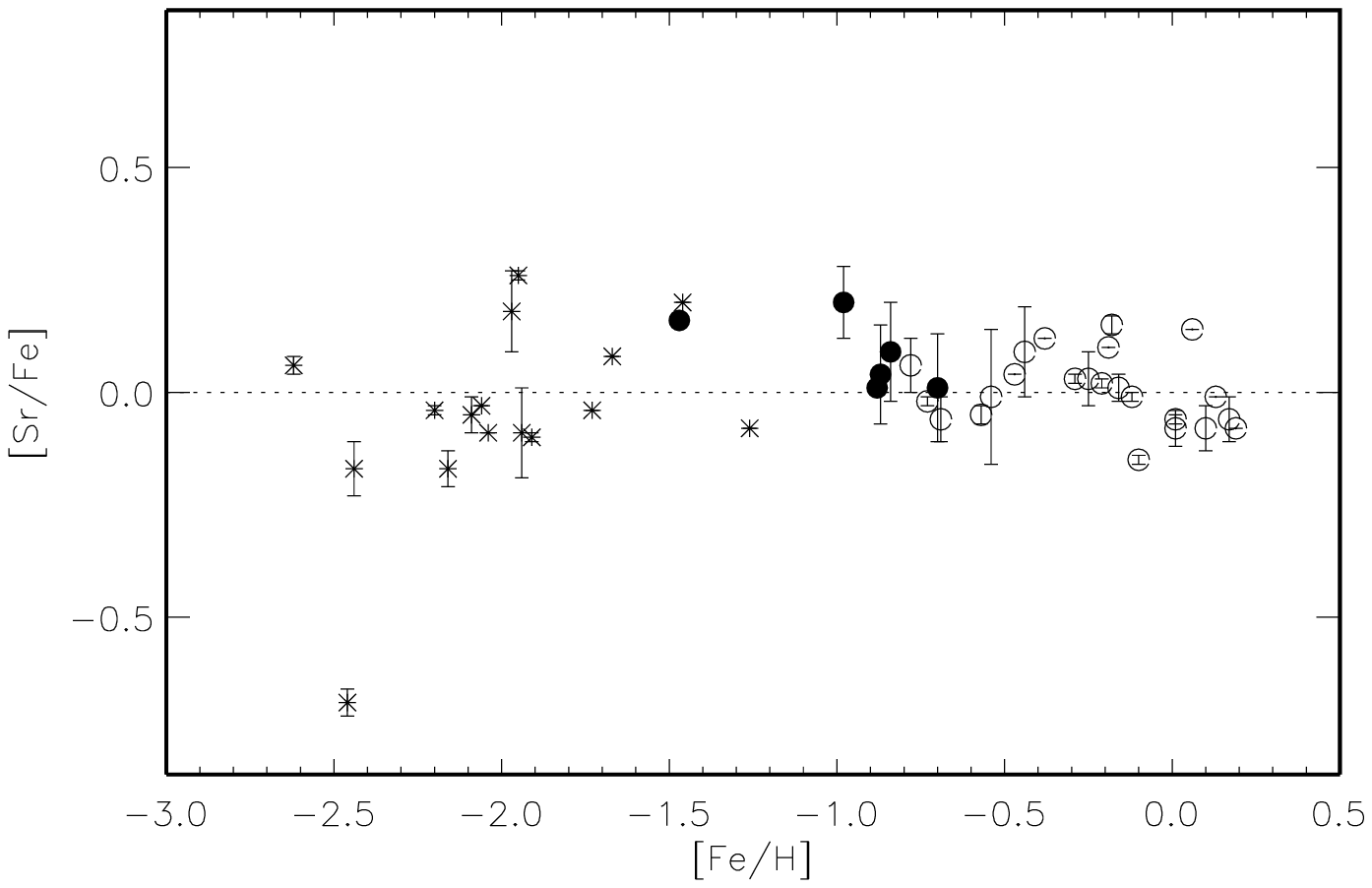}{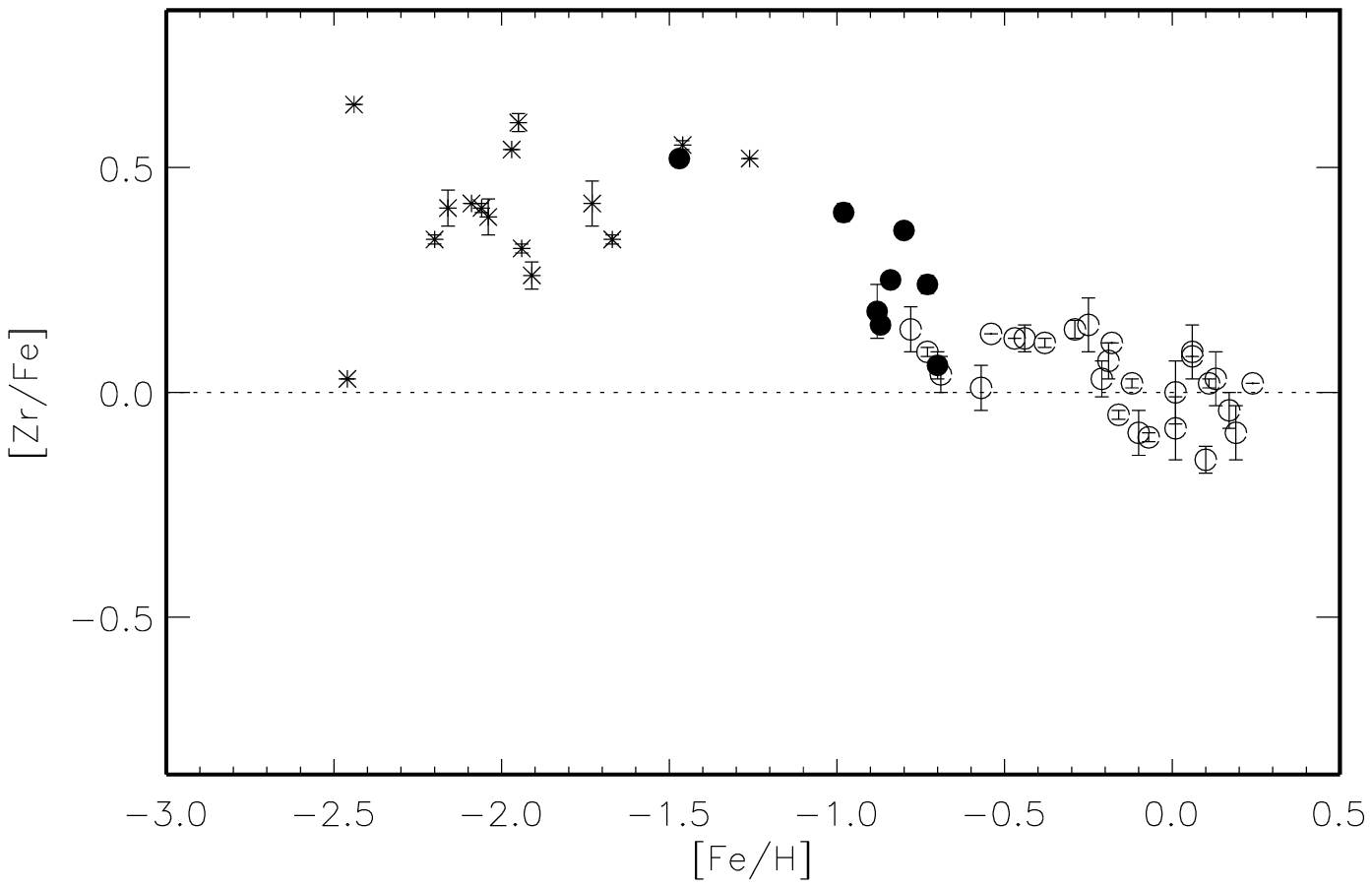}
\plottwo{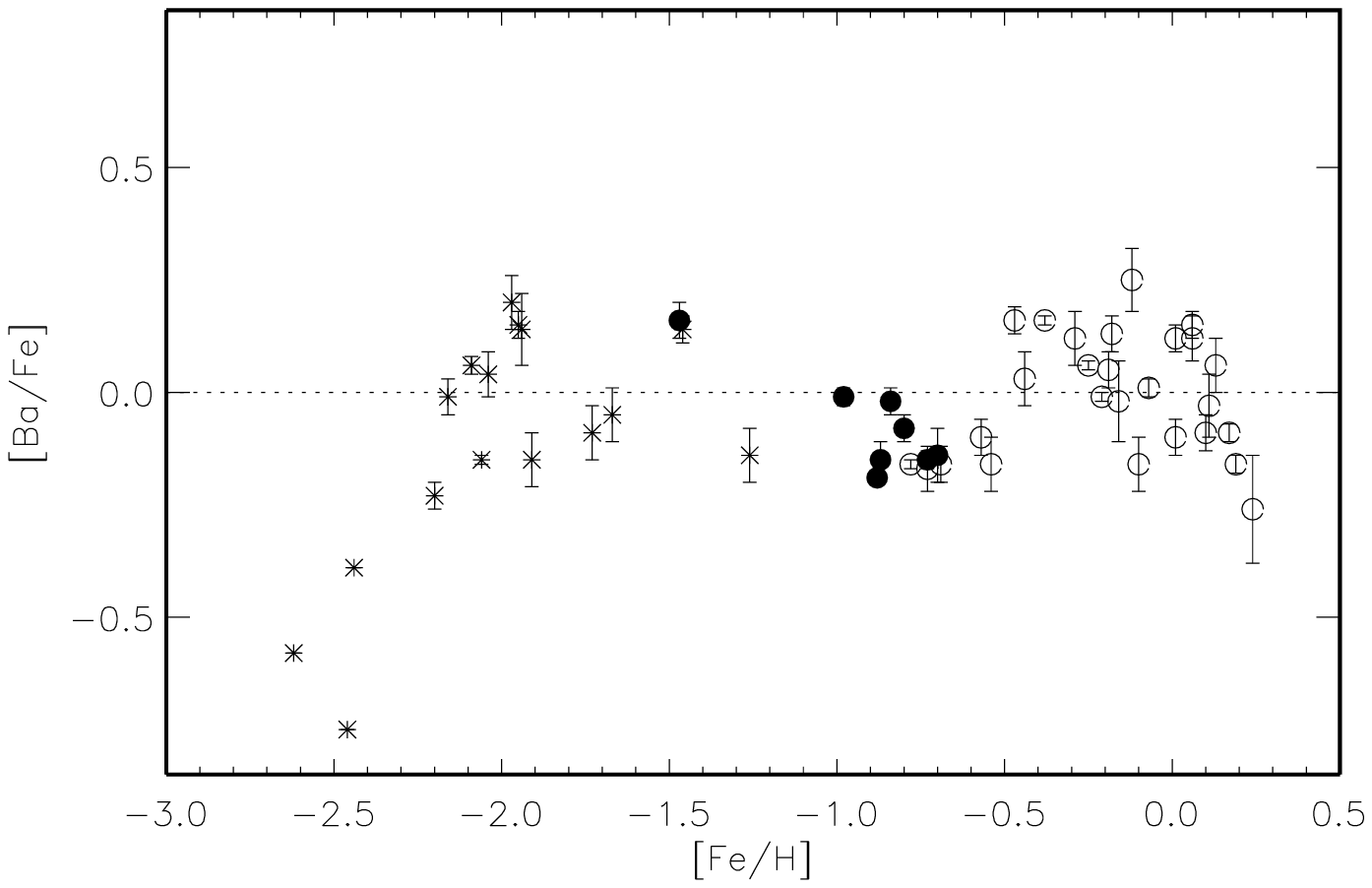}{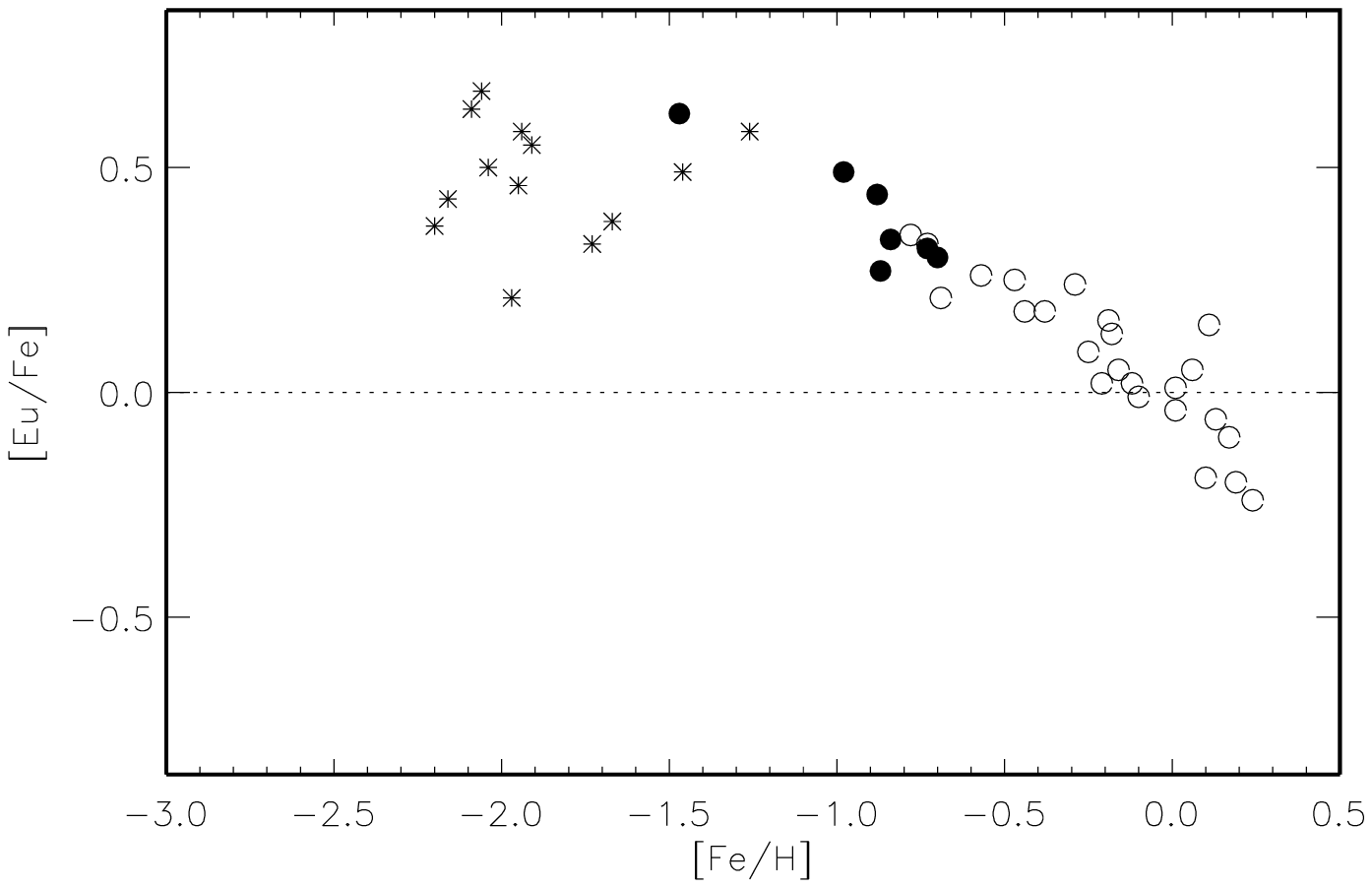}
\caption{The same as in Fig.\,\ref{Fig:alpha} for the neutron-capture elements Sr, Zr, Ba, and Eu. \label{Fig:heavy}}
\end{figure*}

\begin{figure*}
\epsscale{1.0}
\plottwo{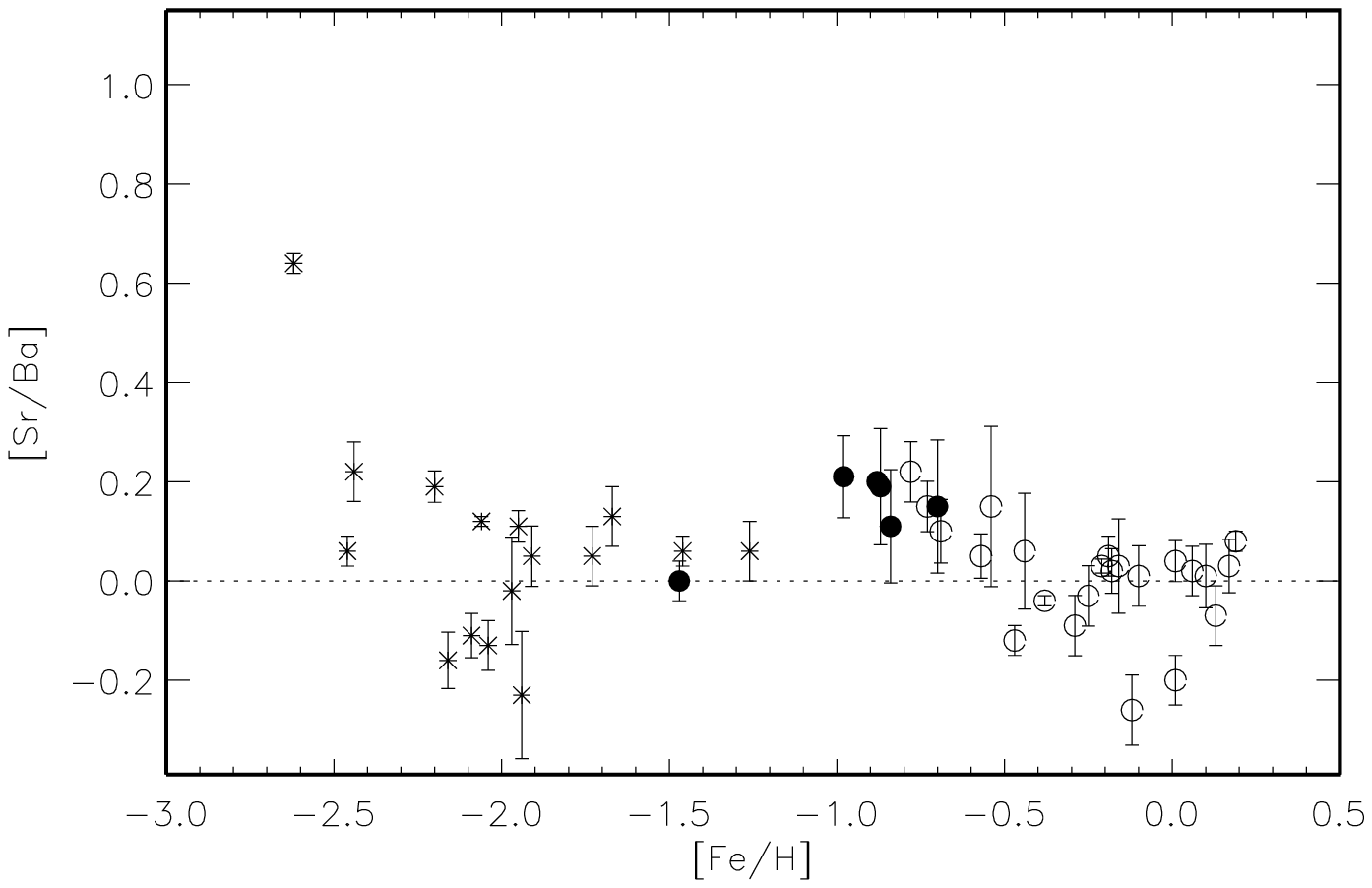}{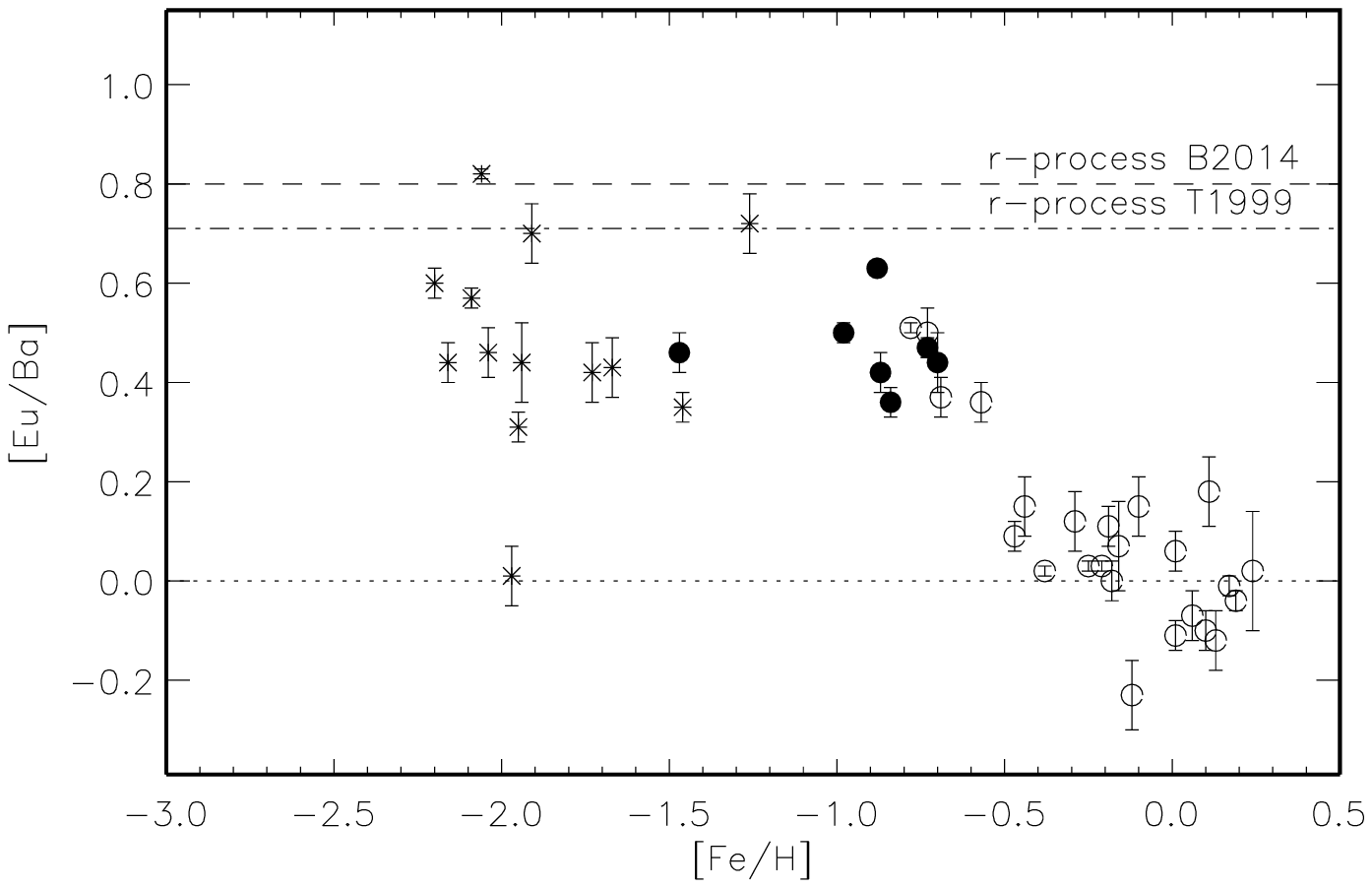}
\plottwo{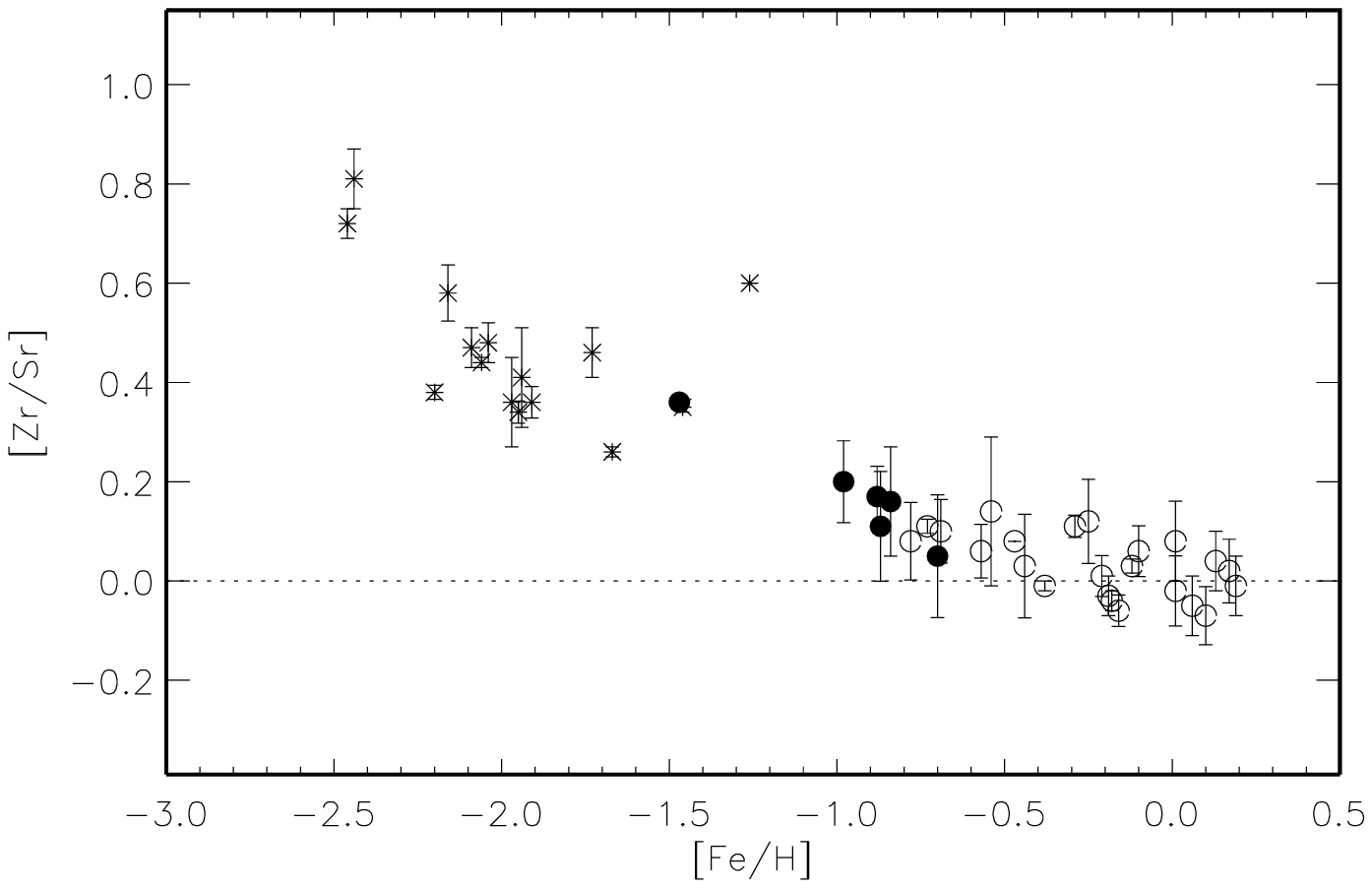}{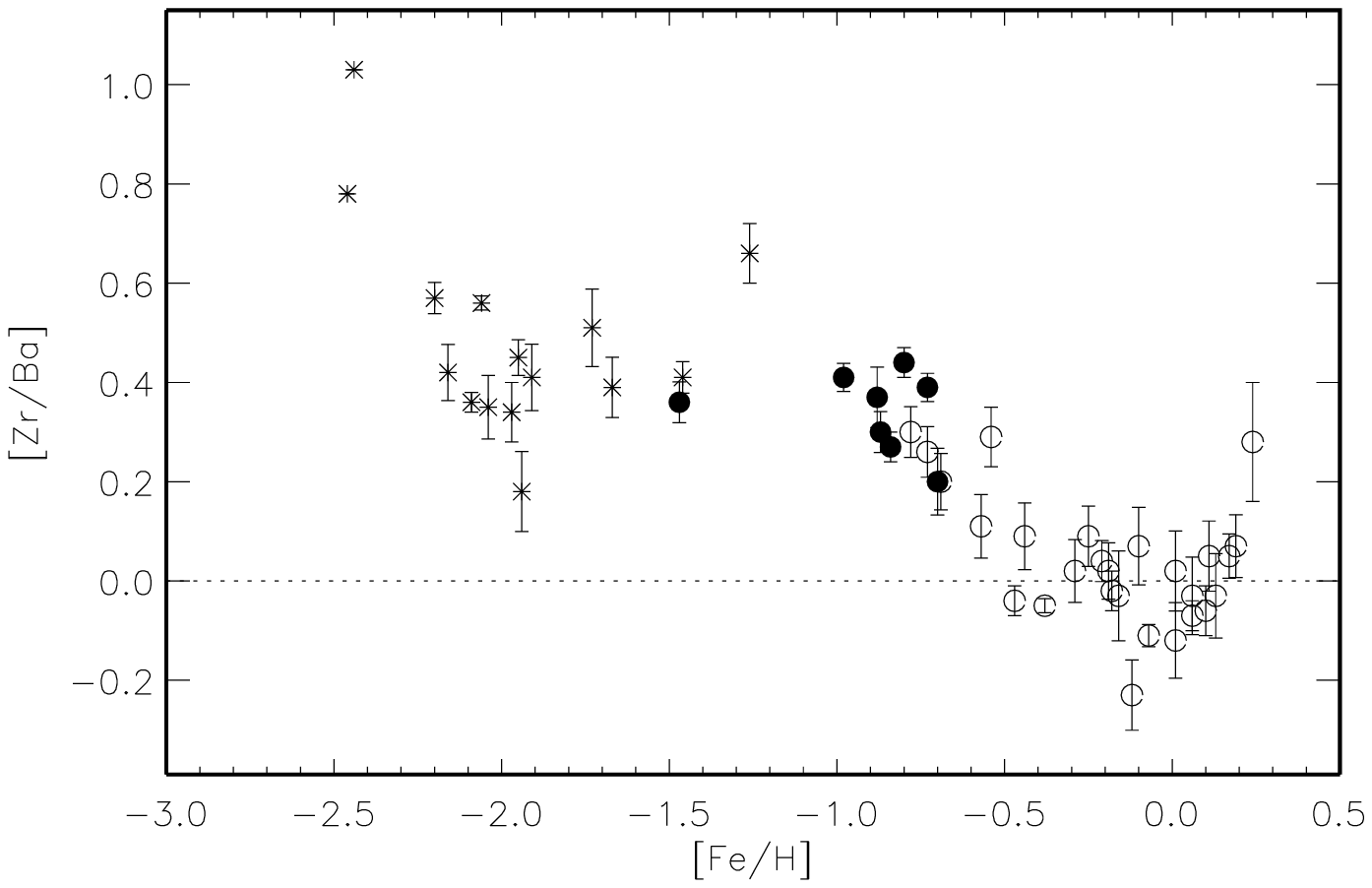}
\caption{Stellar NLTE abundance ratios between the neutron-capture elements. The same symbols are used as in Fig.\,\ref{Fig:li}. The dashed and dash-dotted lines indicate the SSr ratios, as predicted by GCE calculations of \citet{2014arXiv1403.1764B} and \citet{Travaglio1999}, respectively. \label{Fig:ratios3}}
\end{figure*}

{\it Neutron-capture elements: Sr, Zr, Ba, and Eu.}
Here, we concentrate mostly on the $-2.2 \le$ [Fe/H] $\le +0.24$ metallicity range. For none of the three stars at the lower metallicity their Eu abundance was measured, and a big scatter was obtained for Sr/Fe, Zr/Fe, Ba/Fe (Fig.~\ref{Fig:heavy}), and the ratios among the neutron-capture elements (Fig.~\ref{Fig:ratios3}). One of these stars is a well-studied $r$-process poor star HD\,140283 \citep[see][and references therein]{2015A&A...584A..86S} that is strongly underabundant in Sr and Ba relative to Fe and has about 0.4~dex lower Zr/Fe ratio compared with that for the remaining halo stars. In our most MP star, BD\,$-13^\circ$~3442 ([Fe/H] = $-2.62$), abundances of only Sr and Ba were determined, and their ratio deviates strongly from Sr/Ba of the remaining stellar sample. In the $-2.2 \le$ [Fe/H] $\le +0.24$ stars Sr and Ba follow the Fe abundance, although with a substantial scatter of $\pm$0.2~dex. It is worth noting, similar scatter for Ba/Fe was also obtained in the earlier studies \citep[e.g.][]{Edvardsson1993A&A...275..101E,2003A&A...397..275M,2014A&A...562A..71B}. Europium is enhanced relative to Fe in the [Fe/H] $< -1$ stars, with mean [Eu/Fe] $\simeq 0.5$, and downward trend of Eu/Fe, with a rather small scatter of data, is observed at the higher metallicities. Such a behavior is typical of the $r$-process elements, and the knee at [Fe/H] $\simeq -1$ indicates the onset of the Fe production by type Ia supernovae (SNe~Ia). A very similar behavior can be seen in Fig.~\ref{Fig:heavy} for Zr/Fe, although the thin disk stars reveal a less pronounced upward trend compared with that for Eu/Fe.

As seen in Fig.~\ref{Fig:ratios3} the ratios among Sr, Zr, Ba, and Eu reveal the well-defined Galactic trends, with a rather small scatter of data for stars of close metallicity. Barium follows the Sr abundance suggesting their common origin during the period when the Fe abundance of the Galactic matter grew from [Fe/H] $\simeq -2.5$ to the modern value. In the solar system matter 80~\%\ of barium and 80~\%\ of strontium were produced in the slow (s) process of neutron-capture nuclear reactions \citep{Travaglio1999,2004ApJ...601..864T}. For Ba, this is exclusively the main s-process occurring in intermediate-mass stars of $1 - 4\,M_\odot$ during the asymptotic giant branch (AGB) phase, while 9~\%\ of solar Sr originate from the weak s-process occurring in the helium burning core phase of massive stars ($M > 10 M_\odot$). The remaining solar Ba originates from the rapid (r) process. Astrophysical sites for the r-process are still debated, although they are likely associated with explosions of massive stars, with $M > 8 M_\odot$. Analysis of Sr, Y, and Zr in the r-process enhanced stars and extremely MP ([Fe/H] $< -3$) stars led \citet{2004ApJ...601..864T} to suggest the lighter element primary process (LEPP) that in the early Galaxy contributed to the light neutron-capture elements, but did not to the heavy ones, beyond Ba. \citet{2004ApJ...601..864T} estimated empirically the LEPP contribution to solar Sr as 8~\%. Based on our data for stellar Sr/Ba, we infer that, if it existed, the LEPP contribution to galactic Sr did not change during the $-2.5 <$ [Fe/H] $\le +0.24$ epoch.

Europium is enhanced relative to Ba in our halo and thick disk stars, with a scatter of [Eu/Ba] between 0.31 and 0.82. The mean is [Eu/Ba] = 0.50$\pm$0.14. As discussed, a halo star HD~74000 is an outlier, and it was not included in the mean. The thin disk stars reveal the upward trend in Eu/Ba towards lower metallicity. Europium is referred to as an r-process element, because only 6~\%\ of solar Eu originate from the s-process \citep{Travaglio1999}. Theoretical predictions of a pure r-process production of Eu and Ba give [Eu/Ba]$_{\rm r} \simeq$ 0.67 in the classical waiting-point (WP) approximation \citep{Kratz2007} and [Eu/Ba]$_{\rm r} \simeq$ = 0.87 in the large-scale parameterised dynamical network calculations of \citet{Farouqi2010} in the context of an adiabatically expanding high-entropy wind (HEW), as is expected to occur in core-collapse SNe. The solar r-residual, i.e. the difference between solar total and s-abundance, where the s-abundance is deduced from the Galactic chemical evolution models, ranges between [Eu/Ba]$_{\rm r}$ = 0.71 \citep{Travaglio1999} and 0.80 \citep[]{2014arXiv1403.1764B}. Our data on Eu/Ba (top right panel in Fig.~\ref{Fig:ratios3}) provide an evidence for dominant contribution of the r-process to a production of Ba and Eu in the early Galaxy, when the halo and thick disk stellar population formed, and rapidly growing enrichment of the Galactic matter by s-nuclei, when metallicity increased from [Fe/H] $\simeq -0.8$ to solar value.

A behavior similar to that for Eu/Ba is observed also for Zr/Ba, although an enhancement of Zr relative to Ba in the halo and thick disk stars is, on average, smaller, if not to count the two most MP stars. According to \citet{2004ApJ...601..864T}, 67~\%\ of the solar Zr were contributed from the main and weak s-process, and 15~\%\ and 18~\%\ were attributed to the r- and LEPP-process, respectively. Considering contributions from the r- and LEPP-process together, we deduce the solar r-residual [Zr/Ba]$_{\rm r} \simeq$ = 0.22. This is smaller than [Zr/Ba] observed in the halo and thick disk stars.
As expected, the Zr/Sr ratio is close to the solar value in the thin disk stars, but it grows steeply in the thick disk and halo stars, approaching [Zr/Sr] $\simeq 0.8$ at [Fe/H] = $-2.5$. Using predictions of \citet{2004ApJ...601..864T}, we deduced [Zr/Sr]$_{\rm r+LEPP}$ = 0.22 and [Zr/Sr]$_{\rm LEPP}$ = 0.35 for production of Sr and Zr in the r- and LEPP-process together and in a pure LEPP-process. An origin of Zr in the thick disk stars can be attributed to these two processes. However, further efforts should be invested to understand high Zr/Sr ratios observed in the [Fe/H] $< -2$ stars.

Obtained abundances of Sr, Zr, Ba, and Eu support, in general, the literature data in the metallicity range overlapping with ours, despite the fact that most studies were performed under the LTE assumption. This is because the NLTE abundance corrections for lines of \ion{Sr}{2}, \ion{Zr}{2}, and \ion{Eu}{2} are small in the stellar parameter range, with which we concern (see Sect.~\ref{sect:nlte3}). The upward trend in [Zr/Fe] was reported by \citet{Mashonkina2007} and \citet{2013A&A...552A.128M}, although the latter paper studied a narrow range of metallicity, down to [Fe/H] $\simeq -1$. Enhancement of Eu relative to Fe and Ba in the halo and/or thick disk stars was obtained earlier by \citet{mash_eu,Burris2000ApJ...544..302B,HERESII,Bensby2005A&A...433..185B}, and \citet{2013A&A...552A.128M}. With different stellar sample, \citet{Mashonkina2007} found the Galactic trends for Zr/Ba and Zr/Sr that are very similar to ours.

\subsection{Influence of NLTE on the Galactic abundance trends}

\begin{figure*}
\epsscale{0.80}
\plottwo{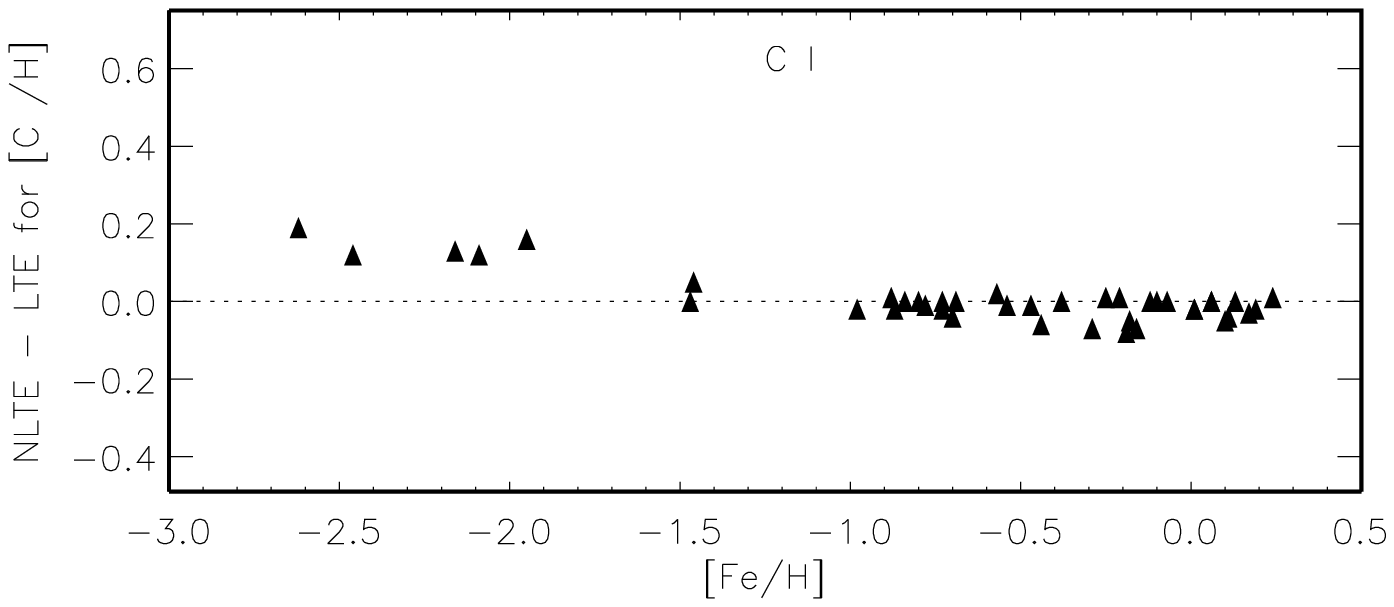}{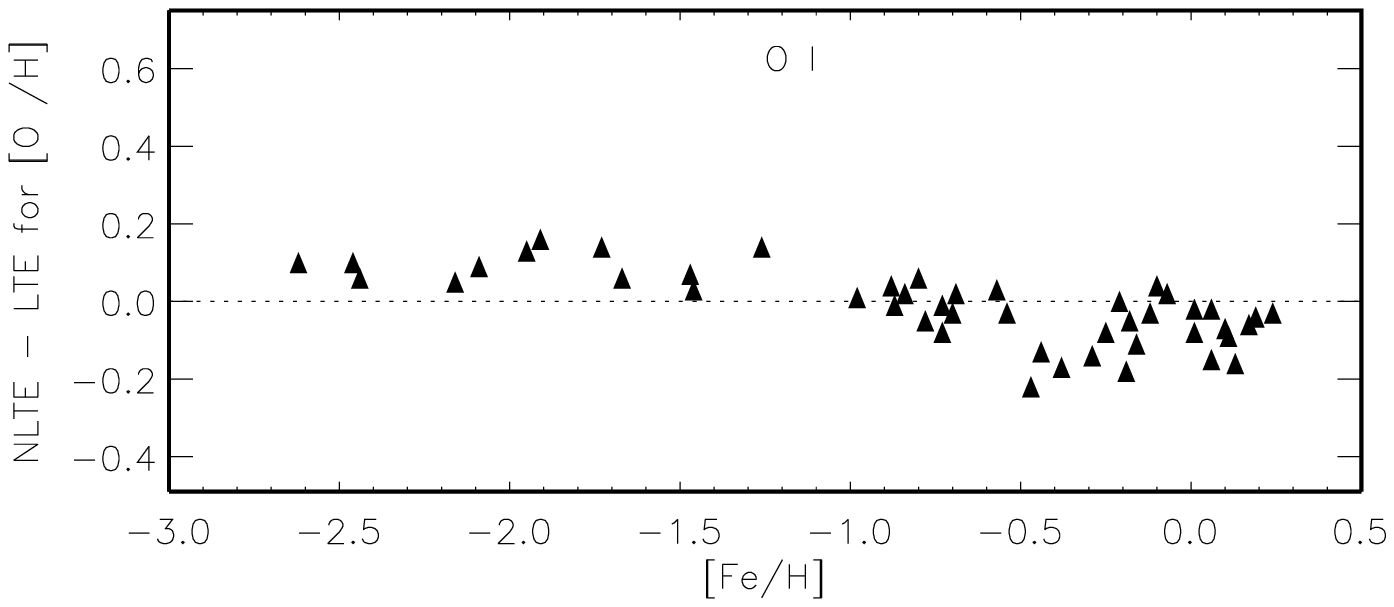}
\plottwo{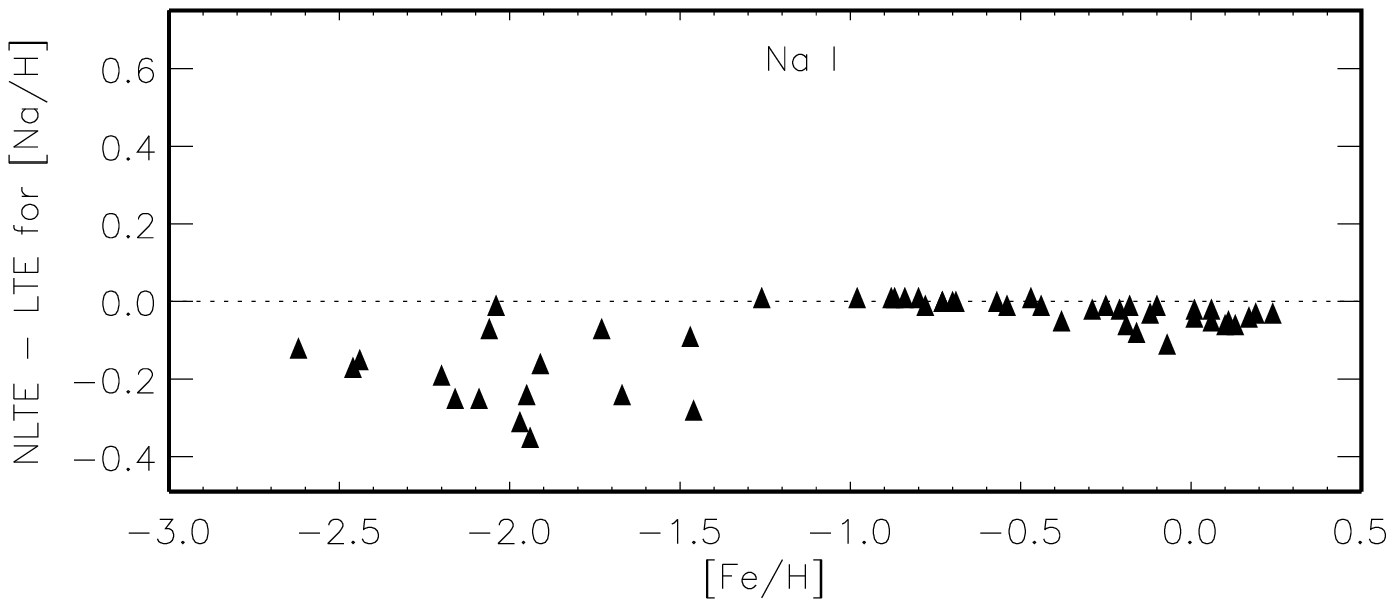}{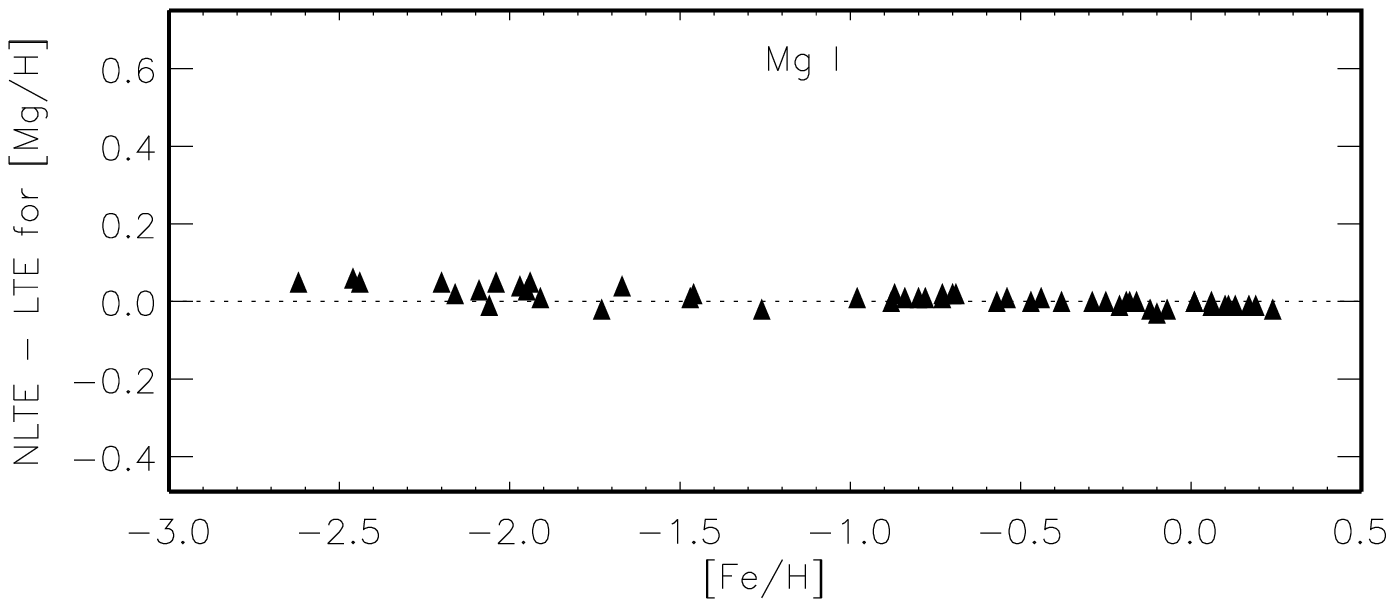}
\plottwo{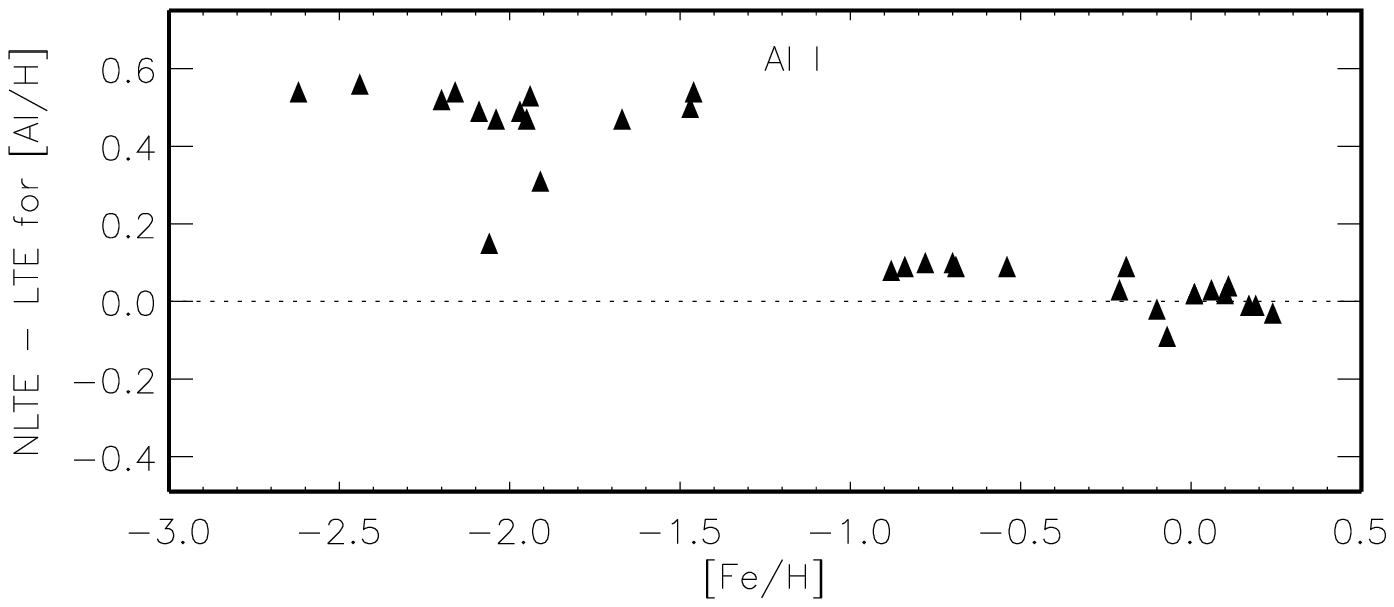}{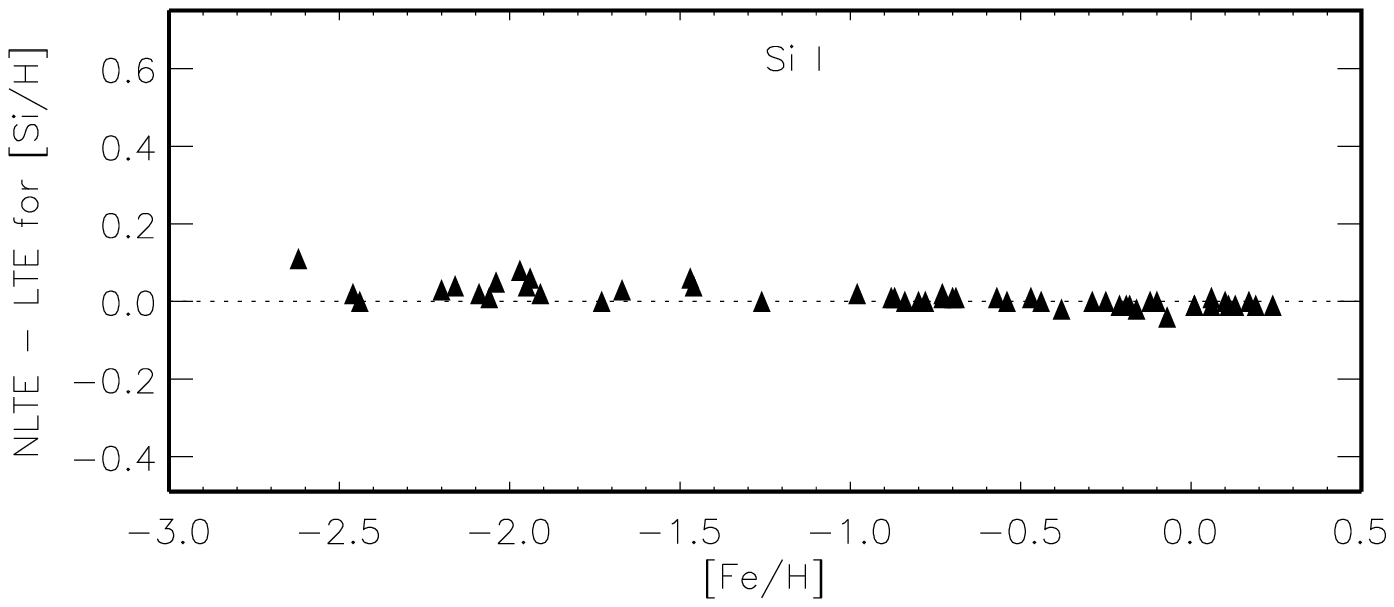}
\plottwo{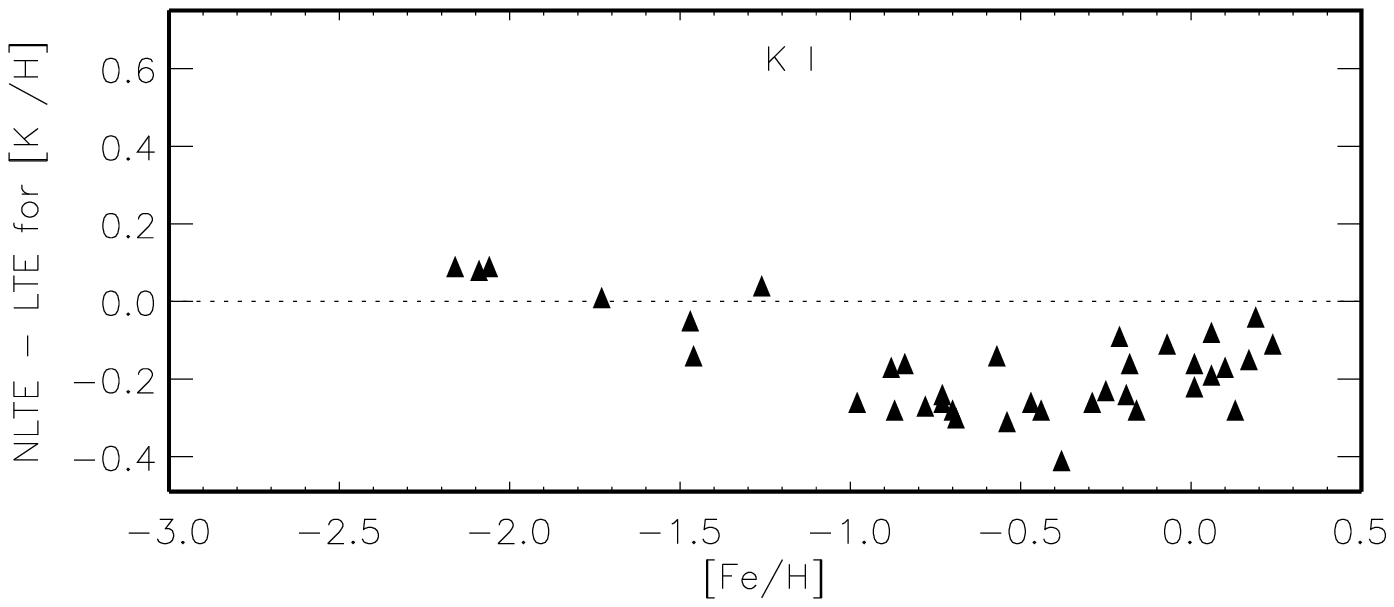}{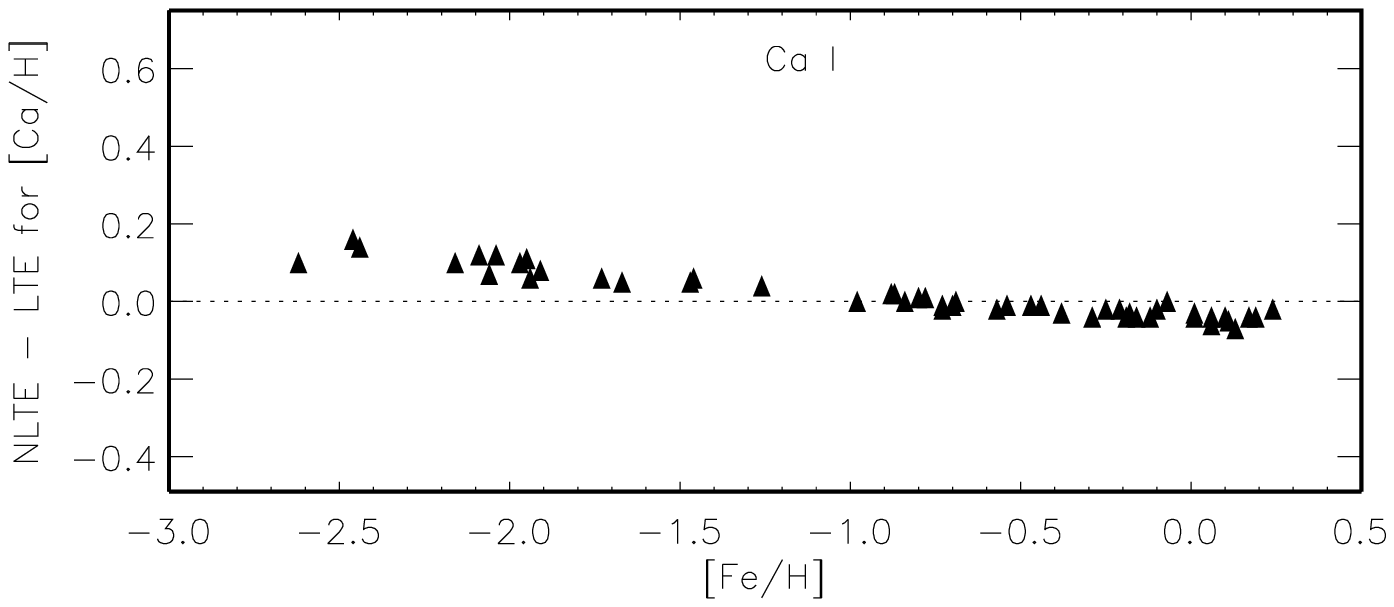}
\plottwo{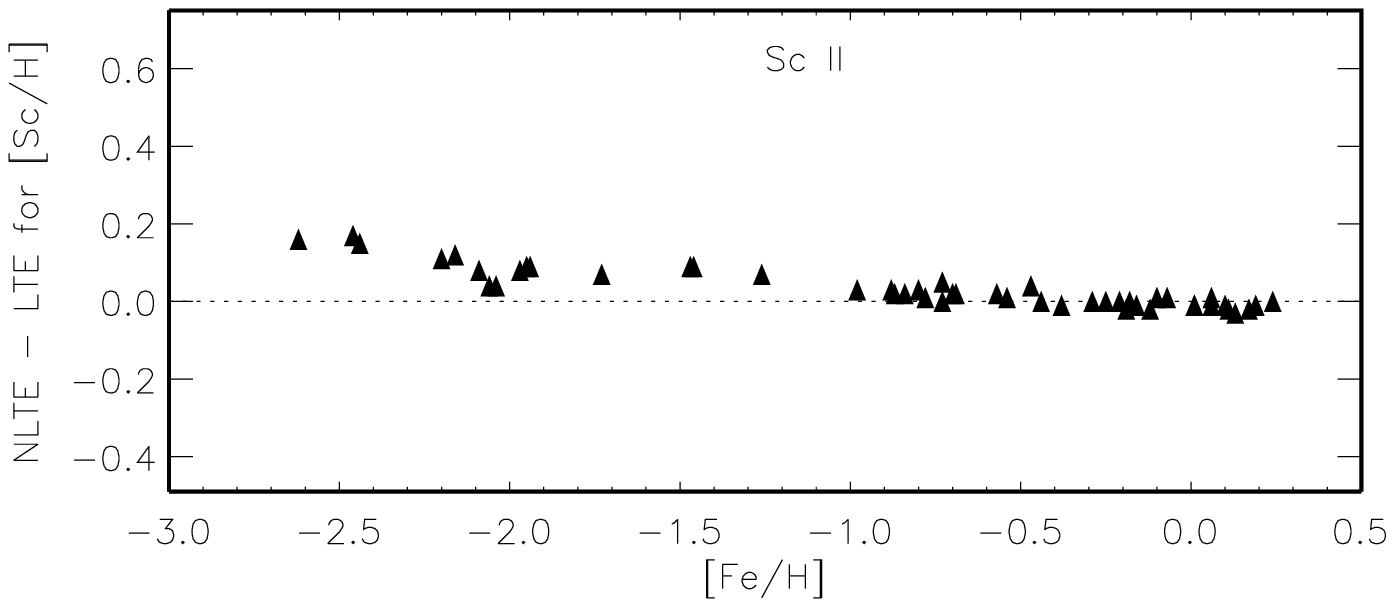}{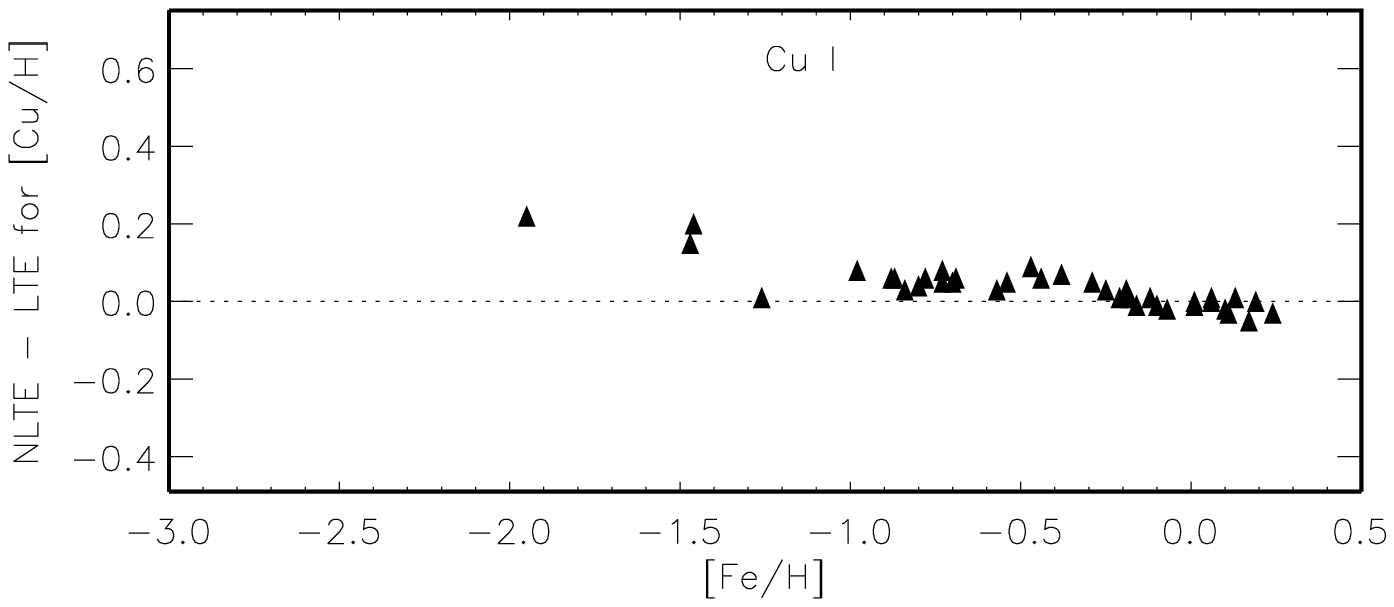}
\plottwo{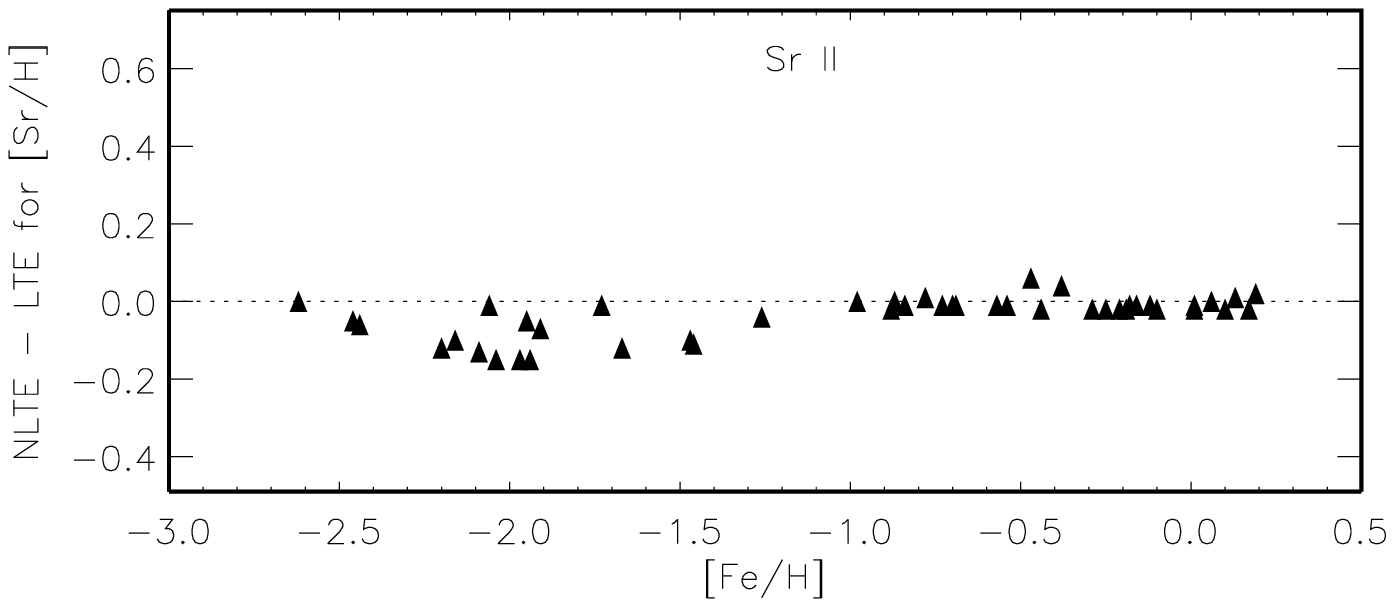}{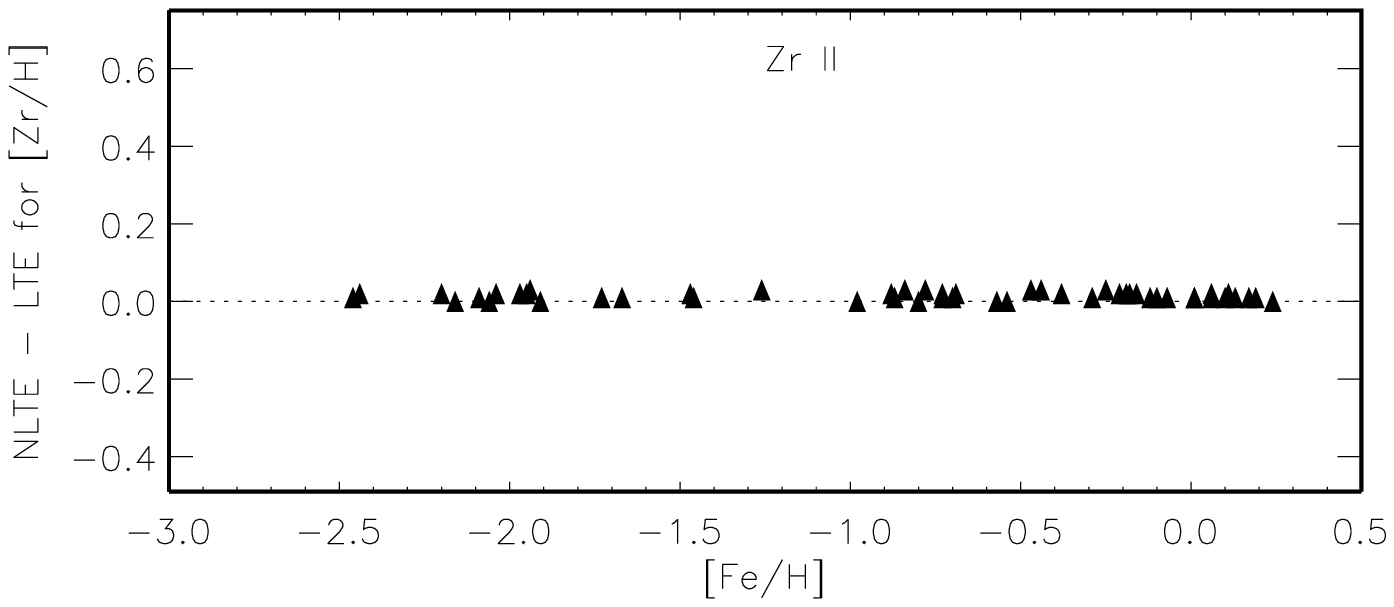}
\plottwo{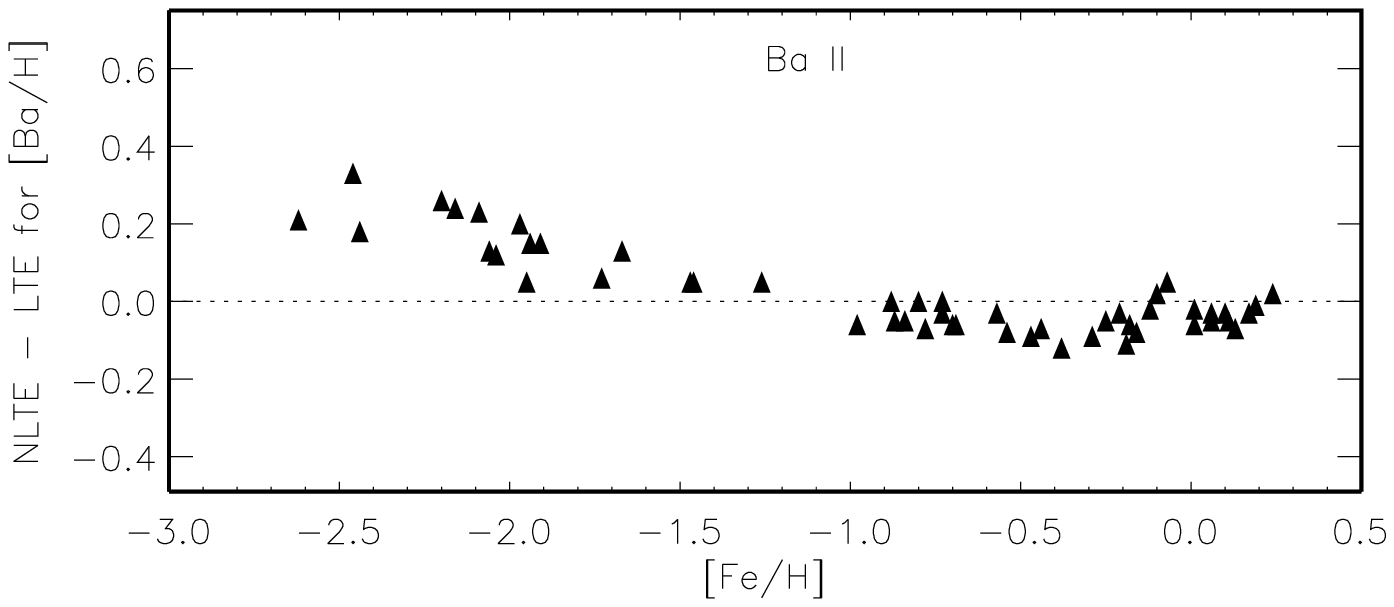}{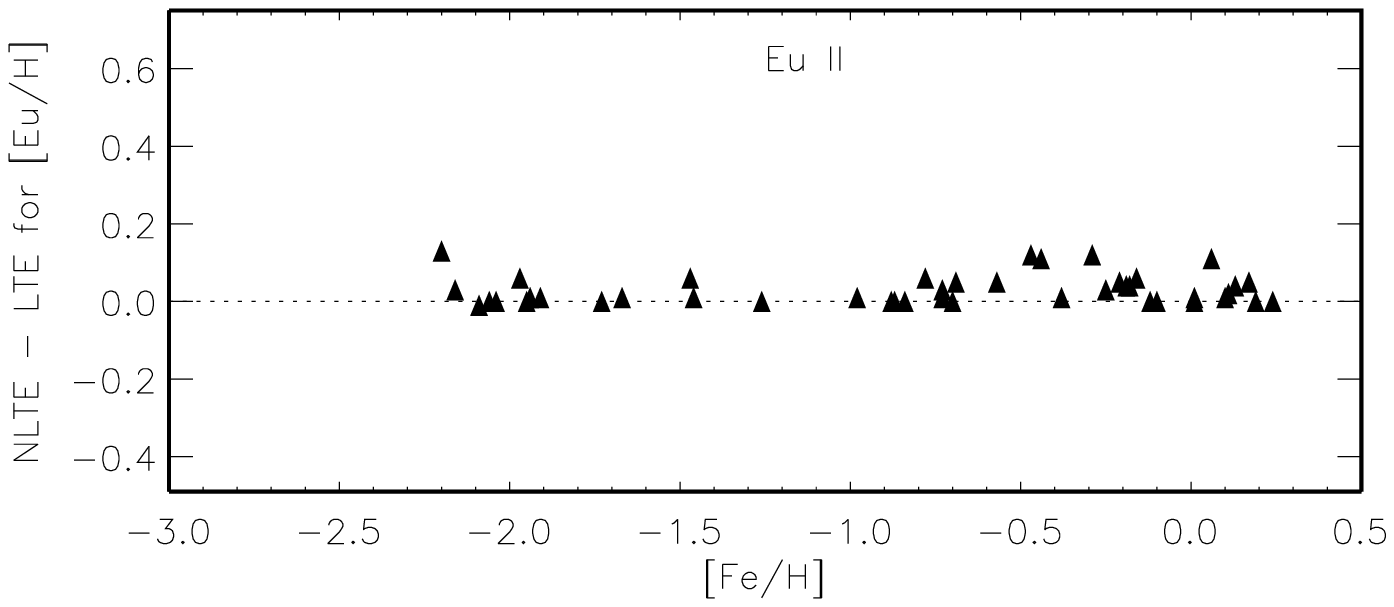}
\caption{Differences in differential abundance [X/H] between NLTE and LTE for the investigated sample. \label{Fig:diff_nlte}}
\end{figure*}

As noted, this study employs a line-by-line differential analysis with respect to the Sun. Here, we discuss an impact of NLTE on determination of the mean element abundances [X/H] and elemental ratios, depending on the star's metallicity. Figure\,\ref{Fig:diff_nlte} displays the differences between the NLTE and LTE [X/H] ratio for the 14 investigated species. We do not show the data for \ion{Li}{1} (see Fig.~\ref{Fig:dnlte} for the NLTE effects), \ion{Ti}{2} due to minor differences, and \ion{Fe}{1}-\ion{Fe}{2}, which were discussed in Paper~I.

A differential approach largely cancels the (NLTE - LTE) differences in [X/H] for most species in the [Fe/H] $> -1$ stars, even if the departures from LTE for individual lines are large. This concerns, in particular, [C/H], [Na/H], [Ca/H], and [Ba/H]. For example, $\Delta_{\rm NLTE}$ for lines of \ion{C}{1} can be up to $-0.4$, while the [C/H] differences between the NLTE and LTE do not exceed 0.1~dex in absolute value. However, notable ($> 0.1$~dex) differences between NLTE and LTE in the [Fe/H] $> -1$ stars remain for [O/H] and [K/H]. It is worth noting that NLTE can also affect the [Eu/Ba] ratios because the differences between NLTE and LTE are negative for [Ba/H], but positive for [Eu/H]. An advantage of NLTE is also proved by the smaller line-to-line scatter obtained for most species and most stars in NLTE compared with LTE (see Table~\ref{Tab:AbundanceSummary}). For example, for HD~49933 (6600/4.15/$-0.47$) LTE leads to [O/H] = $-0.15\pm0.11$, [Ca/H] = $-0.37\pm0.08$, [Ba/H] = $-0.22\pm0.13$, while remarkably smaller statistical errors are obtained in NLTE, with [O/H] = $-0.37\pm0.04$, [Ca/H] = $-0.43\pm0.04$, and [Ba/H] = $-0.31\pm0.03$.

NLTE is a major step forward for studies of stars more metal-poor than [Fe/H] = $-1$. The (NLTE - LTE) differences in [X/H] grow in absolute value towards lower metallicity and, for most species, can reach 0.2~dex and even more. Exceptions are [Mg/H], [Si/H], [K/H], [Zr/H], and [Eu/H], where the departures from LTE are small. NLTE is, in particular, important for elemental ratios involving the species with (NLTE - LTE) of different sign, like [Na/Mg], [Na/Al], [Na/Cu], [Sr/Ba]. For example, the mean for the halo stars, excluding HD~74000 and G090-003, amounts to [Na/Mg] = $-0.47\pm0.10$ in NLTE and $-0.25\pm0.19$ in LTE. NLTE makes Al following Na over the whole metallicity range under investigation, with the mean [Na/Al] = $-0.01\pm$0.14, while LTE finds a large overabundance of Na relative to Al in the halo and thick disk stars ([Na/Al] = 0.58$\pm$0.38) and close-to-solar Na/Al ratios in the thin disk stars.

\subsection{Comparison with the Galactic chemical evolution models}

\begin{figure*}
\epsscale{1.0}
\plottwo{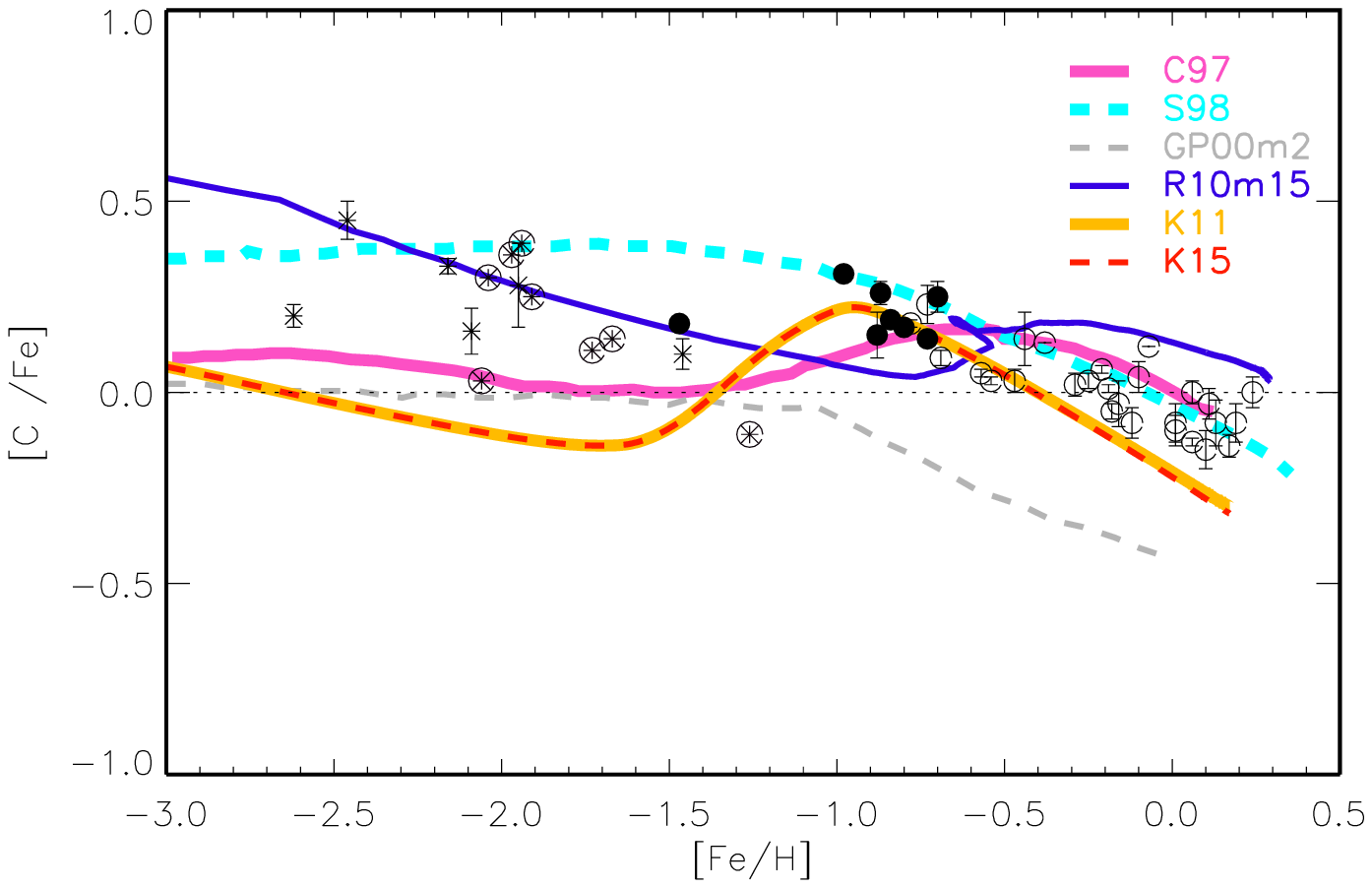}{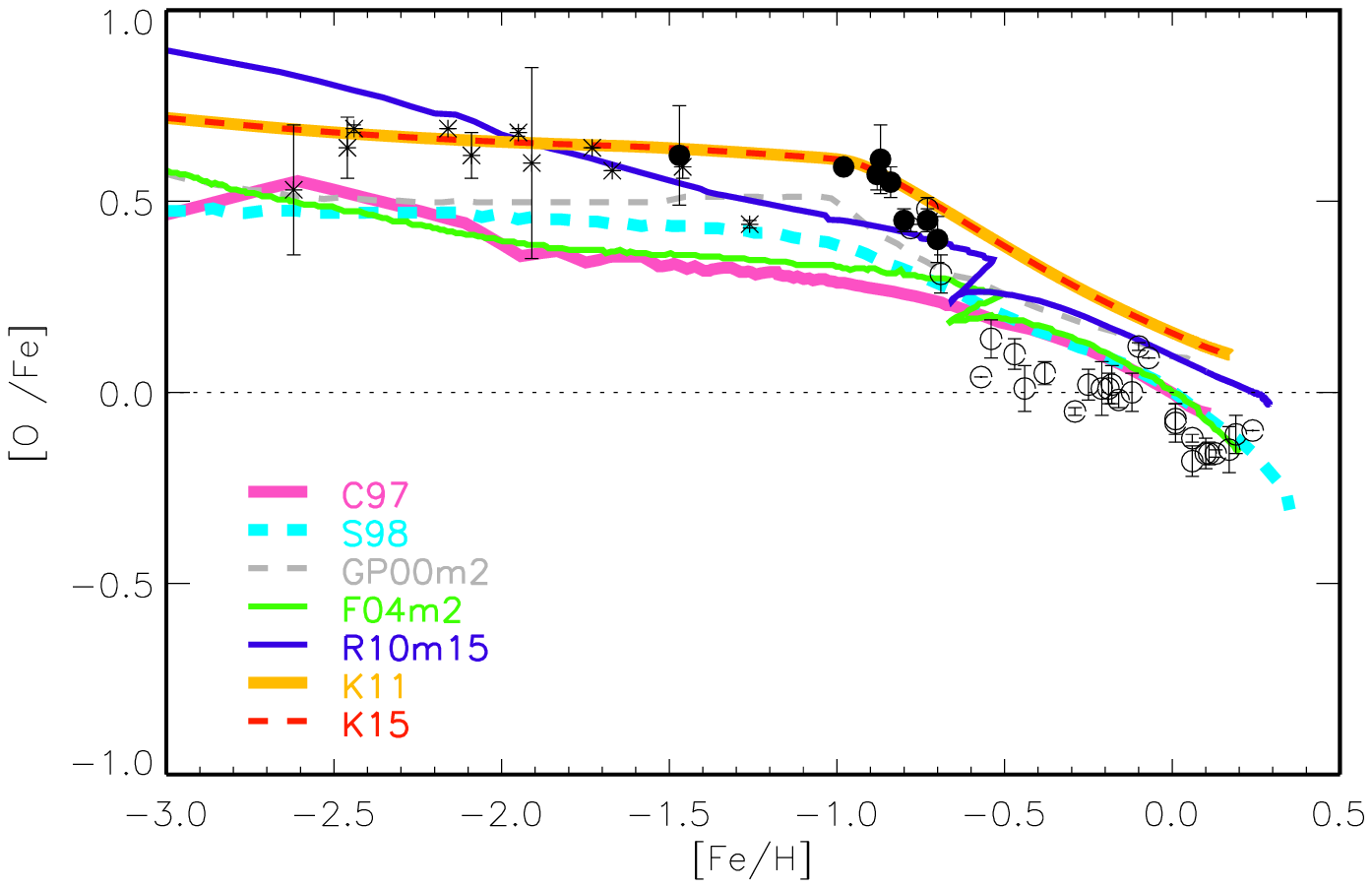}
\plottwo{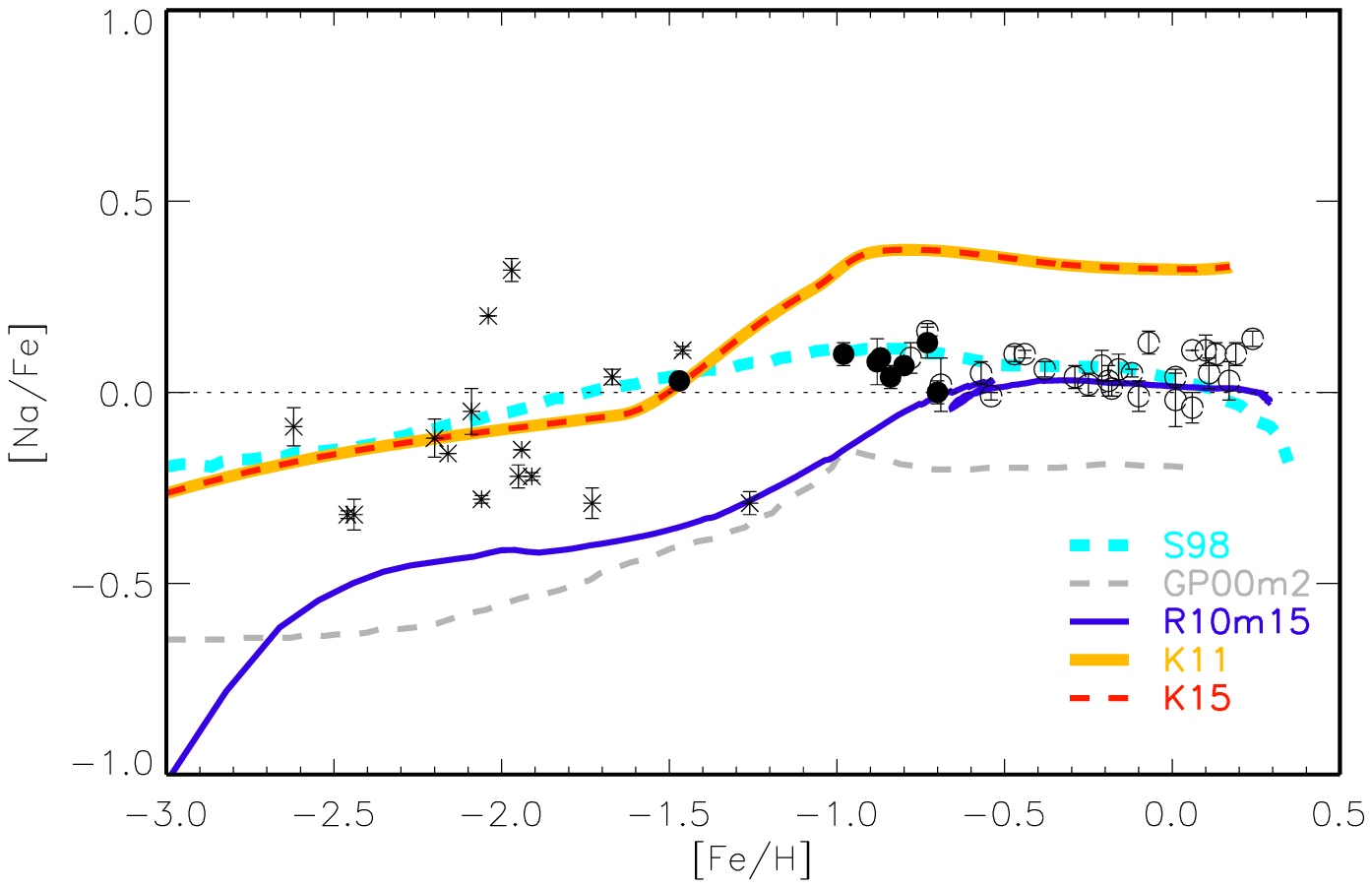}{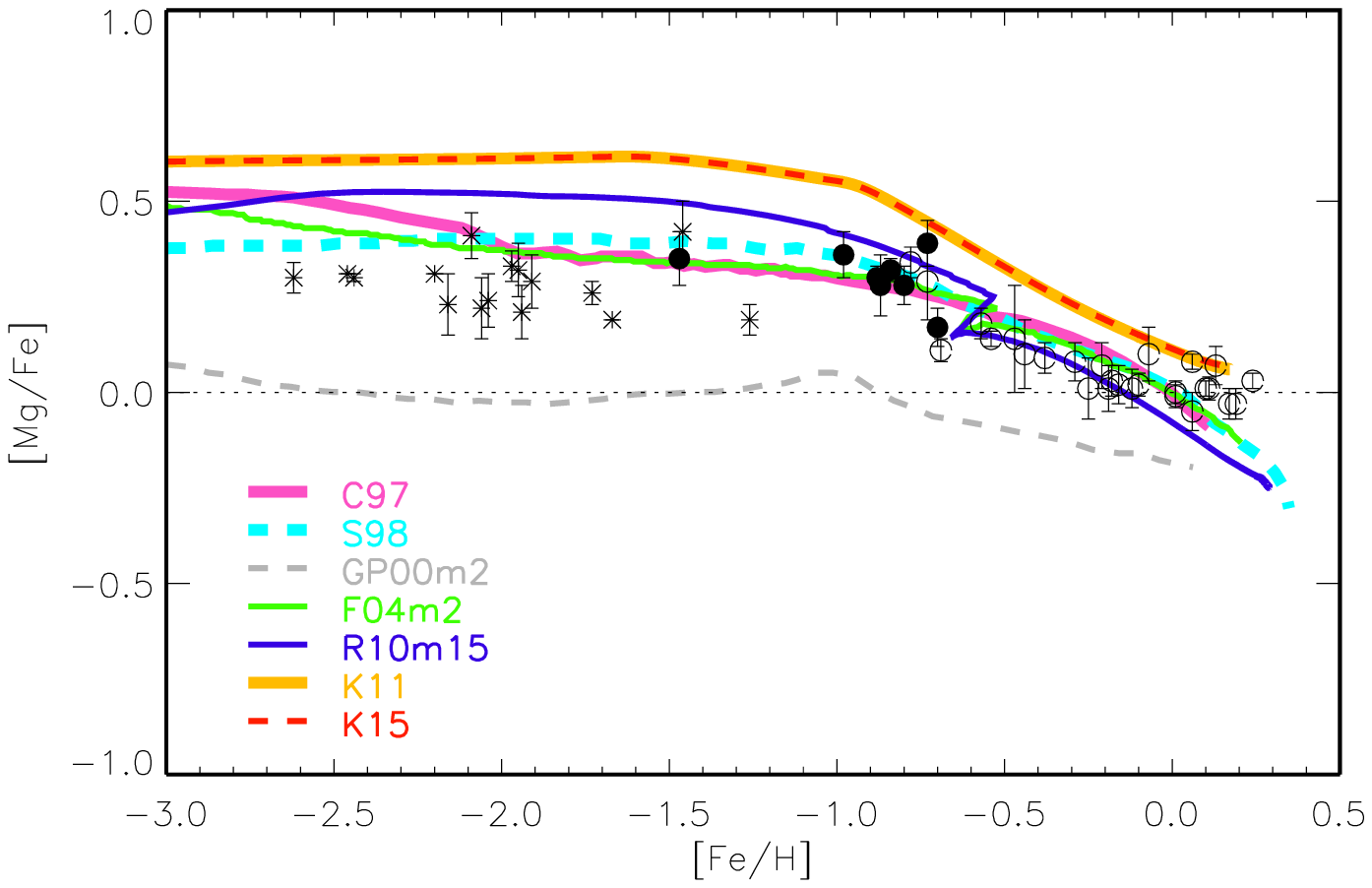}
\plottwo{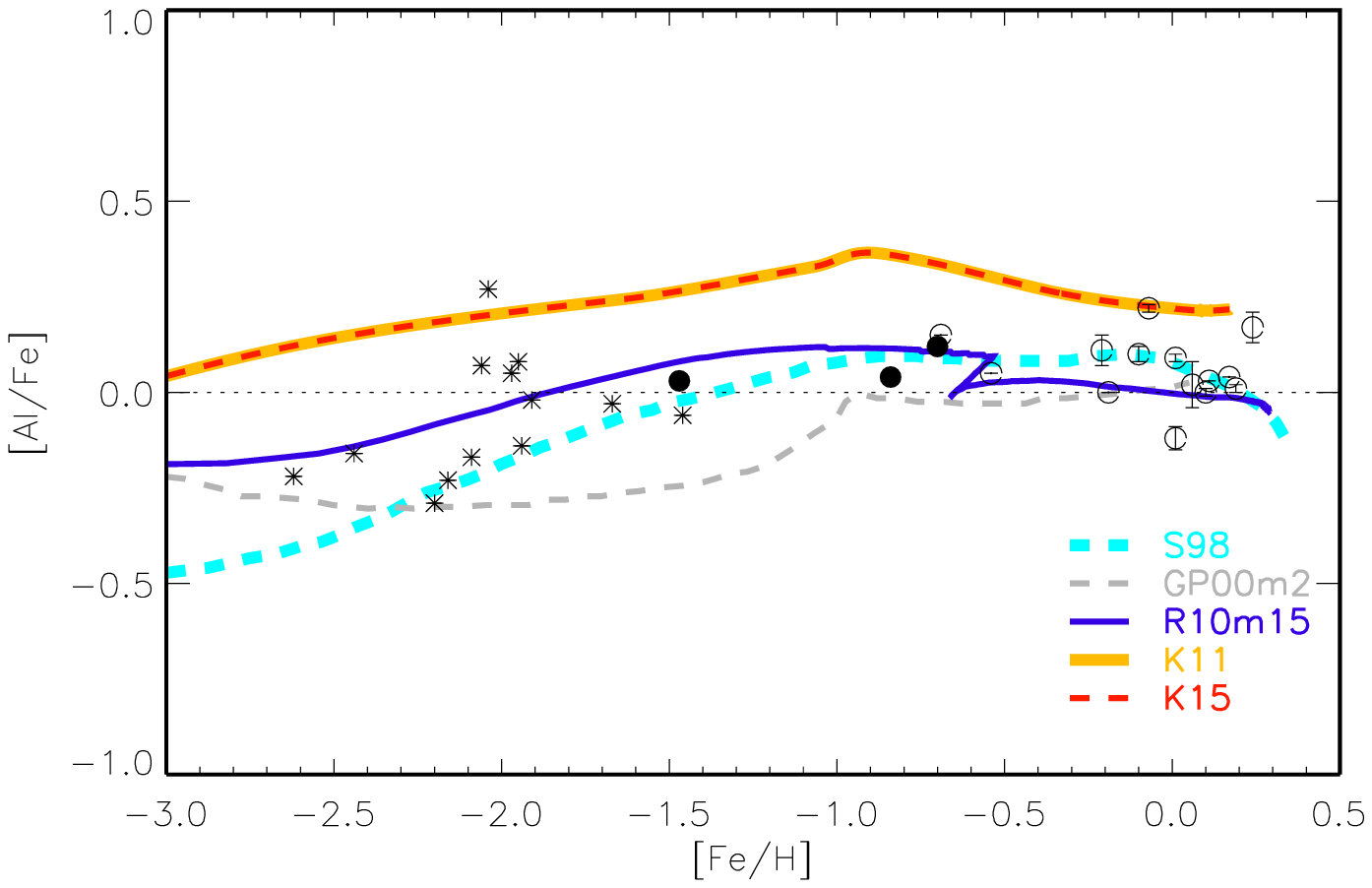}{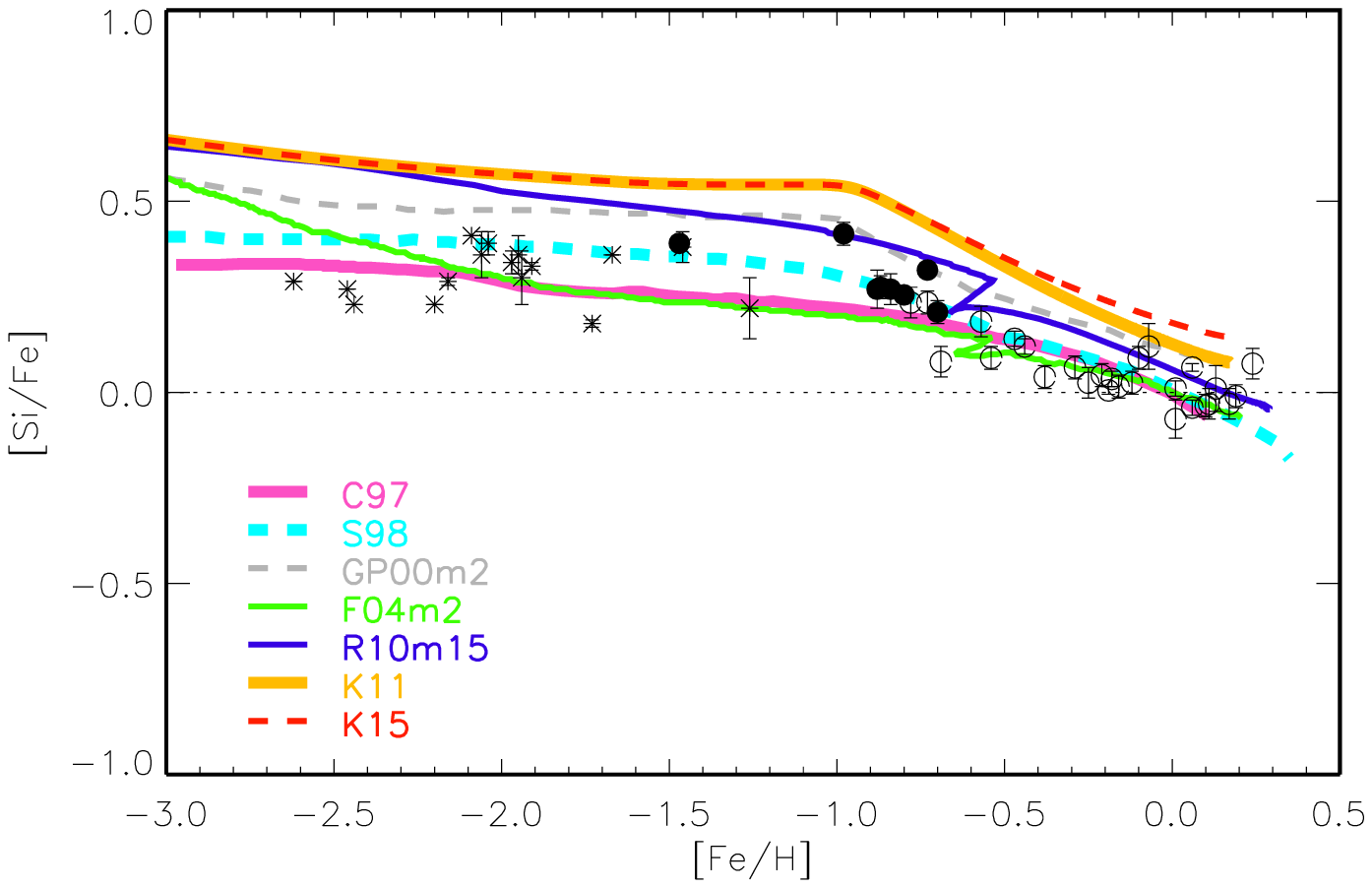}
\plottwo{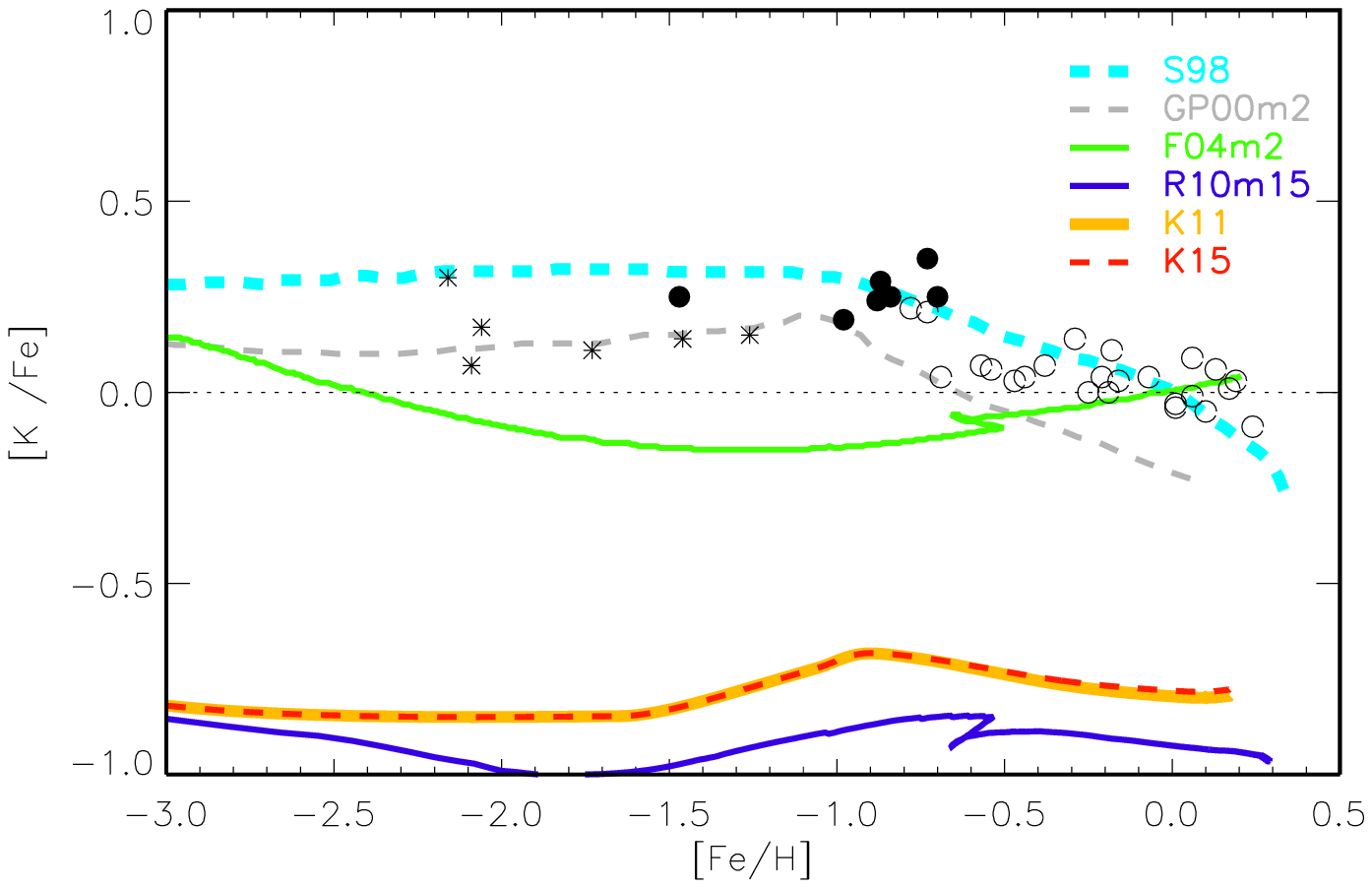}{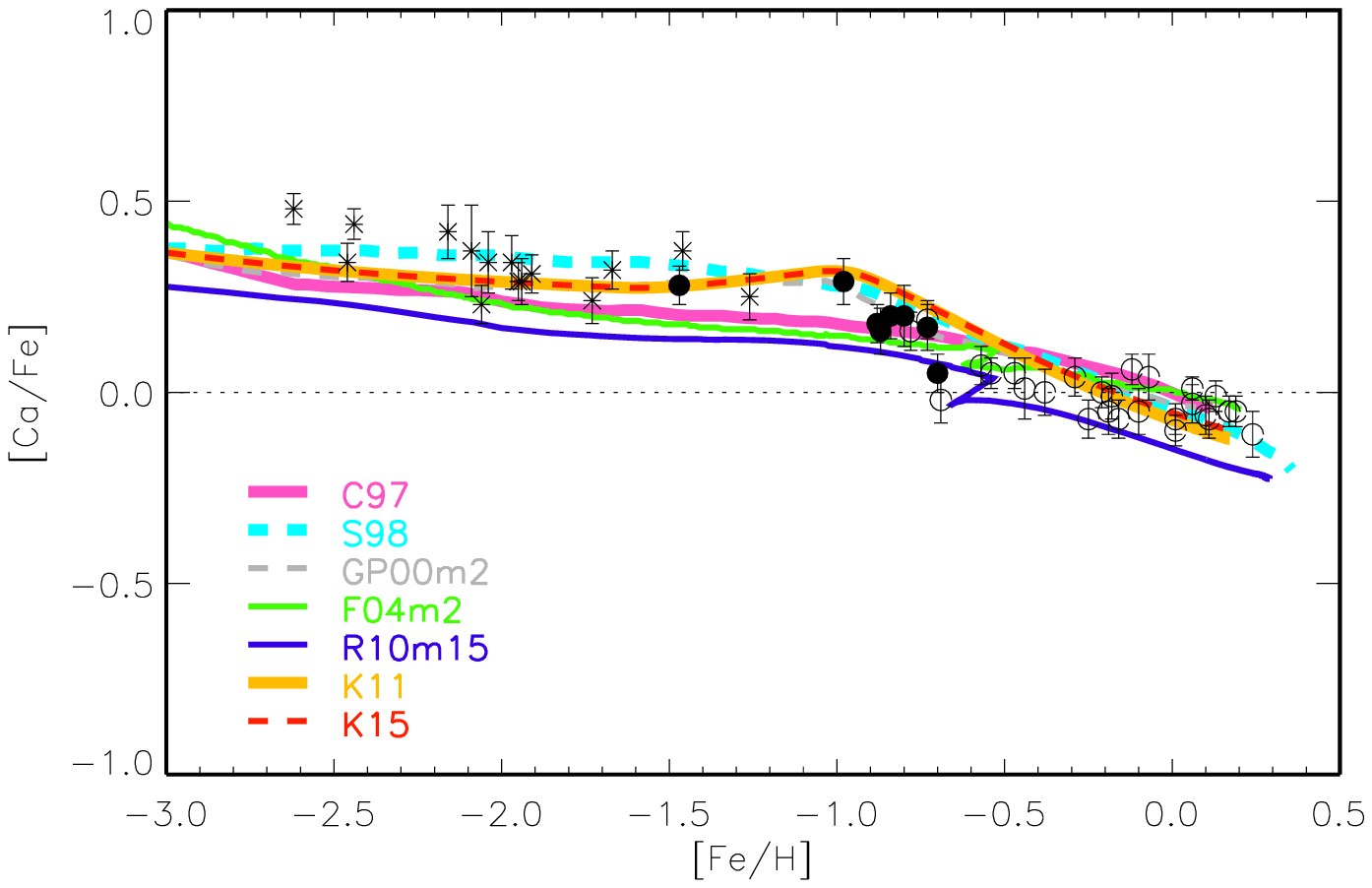}
\caption{Comparison with the Galactic chemical evolution models. The models used in the figure are C97 (violet solid line), S98 (turquoise dashed line), GP00m2 (gray dashed line), F04m2 (green solid line), R10m15 (blue solid line), K11 (orange solid line), and K15 (red dashed line), where GP00m2, F04m2 and R10m15 represents the model of `thick curve' in GP00, the model of Fig4-6 in F04, and the model 15 in R10, respectively.\label{Fig:GCE1}}
\end{figure*}

\begin{figure*}
\epsscale{1.0}
\plottwo{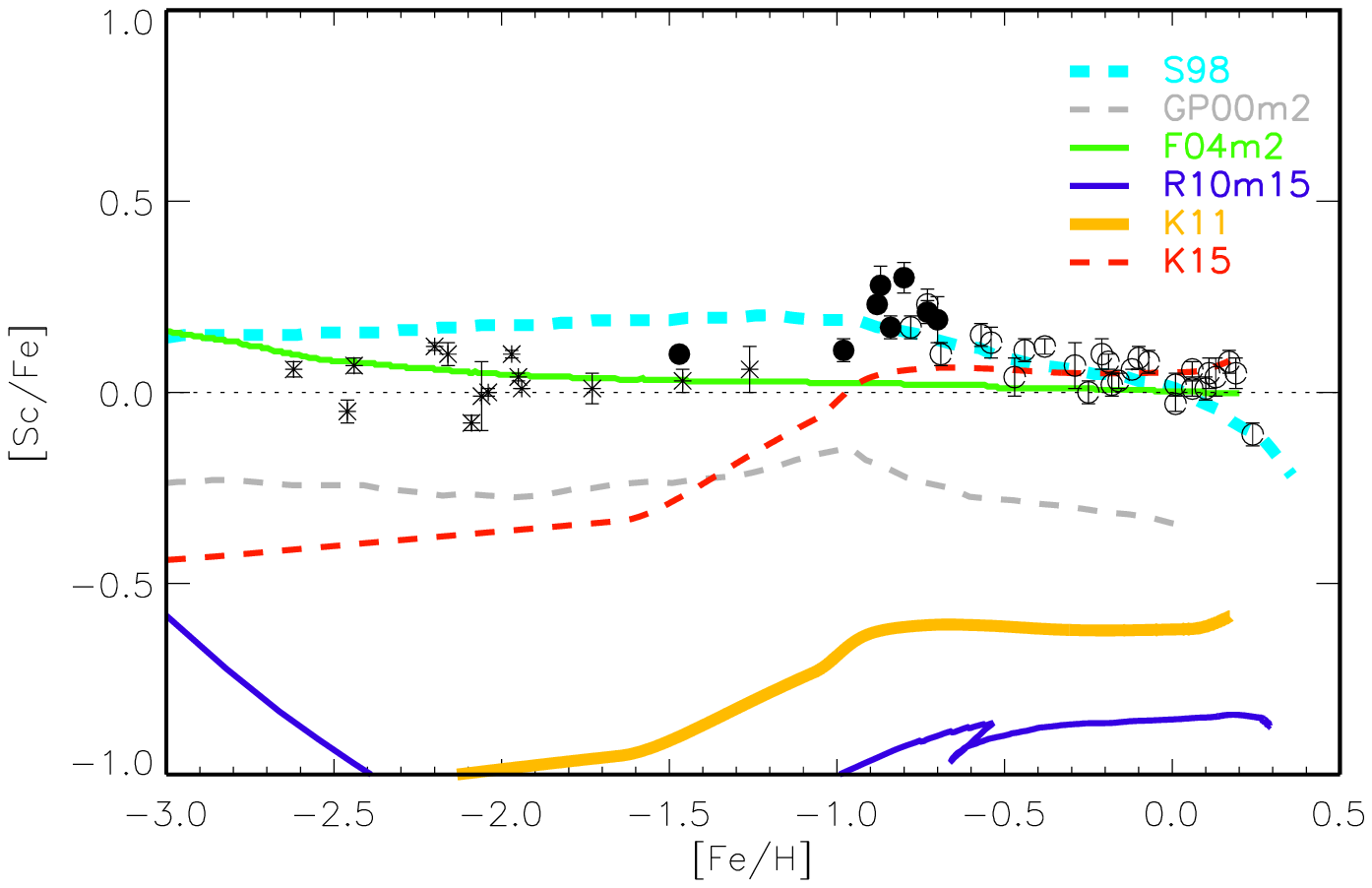}{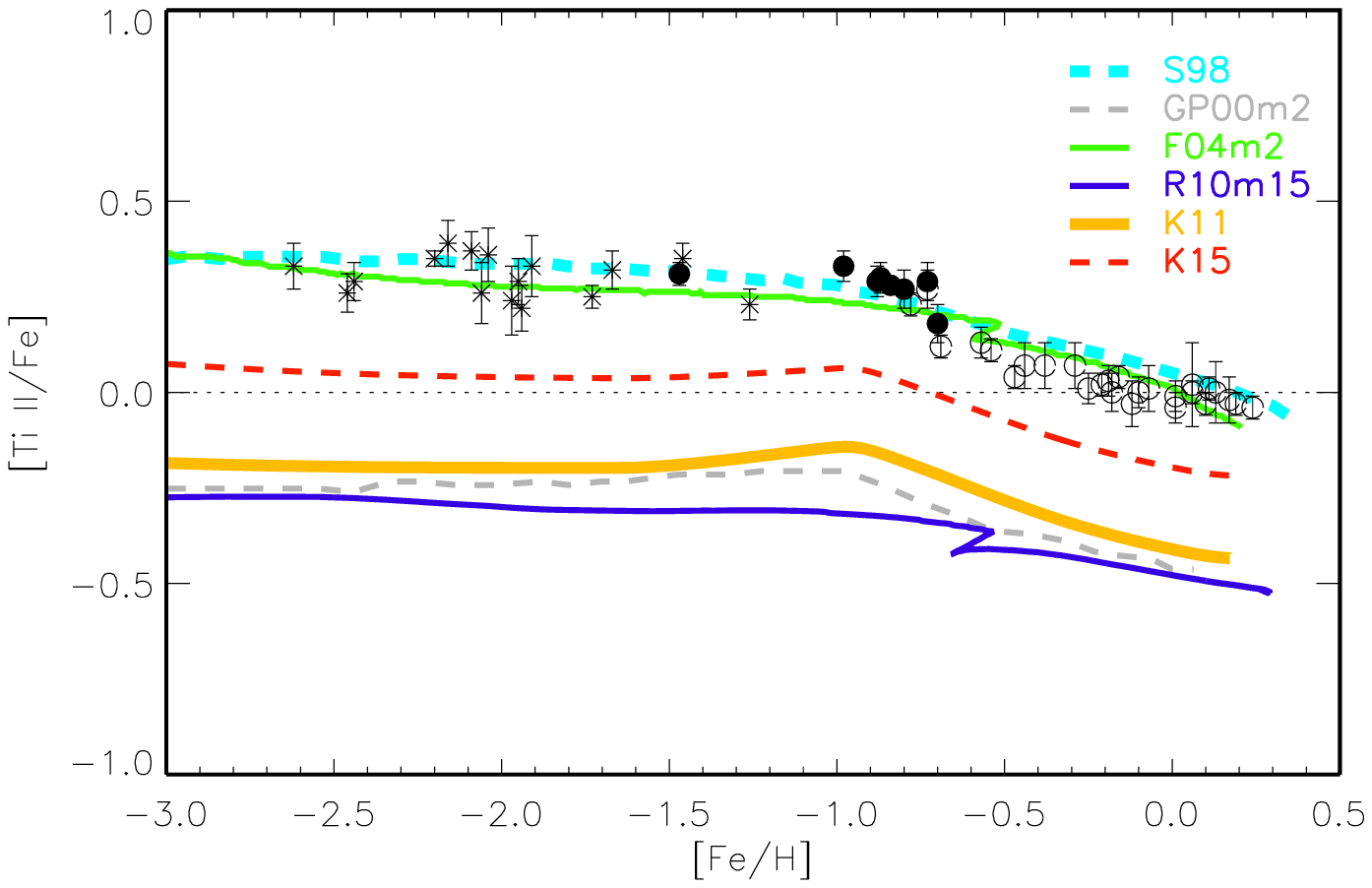}
\plottwo{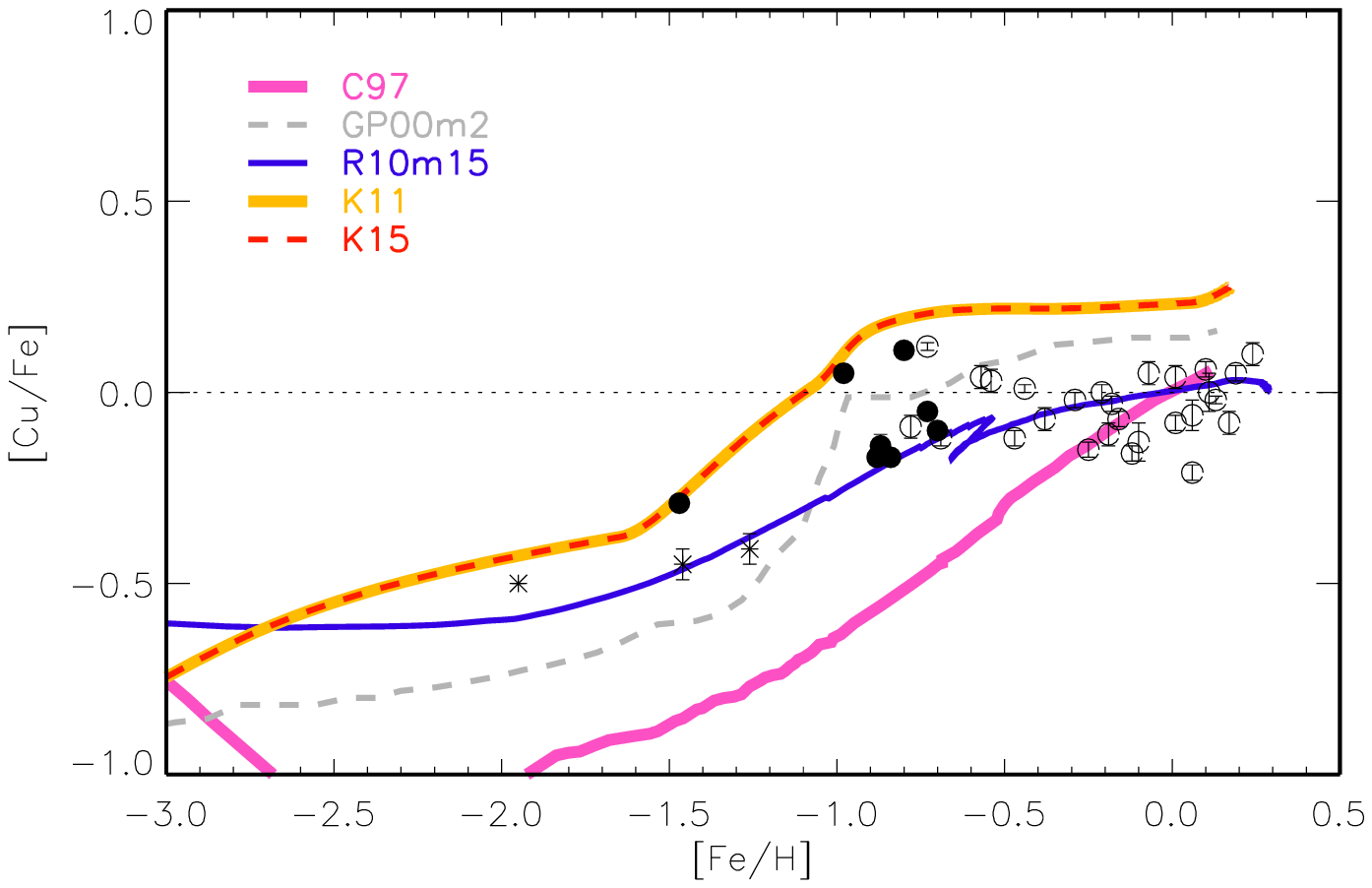}{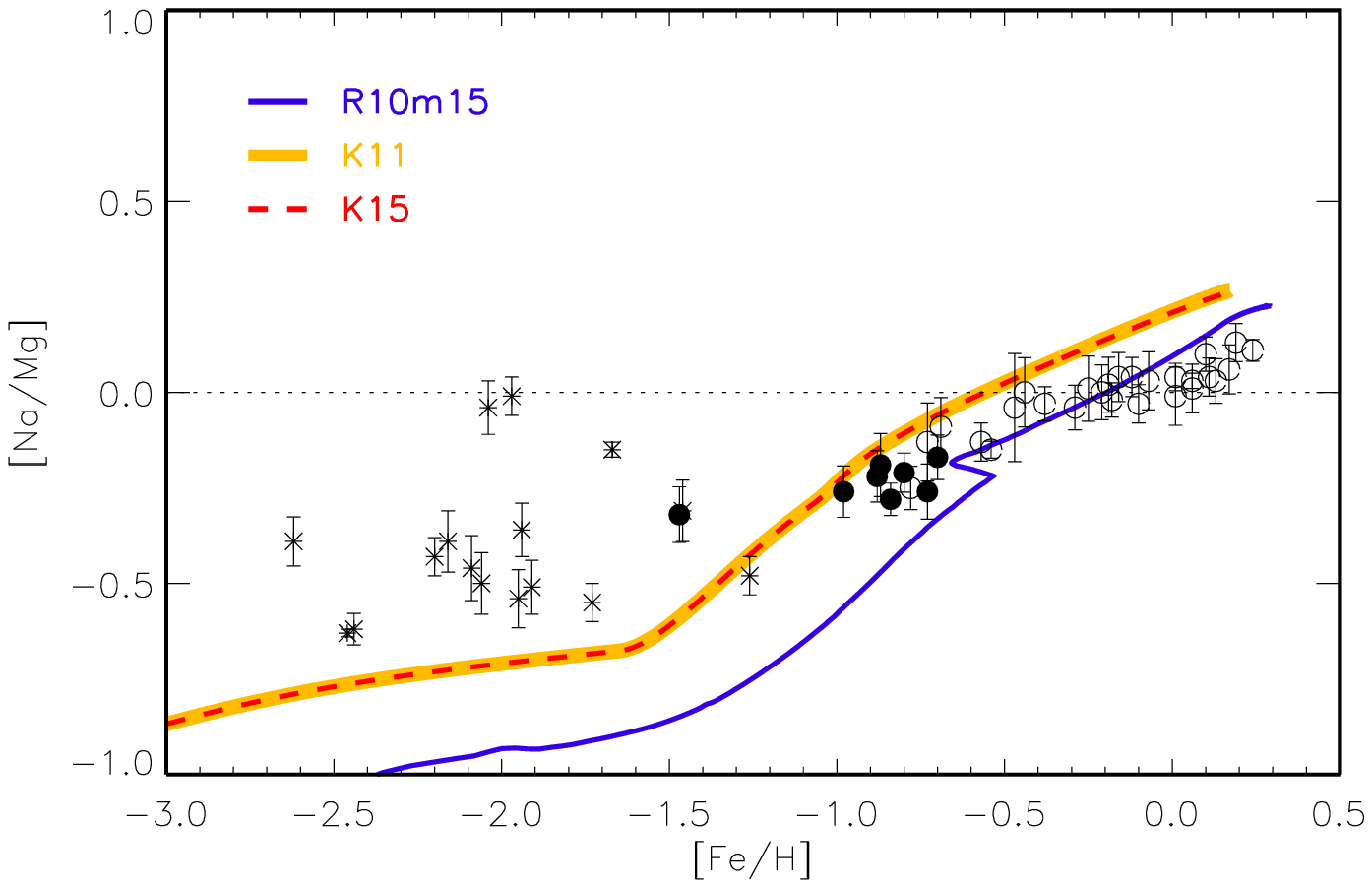}
\plottwo{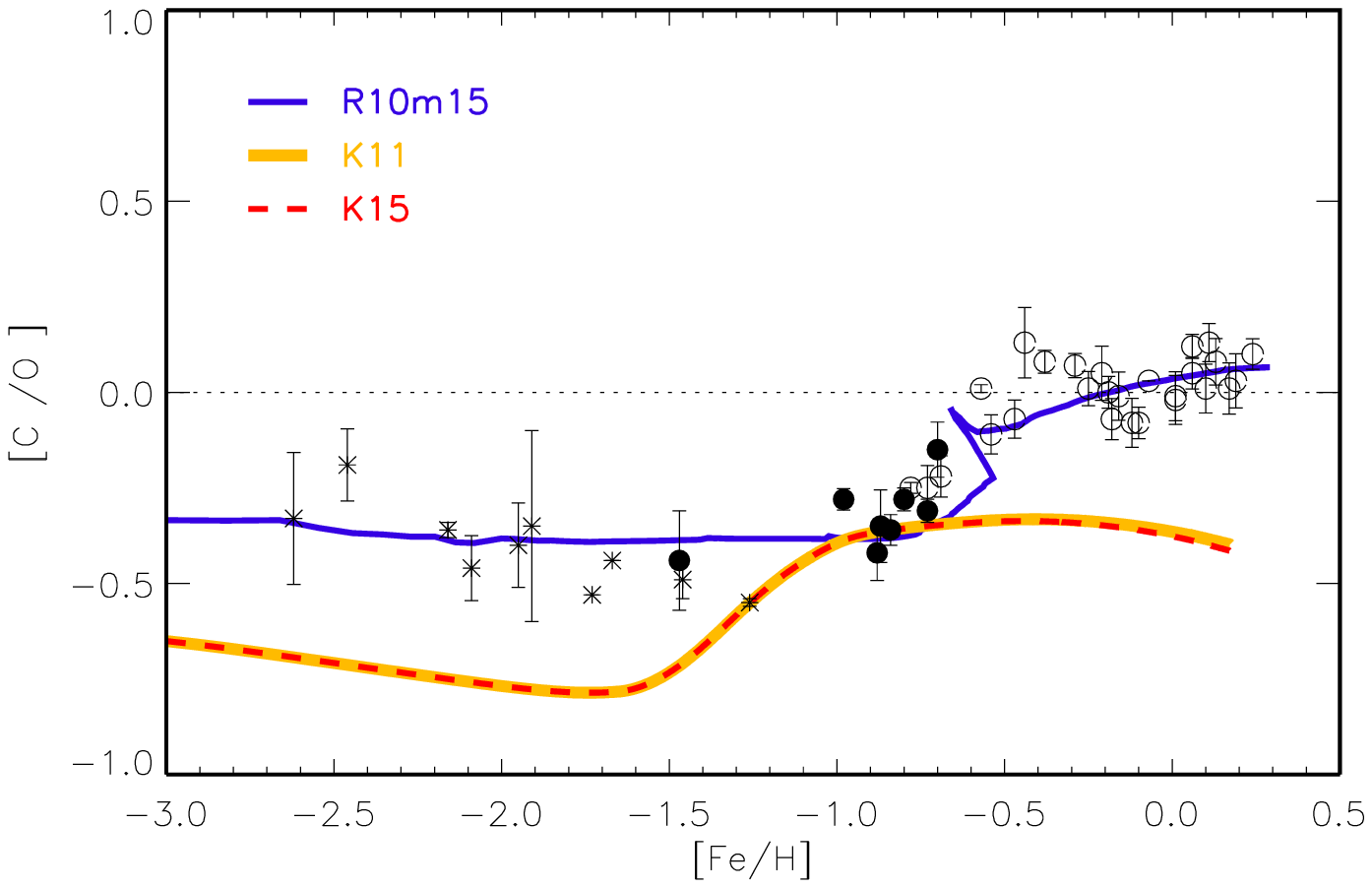}{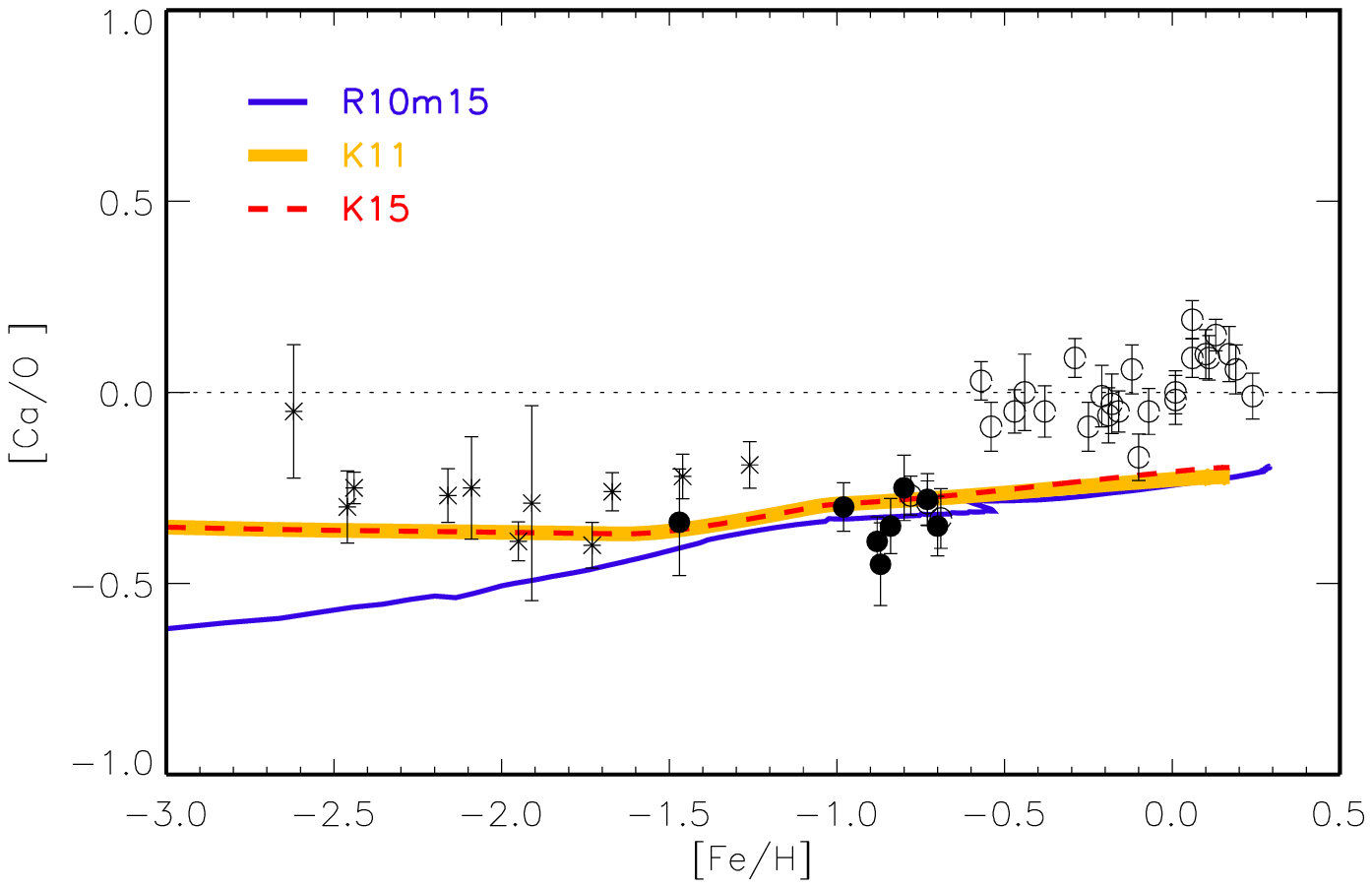}
\plottwo{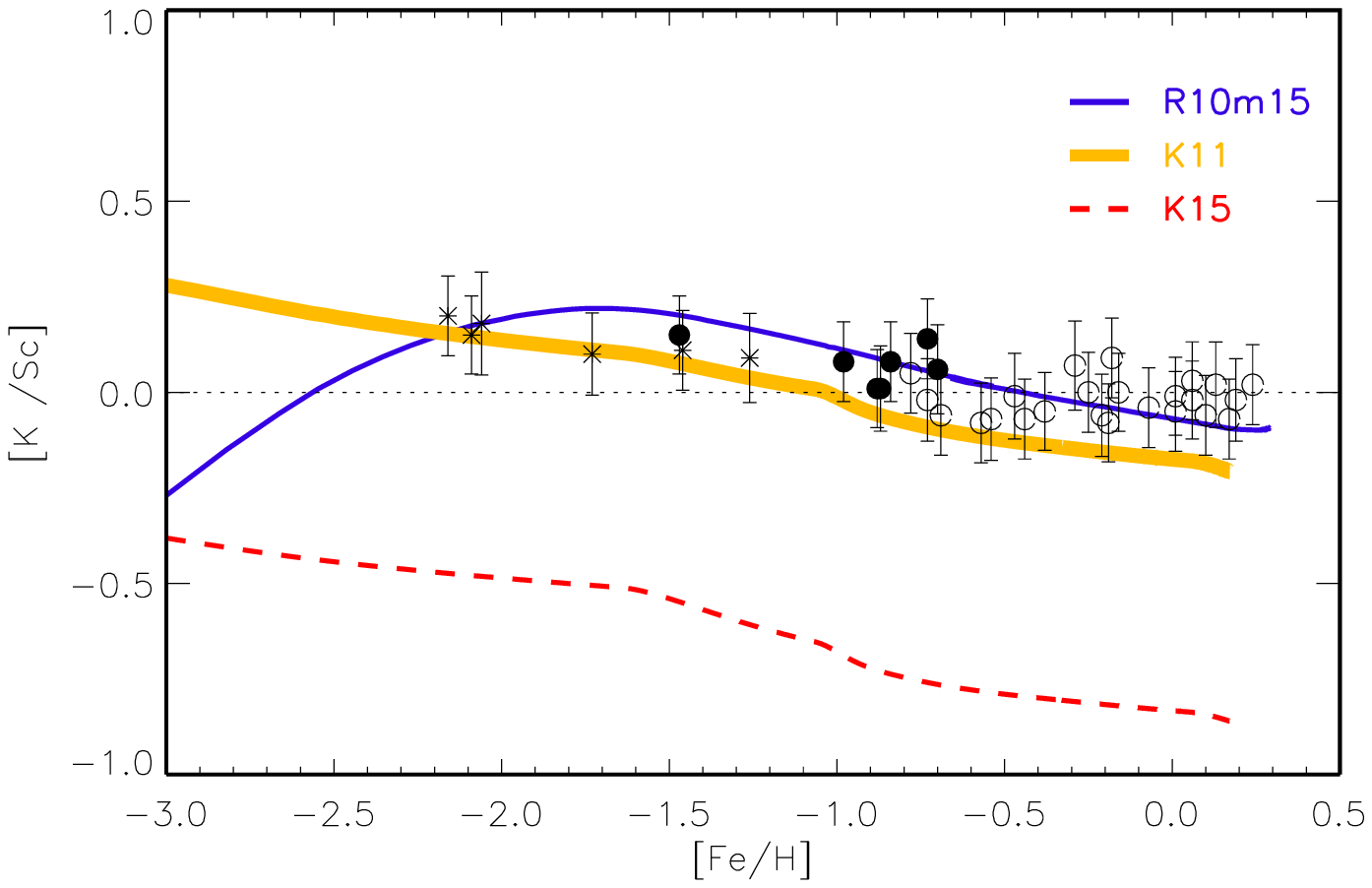}{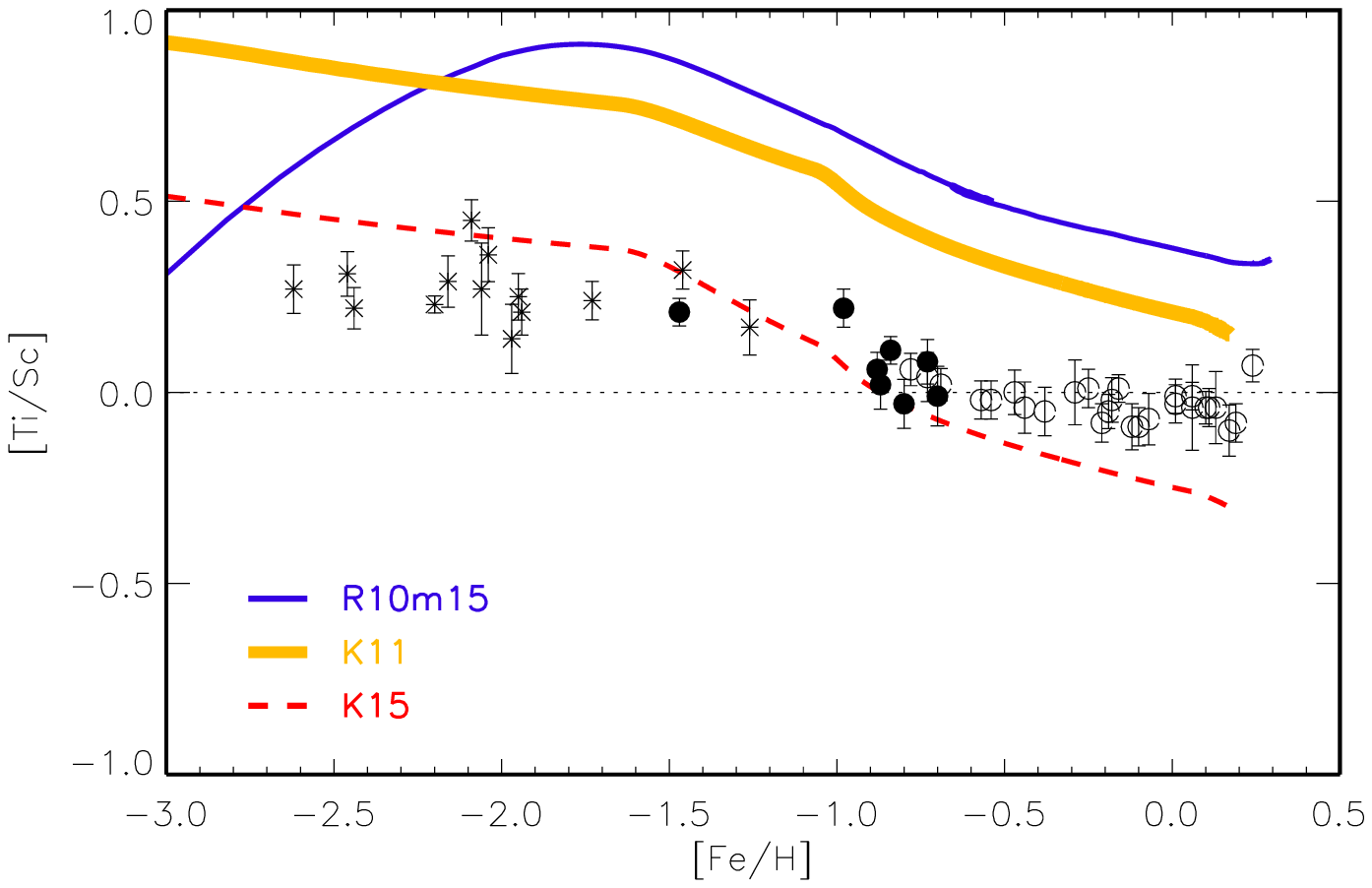}
\caption{Comparison with the Galactic chemical evolution models. The colors are as same as that in Fig. \ref{Fig:GCE1}.\label{Fig:GCE2}}
\end{figure*}

In this section, we compare our observational data with a series of GCE models from literature. We will mainly discuss the models of K11 \citep{2011MNRAS.414.3231K} and its updated version, K15. Their main features include slow infall, no outflow, star formation proportional to gas fraction, the Kroupa (2001) initial mass function (IMF) at $0.01-50M_\odot$, and the Type Ia Supernova (SNe Ia) model based on the single degenerate scenario \citep{2009ApJ...707.1466K} with the metallicity effect \citep{1998ApJ...503L.155K}. The metallicity-dependent nucleosynthesis yields  are taken from \citet[][hereafter K06]{2006ApJ...653.1145K} and \citet{2011MNRAS.414.3231K} for supernovae and hypernovae (with 0.5 fraction of hypernovae at $\ge 20M_\odot$), and from \citet{2010MNRAS.403.1413K} for asymptotic giant branch (AGB) stars, respectively. The yield sets are identical to those in \citet{2013ARA&A..51..457N}. Beside, in the K15 model, the effect of 2D jet-like explosions is applied \citep[see][for the details]{2016ApJ...817...53S}. These GCE results are consistent with the metallicity distribution function, the present star formation rate, and the present gas fraction.

Besides K11 and K15, we also compared our results with other widely used GCE models which are from \citet[][hereafter C97]{1997ApJ...477..765C}, \citet[][hereafter S98]{1998ApJ...496..155S}, \citet[][hereafter GP00]{2000A&A...359..191G}, \citet[][hereafter F04]{2004A&A...421..613F} and \citet[][hereafter R10]{2010A&A...522A..32R}, respectively. C97 presented the model which assumes two main infall episodes that formed the halo/thick-disk and the thin-disk, respectively. S98 developed a chemodynamical model of an isolated disk galaxy to be consistent with the observations deriving empirical yields. The model takes into account the galactic dynamical process for various kinds of stars and ISM. GP00 described an independently evolved halo+disk model, with short timescale outflows for halo and slow infall of the disk. F04 also presented a two-infall model that is similar to C97, but with empirical stellar yields. R10 tested 15 GCE models with various sets of stellar yields from literatures. Fig. \ref{Fig:GCE1} and Fig. \ref{Fig:GCE2} show the comparison between our observational data and the predictions from those GCE models. The offsets of solar abundances from different works have been corrected. 

{\it Carbon} -- The [C/Fe] ratio predicted by K11 and K15 shows a waved line, which slightly decreases from [Fe/H] $\sim -3$ to $-1.7$, due to the smaller envelope mass that contains C of massive progenitor stars. The rapid increase from [Fe/H] $\sim -1.5$ to $-1$ is caused by the delayed enrichment from AGB stars with $\sim 1-4M_\odot$. From [Fe/H] $\sim -1$, [C/Fe] decreases due to the delayed enrichment of SNe Ia. Although the lifetimes of these AGB stars ($0.15-0.2$ Gyr) are comparable to the shortest lifetimes of SNe Ia, the SN Ia contribution appears after the AGB contribution because of the metallicity dependence of the SN Ia lifetimes \citep{1998ApJ...503L.155K,2009ApJ...707.1466K}. These trends are characteristic with K11 and K15 models, and are in excellent agreement with our observational data at [Fe/H] $\gtsim -1.5$. At [Fe/H] $\ltsim -1.5$, the models predict $\sim$ 0.1$-$0.2 dex lower [C/Fe]. This is due to the input C yields, and this amount of offset can easily be solved with normal rotation of stars or the convective mixing of hydrogen into the He-burning layer (without rotation). The waved line can be also seen in the model 15 of R10.

{\it Oxygen} -- The observed [O/Fe] trend is in good agreement with the K11 and K15 models, where the plateau at [O/Fe] $\sim 0.6$ is caused by core-collapse supernovae, while the decreasing trend from [Fe/H] $\sim -1$ is caused by SNe Ia. This trend should exist for all $\alpha$ elements, i.e., O, Mg, Si, S, and Ca. It is very important that this evolutionary change appears sharply, and there is no stars with [O/Fe] $\ltsim 0.5$ at [Fe/H] $= -1$. As in Figure 6 of K11, without the metallicity effect of SNe Ia, the evolutionary change occurs much more gradually, being inconsistent with our new observational data. Other models did not show the sharply changing at [Fe/H]$= -1$.

{\it Sodium} -- In the K11 and K15 models, Na production highly depends on the metallicity of progenitors of core-collapse supernovae, which causes the increasing trend from [Fe/H] $\sim -3$ to $\sim -1.5$. This agrees with the observational data very well. From [Fe/H] $\sim -1.5$, [Na/Fe] ratios increase quickly due to AGB stars. Note that with the updated reaction rates, the Na yields of AGB stars have been reduced. However, Na is still over-produced by AGB stars. S98 may predict a better trend, but it uses empirical yields that are determined from the observations.

{\it Magnesium} -- The observed [Mg/Fe] ratios show the same trend as [O/Fe], but there is a $\sim 0.25$ dex offset between the observations and K11/K15 models. This means that [O/Mg] is not zero at a wide range of metallicity. This could be partially solved with the mass dependence of core-collapse supernovae, where [O/Mg] is slightly higher for more massive supernovae ($\gtsim 30M_\odot$, see Fig 1-4 of K06). This could also be solved by uncertain reactions rates in the hydrostatic burning of progenitor stars, as shown in Figure 9 of K06. Note that O and Mg are synthesized roughly the same region of supernova ejecta, and hence [O/Mg] should not depend on the parameters of supernova explosions very much. The Mg production of AGB stars is negligible in GCE models (except for the isotopic ratios, K11).

{\it Aluminum} -- Similar to Na, the trend predicted by K11 and K15 is consistent with the observational data, but similar to Mg, there is a $\sim 0.25$ dex offset. These could be due to the reaction rates, the rotational/convective mixing, or the combination of both. The trend predicted by the model 15 of R10 is in good agreement with the observation.

{\it Silicon} -- Similar to Mg, the observed trend is well reproduced with the K11 and K15 models, but the model is $\sim 0.2-0.3$ dex higher than observed. Si yields depends on the progenitor mass and the explosion energy of core-collapse supernovae, so some combination of these could reduce [Si/O] ratios. Note that if there is pre-enrichment from pair-instability supernovae, which should occur if $140-300M_\odot$ stars exist, the [Si/Fe] ratios become much higher, being inconsistent with the observations \citep[e.g.,][]{2004A&A...416.1117C}.

{\it Potassium} -- The underproduction of K in K11 and K15 is, at least partially, due to the lack of the neutrino process \citep{2011ApJ...739L..57K}. This element has not been well studied before because of the uncertainty of the NLTE effect, but our observational data can give strong constraints on supernova nucleosynthesis. Similar underproduction of K is seen in the model 15 of R10.

{\it Calcium} -- Similar to [O/Fe], the K11 and K15 models excellently reproduces the observed [Ca/Fe] ratios, and the sharp evolutionary change at [Fe/H] $=-1$ strongly supports the metallicity effect of SNe Ia.

{\it Scandium} -- In K15 the Sc abundance can be increased by the 2D jet-like explosions \citep{2003ApJ...598.1163M}, as shown in Fig. \ref{Fig:GCE2} by dashed line. This could also be enhanced by the neutrino process as for K. It worth noting, all the models except S98 failed to reproduce the observed [Sc/Fe] trend in the whole metallicity range.

{\it Titanium} -- This is the long-standing Ti problem in GCE, where the predicted [Ti/Fe] ratios are much lower than observed. The 2D jet effect should increase Ti abundances (dashed lines in Fig. \ref{Fig:GCE2}), but may not enough to solve this problem. Ti is produced in almost the same region as Fe in supernova ejecta, and hence it is categorized not $\alpha$ element but an iron-peak element. Nucleosynthsis with multidimensional explosions is necessary to understand the Ti production.

{\it Copper} -- The trends predicted by K11 and K15 models are similar with those of observed, although the model is a bit higher at [Fe/H] $> -1.0$. This is caused by the star formation rates connected with IMF. [Cu/Fe] increases toward higher metallicity because Cu is an odd-Z element, the production of which depends on the progenitor metallicity. This agreement suggests that the main producer of Cu is core-collapse supernovae, not the weak slow neutron-capture process suggested by \citet{2010ApJ...710.1557P}. 

Without normalized with respect to Fe, it may be possible to constrain uncertain processes that are important for some specific elements. [Na/Mg] ratio in \ref{Fig:GCE2} implies that the metallicity dependence may be smaller than those in K11 and K15. The [C/O] ratio may suggest that the mixing and/or rotation may be more important than those in K11 and K15. Note that in the model 15 of R10, the C yields in stellar winds are added, but a part of which have already been included in supernova yields, so the high [C/O] ratio should be due to the double-count of C production. In K15, the [Ca/O] ratios are consistent with the observed ones at [Fe/H] $\ltsim -0.5$, but are lower at higher metallicity, which may be due to the contribution to observed Ca from SNe Ia. [K/Sc] and [Ti/Sc] are in particular interesting since [(K,Sc,Ti)/Fe] is underabundant in K11. The low [K/Sc] may suggest the importance of $\nu$ process, and the [Ti/Sc] support the 2D effect applied in K15 to some extent. These figures should be used to test the next generation of nucleosynthesis yields with multi-dimensional calculations.

\section{Conclusions}\label{sect:conclusions}

Using accurate atmospheric parameters determined in Paper~I and high-resolution (R $\simeq$ 60\,000) stellar spectra observed for our project with the Shane/Hamilton spectrograph (the Lick observatory) and also taken from the archives, we calculated the NLTE abundances for 17 elements in a sample of stars uniformly distributed over the $-2.62 \le$ [Fe/H] $\le +0.24$ metallicity range. The star sample has been kinematically selected to trace the Galactic thin and thick disks and halo. This is the first extensive NLTE study of the stellar sample suitable for the Galactic chemical evolution research.

We derive differential abundances relative to the Sun, and such an approach largely cancels the difference between NLTE and LTE for [C/H], [Na/H], [Ca/H], and [Ba/H] for the [Fe/H] $> -1.0$ stars. However, notable ($> 0.1$~dex) differences in the same stars were found for [O/H], [K/H], and [Eu/Ba]. An advantage of NLTE is proved by the smaller line-to-line scatter obtained for most species in most stars in NLTE compared with that for LTE. The (NLTE - LTE) abundance differences grow towards lower metallicity, and NLTE is essential for stars more metal-poor than [Fe/H] = $-1$, in particular, for elemental ratios involving the species with (NLTE - LTE) of different sign, like [Na/Mg], [Na/Al], [Na/Cu], [Sr/Ba].

In line with the earlier studies, we obtained that the halo dwarf stars, which are expected to keep the pristine Li abundance, reveal a clear temperature dependence of their Li abundance. In our warm ($\Teff \ge 5800$~K) stars, the mean is $\eps{Li}$  = 2.2, which is consistent with $\eps{Li} = 2.177\pm0.071$ deduced by \citet{2005A&A...442..961C} for the [Fe/H] $\le -1.5$ stars with $\Teff \ge  5700$~K. Further theoretical studies of stellar physics and evolution are needed to understand source(s) of discrepancy with the standard Big Bang Nucleosynthesis that predicts $\eps{Li}$ = 2.72$\pm$0.06 \citep{2012ApJ...744..158C} and discrepancy between the warm and cool ($\Teff < 5800$~K) halo dwarf stars.

Most Galactic abundance trends obtained for elements of stellar origin have a rather small scatter of data for stars of close metallicity. It was found that the element-to-iron ratios reveal a common behavior for O, Mg, Si, Ca, and Ti, with a MP plateau, the knee at [Fe/H] $\simeq -0.8$, and a downward trend for higher metallicity. In the halo and thick disk stars [O/Fe] = 0.61 and a 0.3~dex lower [X/Fe] ratio was obtained for Mg, Si, Ca, and Ti. An upward trend of [C/Fe] with decreasing metallicity is observed in the thin disk stars and a very similar value of [C/Fe] = 0.21 in the thick disk stars. In contrast to [C/Fe] that reveals a substantial scatter in the halo stars, a well defined trend was obtained for the C/O ratios, with the upturn at [Fe/H] $\simeq -1.2$, in line with the earlier finding of \citet{2004A&A...414..931A}. We obtained no systematic shift between the NLTE abundances from lines of \ion{C}{1} and the CH-based abundances over the entire metallicity range under investigation.

A rather large scatter of data is observed when comparing abundances of the odd-$Z$ elements Na, K, and Sc with iron, however, it is largely removed in the ratios between these elements, like K/Sc, and in the ratios between nearby odd-$Z$ and even-$Z$ elements, like Na/Mg and Sc/Ti, suggesting a common production site for Na to Ti. We find a nearly constant underabundance of Na relative to Mg in the [Fe/H] $< -1$ stars, with the mean [Na/Mg] $\simeq -0.5$.

The light neutron-capture elements Sr and Zr reveal a different behavior, namely Sr follows the Fe abundance down to [Fe/H] $\simeq -2.5$, but Zr is enhanced relative to Fe and Sr in the MP stars. The [Zr/Sr] ratio is close to the solar value in the thin disk stars, grows to 0.4 at [Fe/H] = -2, and approaches to 0.8 at [Fe/H] = $-2.5$. In line with the earlier studies, the upward trend in [Eu/Fe] exists for [Fe/H] $> -1$, and europium is enhanced relative to Fe by more than 0.3~dex in the halo stars. A plateau of [Eu/Ba] at 0.50 is formed by the halo and thick disk stars, the knee occurs at [Fe/H] $\simeq -0.8$, and the downward trend in [Eu/Ba] is observed for higher metallicities.

The use of the NLTE element abundances raises credit to the interpretation of the data in the context of the chemical evolution of the Galaxy. Although GCE models are not calibrated with our NLTE abundances in this paper, K15 model predictions are in good agreement for C, O, Ca, and Fe in some metallicity coverages and the overall shapes. The underproduction of K, Sc, and Ti is somewhat known, and is due to the lack of $\nu$ processes \citep{2016ApJ...817...53S}. The offsets in odd-Z elements (i.e., Na, Al, Cu) give important constraints on the uncertain processes such as mixing. Despite the agreement for O, the offsets for Mg may be the most problematic since both elements have formed in relatively robust stellar evolution phase. If [Mg/Fe] is as low as in our NLTE analysis, that requires a different C/O ratio due to the mixing, mass-loss, and/or reaction rates in the progenitor stars, which should be studied in future works.

\begin{acknowledgements}
The authors thank Klaus Fuhrmann and Thomas Gehren for providing the FOCES spectra at our disposal, Oleg Kochukhov and Vadim Tsymbal for providing the codes {\sc binmag3} and {\sc synthV-NLTE}, Donatella Romano for providing the data of GCE models.
This study was supported by the Russian Foundation for Basic Research (grant 14-02-91153 and 16-32-00695), the National Natural Science Foundation of China (grants 11390371, 11233004, 11222326, 11103034, 11473033, 11473001), the National Basic Research Program of China (grant 2014CB845701/02/03), and the Swiss National Science Foundation (SCOPES project No. IZ73Z0-152485).
We made use the Simbad, MARCS, and VALD databases.
\end{acknowledgements}

\clearpage

\begin{deluxetable*}{lrrlclcc|lrrlclcc}
\tabletypesize{\scriptsize}
\tablecaption{Line data, references to their sources and the obtained solar element LTE and NLTE abundances, $\eps{\odot}$. \label{Tab:lines}}
\tablewidth{\textwidth}
\setlength{\tabcolsep}{4pt}
\renewcommand\arraystretch{1.0}
\tablehead{
\colhead{$\lambda$, \AA}  & \colhead{$\eexc$} &  \colhead{log $gf$} & Ref. & \colhead{log $C_6$}& Ref. & \multicolumn{2}{c}{$\eps{\odot}$}
& \colhead{$\lambda$, \AA}  & \colhead{$\eexc$} &  \colhead{log $gf$} & Ref. & \colhead{log $C_6$}& Ref. & \multicolumn{2}{c}{$\eps{\odot}$} \\
 \colhead{ } & \colhead{(eV)} &  \colhead{ } & \colhead{ } & & & \colhead{LTE}& \colhead{NLTE}
& \colhead{ } & \colhead{(eV)} &  \colhead{ } & \colhead{ } & & & \colhead{LTE}& \colhead{NLTE}
}
\startdata
\ion{Li}{1}   & & & & & & &                                                     & \ion{Ca}{1}  & & & & & & &							  \\
 6707.80$^1$ &   0.00 &   0.17 & NIST   &  -31.22 & ABO    &        &    &  5512.98 &	2.93 &  -0.46 &  S88   &  -30.61 &  S81   &   6.45 &   6.42 \\
             &        &        &        &         &        &        &    &  5588.75 &	2.53 &   0.36 &  SR81  &  -31.39 &  S81   &   6.35 &   6.29 \\
\ion{C}{1}   & & & & & & &                                                      &  5590.12 &	2.51 &  -0.57 &  SR81  &  -31.39 &  S81   &   6.45 &   6.37 \\
 4932.05 &   7.68 &  -1.66 & NIST   &  -30.50 &  K12   &   8.48 &   8.45 &  5857.45 &	2.93 &   0.24 &  SR81  &  -30.61 &  S81   &   6.42 &   6.38 \\
 5052.14 &   7.68 &  -1.30 & NIST   &  -30.60 &  K12   &   8.49 &   8.46 &  5867.57 &	2.93 &  -1.57 &  S88   &  -30.97 &  K07   &   6.36 &   6.38 \\
 5380.33 &   7.68 &  -1.62 & NIST   &  -30.75 &  K12   &   8.46 &   8.43 &  6161.29 &	2.51 &  -1.27 &  SR81  &  -30.48 &  S81   &   6.36 &   6.36 \\
 6587.61 &   8.54 &  -1.00 & NIST   &  -30.47 &  K12   &   8.45 &   8.43 &  6162.17 &	1.90 &  -0.09 &  SN75  &  -30.30 &  ABO   &   6.29 &   6.26 \\
\enddata
\tablecomments{ABO   = \citet{BPM}, BG96  = \citet{Baumueller_al1}, B00 = Keith Butler (private communication), B75   = \citet{1975JQSRT..15..463B}, BM98 = \citet{1998MNRAS.300..863B}, GLS   = \citet{mg_c6}, K07, K12 = R. Kurucz's website {\tt http://kurucz.harvard.edu/atoms.html}, LD89 = \citet{1989JOSAB...6.1457L}, LNA = \citet{LNAJ}, LWD = \citet{Lawler_Eu}, MZG   = \citet{Mashonkina2008}, NIST = \citet{NIST08}, RCW   = \citet{1980wtpa.book.....R}, S81 = \citet{1981A\string&A...103..351S}, S88 = \citet{1988JPhB...21.2827S}, SGM   = \citet{2009A\string&A...503..533S}, SN75 = \citet{1975A\string&A....38....1S}, SR81 = \citet{1981JPhB...14.4015S}, Sun = the solar line profile fits,  WLS   = \citet{2013_gf_ti2}, ZGZ   = \citet{Zhang2008_sc}; $^1$IS, \citet{1995PhRvA..52.2682S}; $^2$HFS, ZGZ; $^3$HFS, \citet{cu1_nlte_shi}; $^4$HFS, \citet{1983HyInt..15..177B}; $^5$HFS, K07; $^6$HFS, LWD.}
\end{deluxetable*}
\null

\clearpage
\begin{deluxetable*}{rlrrrrrcrlrrrrr}
\tabletypesize{\scriptsize}
\tablecaption{Summary of the obtained stellar abundances. \label{Tab:AbundanceSummary}}
\tablewidth{0pt}
\tablehead{
\colhead{Z}  & \colhead{Species} &  \colhead{N} & \multicolumn{2}{c}{LTE} & \multicolumn{2}{c}{NLTE} & \ \ & \colhead{Z}  & \colhead{Species} &  \colhead{N} & \multicolumn{2}{c}{LTE} & \multicolumn{2}{c}{NLTE} \\
 \colhead{ } & \colhead{ } &  \colhead{ } & \colhead{[X/H]} & \colhead{[X/Fe]}& \colhead{[X/H]} & \colhead{[X/Fe]} & & \colhead{ } & \colhead{ } &  \colhead{ } & \colhead{[X/H]} & \colhead{[X/Fe]}& \colhead{[X/H]} & \colhead{[X/Fe]}
}
\startdata
 \multicolumn{7}{c}{\underline{HD  19373 \ \ \ 6045/4.24/ 0.10, $\xi_t$ = 1.2      \ \ \ Thin disk} } & & \multicolumn{7}{c}{\underline{HD  22484 \ \ \ 6000/4.07/ 0.01, $\xi_t$ = 1.1	\ \ \ Thin disk} } \\
\noalign{\smallskip}
  3 & Li I  &    1 &    2.50          &         &    2.47          &         & & 	          3 & Li I  &    1 &    2.37          &	      &    2.33          &         \\
  6 & C I   &    5 &   -0.00$\pm$0.04 &   -0.10 &   -0.05$\pm$0.05 &   -0.15 & &  	        6 & C I   &    5 &   -0.07$\pm$0.04 &	-0.08 &   -0.09$\pm$0.04 &   -0.10 \\
  - & CH    &   13 &    0.10$\pm$0.06 &    0.00 &          \nodata & \nodata & &  	        - & CH    &   13 &    0.07$\pm$0.03 &	 0.06 &          \nodata & \nodata \\
  8 & O I   &    4 &    0.01$\pm$0.01 &   -0.09 &   -0.06$\pm$0.04 &   -0.16 & &  	        8 & O I   &    4 &    0.01$\pm$0.06 &	 0.00 &   -0.07$\pm$0.05 &   -0.08 \\
 11 & Na I  &    4 &    0.27$\pm$0.06 &    0.17 &    0.21$\pm$0.04 &    0.11 & &  	       11 & Na I  &    4 &    0.03$\pm$0.09 &	 0.02 &   -0.01$\pm$0.07 &   -0.02 \\
 12 & Mg I  &    5 &    0.12$\pm$0.02 &    0.02 &    0.11$\pm$0.02 &    0.01 & &  	       12 & Mg I  &    5 &    0.00$\pm$0.04 &	-0.01 &    0.00$\pm$0.03 &   -0.01 \\
 13 & Al I  &    2 &    0.08$\pm$0.02 &   -0.02 &    0.10$\pm$0.01 &    0.00 & &  	       13 & Al I  &    2 &   -0.13$\pm$0.04 &	-0.14 &   -0.11$\pm$0.03 &   -0.12 \\
\enddata
\end{deluxetable*}

\bibliography{references}

\begin{thebibliography}{}
\expandafter\ifx\csname natexlab\endcsname\relax\def\natexlab#1{#1}\fi

\bibitem[{{Adibekyan} {et~al.}(2012){Adibekyan}, {Sousa}, {Santos}, {Delgado
  Mena}, {Gonz{\'a}lez Hern{\'a}ndez}, {Israelian}, {Mayor}, \&
  {Khachatryan}}]{2012A&A...545A..32A}
{Adibekyan}, V.~Z., {Sousa}, S.~G., {Santos}, N.~C., {et~al.} 2012, \aap, 545,
  A32

\bibitem[{{Akerman} {et~al.}(2004){Akerman}, {Carigi}, {Nissen}, {Pettini}, \&
  {Asplund}}]{2004A&A...414..931A}
{Akerman}, C.~J., {Carigi}, L., {Nissen}, P.~E., {Pettini}, M., \& {Asplund},
  M. 2004, \aap, 414, 931

\bibitem[{{Alexeeva} \& {Mashonkina}(2015)}]{2015MNRAS.453.1619A}
{Alexeeva}, S.~A., \& {Mashonkina}, L.~I. 2015, \mnras, 453, 1619

\bibitem[{{Allende Prieto} {et~al.}(2004){Allende Prieto}, {Barklem},
  {Lambert}, \& {Cunha}}]{2004A&A...420..183A}
{Allende Prieto}, C., {Barklem}, P.~S., {Lambert}, D.~L., \& {Cunha}, K. 2004,
  \aap, 420, 183

\bibitem[{{Amarsi} {et~al.}(2015){Amarsi}, {Asplund}, {Collet}, \&
  {Leenaarts}}]{2015MNRAS.454L..11A}
{Amarsi}, A.~M., {Asplund}, M., {Collet}, R., \& {Leenaarts}, J. 2015, \mnras,
  454, L11

\bibitem[{{Andrievsky} {et~al.}(2010){Andrievsky}, {Spite}, {Korotin}, {Spite},
  {Bonifacio}, {Cayrel}, {Fran{\c c}ois}, \& {Hill}}]{2010A&A...509A..88A}
{Andrievsky}, S.~M., {Spite}, M., {Korotin}, S.~A., {et~al.} 2010, \aap, 509,
  A88

\bibitem[{{Bagnulo} {et~al.}(2003){Bagnulo}, {Jehin}, {Ledoux}, {Cabanac},
  {Melo}, {Gilmozzi}, \& {ESO Paranal Science Operations
  Team}}]{2003Msngr.114...10B}
{Bagnulo}, S., {Jehin}, E., {Ledoux}, C., {et~al.} 2003, The Messenger, 114, 10

\bibitem[{{Barklem} {et~al.}(2012){Barklem}, {Belyaev}, {Spielfiedel},
  {Guitou}, \& {Feautrier}}]{mg_hyd2012}
{Barklem}, P.~S., {Belyaev}, A.~K., {Spielfiedel}, A., {Guitou}, M., \&
  {Feautrier}, N. 2012, \aap, 541, A80

\bibitem[{{Barklem} \& {O'Mara}(1998)}]{1998MNRAS.300..863B}
{Barklem}, P.~S., \& {O'Mara}, B.~J. 1998, \mnras, 300, 863

\bibitem[{{Barklem} {et~al.}(2000){Barklem}, {Piskunov}, \& {O'Mara}}]{BPM}
{Barklem}, P.~S., {Piskunov}, N., \& {O'Mara}, B.~J. 2000, Astron. and
  Astrophys. Suppl. Ser., 142, 467

\bibitem[{{Barklem} {et~al.}(2005){Barklem}, {Christlieb}, {Beers}, {Hill},
  {Bessell}, {Holmberg}, {Marsteller}, {Rossi}, {Zickgraf}, \&
  {Reimers}}]{HERESII}
{Barklem}, P.~S., {Christlieb}, N., {Beers}, T.~C., {et~al.} 2005, \aap, 439,
  129

\bibitem[{{Baumueller} \& {Gehren}(1996)}]{Baumueller_al1}
{Baumueller}, D., \& {Gehren}, T. 1996, \aap, 307, 961

\bibitem[{{Bautista} {et~al.}(2002){Bautista}, {Gull}, {Ishibashi}, {Hartman},
  \& {Davidson}}]{2002MNRAS.331..875B}
{Bautista}, M.~A., {Gull}, T.~R., {Ishibashi}, K., {Hartman}, H., \&
  {Davidson}, K. 2002, \mnras, 331, 875

\bibitem[{{Belyaev}(2013)}]{Belyaev2013_Al}
{Belyaev}, A.~K. 2013, \aap, 560, A60

\bibitem[{{Belyaev} \& {Barklem}(2003)}]{belyaev03_Li}
{Belyaev}, A.~K., \& {Barklem}, P.~S. 2003, \pra, 68, 062703

\bibitem[{{Belyaev} {et~al.}(2014){Belyaev}, {Yakovleva}, \&
  {Barklem}}]{Belyaev2014_Si}
{Belyaev}, A.~K., {Yakovleva}, S.~A., \& {Barklem}, P.~S. 2014, \aap, 572, A103

\bibitem[{{Bensby} {et~al.}(2005){Bensby}, {Feltzing}, {Lundstr{\"o}m}, \&
  {Ilyin}}]{Bensby2005A&A...433..185B}
{Bensby}, T., {Feltzing}, S., {Lundstr{\"o}m}, I., \& {Ilyin}, I. 2005, \aap,
  433, 185

\bibitem[{{Bensby} {et~al.}(2014){Bensby}, {Feltzing}, \&
  {Oey}}]{2014A&A...562A..71B}
{Bensby}, T., {Feltzing}, S., \& {Oey}, M.~S. 2014, \aap, 562, A71

\bibitem[{{Bergemann} \& {Cescutti}(2010)}]{2010A&A...522A...9B}
{Bergemann}, M., \& {Cescutti}, G. 2010, \aap, 522, A9

\bibitem[{{Bergemann} \& {Gehren}(2008)}]{2008A&A...492..823B}
{Bergemann}, M., \& {Gehren}, T. 2008, \aap, 492, 823

\bibitem[{{Bergemann} {et~al.}(2010){Bergemann}, {Pickering}, \&
  {Gehren}}]{2010MNRAS.401.1334B}
{Bergemann}, M., {Pickering}, J.~C., \& {Gehren}, T. 2010, \mnras, 401, 1334

\bibitem[{{Bielski}(1975)}]{1975JQSRT..15..463B}
{Bielski}, A. 1975, \jqsrt, 15, 463

\bibitem[{{Bisterzo} {et~al.}(2014){Bisterzo}, {Travaglio}, {Gallino},
  {Wiescher}, \& {K{\"a}ppeler}}]{2014arXiv1403.1764B}
{Bisterzo}, S., {Travaglio}, C., {Gallino}, R., {Wiescher}, M., \&
  {K{\"a}ppeler}, F. 2014, \apj, 787, 10

\bibitem[{{Borghs} {et~al.}(1983){Borghs}, {de Bisschop}, {van Hove}, \&
  {Silverans}}]{1983HyInt..15..177B}
{Borghs}, G., {de Bisschop}, P., {van Hove}, M., \& {Silverans}, R.~E. 1983,
  Hyperfine Interactions, 15, 177

\bibitem[{{Bruls} {et~al.}(1992){Bruls}, {Rutten}, \&
  {Shchukina}}]{1992A&A...265..237B}
{Bruls}, J.~H.~M.~J., {Rutten}, R.~J., \& {Shchukina}, N.~G. 1992, \aap, 265,
  237

\bibitem[{{Burris} {et~al.}(2000){Burris}, {Pilachowski}, {Armandroff},
  {Sneden}, {Cowan}, \& {Roe}}]{Burris2000ApJ...544..302B}
{Burris}, D.~L., {Pilachowski}, C.~A., {Armandroff}, T.~E., {et~al.} 2000,
  \apj, 544, 302

\bibitem[{{Butler} \& {Giddings}(1985)}]{detail}
{Butler}, K., \& {Giddings}, J. 1985, Newsletter on the analysis of
  astronomical spectra, No. 9, University of London

\bibitem[{{Carbon} {et~al.}(1987){Carbon}, {Barbuy}, {Kraft}, {Friel}, \&
  {Suntzeff}}]{1987PASP...99..335C}
{Carbon}, D.~F., {Barbuy}, B., {Kraft}, R.~P., {Friel}, E.~D., \& {Suntzeff},
  N.~B. 1987, \pasp, 99, 335

\bibitem[{{Cayrel} {et~al.}(2004){Cayrel}, {Depagne}, {Spite}, {Hill}, {Spite},
  {Fran{\c c}ois}, {Plez}, {Beers}, {Primas}, {Andersen}, {Barbuy},
  {Bonifacio}, {Molaro}, \& {Nordstr{\"o}m}}]{2004A&A...416.1117C}
{Cayrel}, R., {Depagne}, E., {Spite}, M., {et~al.} 2004, \aap, 416, 1117

\bibitem[{{Charbonnel} \& {Primas}(2005)}]{2005A&A...442..961C}
{Charbonnel}, C., \& {Primas}, F. 2005, \aap, 442, 961

\bibitem[{{Chiappini} {et~al.}(1997){Chiappini}, {Matteucci}, \&
  {Gratton}}]{1997ApJ...477..765C}
{Chiappini}, C., {Matteucci}, F., \& {Gratton}, R. 1997, \apj, 477, 765, (C97)

\bibitem[{{Coc} {et~al.}(2012){Coc}, {Goriely}, {Xu}, {Saimpert}, \&
  {Vangioni}}]{2012ApJ...744..158C}
{Coc}, A., {Goriely}, S., {Xu}, Y., {Saimpert}, M., \& {Vangioni}, E. 2012,
  \apj, 744, 158

\bibitem[{{Edvardsson} {et~al.}(1993){Edvardsson}, {Andersen}, {Gustafsson},
  {Lambert}, {Nissen}, \& {Tomkin}}]{Edvardsson1993A&A...275..101E}
{Edvardsson}, B., {Andersen}, J., {Gustafsson}, B., {et~al.} 1993, \aap, 275,
  101

\bibitem[{{Fabbian} {et~al.}(2006){Fabbian}, {Asplund}, {Carlsson}, \&
  {Kiselman}}]{2006A&A...458..899F}
{Fabbian}, D., {Asplund}, M., {Carlsson}, M., \& {Kiselman}, D. 2006, \aap,
  458, 899, (F06)

\bibitem[{{Farouqi} {et~al.}(2010){Farouqi}, {Kratz}, {Pfeiffer}, {Rauscher},
  {Thielemann}, \& {Truran}}]{Farouqi2010}
{Farouqi}, K., {Kratz}, K.-L., {Pfeiffer}, B., {et~al.} 2010, \apj, 712, 1359

\bibitem[{{Fran{\c c}ois} {et~al.}(2004){Fran{\c c}ois}, {Matteucci}, {Cayrel},
  {Spite}, {Spite}, \& {Chiappini}}]{2004A&A...421..613F}
{Fran{\c c}ois}, P., {Matteucci}, F., {Cayrel}, R., {et~al.} 2004, \aap, 421,
  613, (F04)

\bibitem[{{Fu} {et~al.}(2015){Fu}, {Bressan}, {Molaro}, \&
  {Marigo}}]{2015MNRAS.452.3256F}
{Fu}, X., {Bressan}, A., {Molaro}, P., \& {Marigo}, P. 2015, \mnras, 452, 3256

\bibitem[{{Fuhrmann}(2008)}]{2008MNRAS.384..173F}
{Fuhrmann}, K. 2008, \mnras, 384, 173

\bibitem[{{Fuhrmann}(2011)}]{2011MNRAS.414.2893F}
---. 2011, \mnras, 414, 2893

\bibitem[{{Gallagher} {et~al.}(2016){Gallagher}, {Caffau}, {Bonifacio},
  {Ludwig}, {Steffen}, \& {Spite}}]{2016arXiv160507215G}
{Gallagher}, A.~J., {Caffau}, E., {Bonifacio}, P., {et~al.} 2016, ArXiv
  e-prints, arXiv:1605.07215

\bibitem[{{Gehren} {et~al.}(2004){Gehren}, {Liang}, {Shi}, {Zhang}, \&
  {Zhao}}]{mg_c6}
{Gehren}, T., {Liang}, Y.~C., {Shi}, J.~R., {Zhang}, H.~W., \& {Zhao}, G. 2004,
  \aap, 413, 1045

\bibitem[{{Gehren} {et~al.}(2006){Gehren}, {Shi}, {Zhang}, {Zhao}, \&
  {Korn}}]{2006A&A...451.1065G}
{Gehren}, T., {Shi}, J.~R., {Zhang}, H.~W., {Zhao}, G., \& {Korn}, A.~J. 2006,
  \aap, 451, 1065

\bibitem[{{Goswami} \& {Prantzos}(2000)}]{2000A&A...359..191G}
{Goswami}, A., \& {Prantzos}, N. 2000, \aap, 359, 191, (GP00)

\bibitem[{{Gray}(1977)}]{1977ApJ...218..530G}
{Gray}, D.~F. 1977, \apj, 218, 530

\bibitem[{Gustafsson {et~al.}(2008)Gustafsson, Edvardsson, Eriksson, Jorgensen,
  Nordlund, \& Plez}]{Gustafssonetal:2008}
Gustafsson, B., Edvardsson, B., Eriksson, K., {et~al.} 2008, A\&A, 486, 951

\bibitem[{{Ishigaki} {et~al.}(2013){Ishigaki}, {Aoki}, \&
  {Chiba}}]{2013ApJ...771...67I}
{Ishigaki}, M.~N., {Aoki}, W., \& {Chiba}, M. 2013, \apj, 771, 67

\bibitem[{{Karakas}(2010)}]{2010MNRAS.403.1413K}
{Karakas}, A.~I. 2010, \mnras, 403, 1413

\bibitem[{{Kobayashi} {et~al.}(2011{\natexlab{a}}){Kobayashi}, {Izutani},
  {Karakas}, {Yoshida}, {Yong}, \& {Umeda}}]{2011ApJ...739L..57K}
{Kobayashi}, C., {Izutani}, N., {Karakas}, A.~I., {et~al.} 2011{\natexlab{a}},
  \apjl, 739, L57

\bibitem[{{Kobayashi} {et~al.}(2011{\natexlab{b}}){Kobayashi}, {Karakas}, \&
  {Umeda}}]{2011MNRAS.414.3231K}
{Kobayashi}, C., {Karakas}, A.~I., \& {Umeda}, H. 2011{\natexlab{b}}, \mnras,
  414, 3231, (K11)

\bibitem[{{Kobayashi} \& {Nomoto}(2009)}]{2009ApJ...707.1466K}
{Kobayashi}, C., \& {Nomoto}, K. 2009, \apj, 707, 1466

\bibitem[{{Kobayashi} {et~al.}(1998){Kobayashi}, {Tsujimoto}, {Nomoto},
  {Hachisu}, \& {Kato}}]{1998ApJ...503L.155K}
{Kobayashi}, C., {Tsujimoto}, T., {Nomoto}, K., {Hachisu}, I., \& {Kato}, M.
  1998, \apjl, 503, L155

\bibitem[{{Kobayashi} {et~al.}(2006){Kobayashi}, {Umeda}, {Nomoto}, {Tominaga},
  \& {Ohkubo}}]{2006ApJ...653.1145K}
{Kobayashi}, C., {Umeda}, H., {Nomoto}, K., {Tominaga}, N., \& {Ohkubo}, T.
  2006, \apj, 653, 1145, (K06)

\bibitem[{{Korn} {et~al.}(2006){Korn}, {Grundahl}, {Richard}, {Barklem},
  {Mashonkina}, {Collet}, {Piskunov}, \& {Gustafsson}}]{2006Natur.442..657K}
{Korn}, A.~J., {Grundahl}, F., {Richard}, O., {et~al.} 2006, \nat, 442, 657

\bibitem[{{Kratz} {et~al.}(2007){Kratz}, {Farouqi}, {Pfeiffer}, {Truran},
  {Sneden}, \& {Cowan}}]{Kratz2007}
{Kratz}, K.-L., {Farouqi}, K., {Pfeiffer}, B., {et~al.} 2007, \apj, 662, 39

\bibitem[{{Kurucz} {et~al.}(1984){Kurucz}, {Furenlid}, {Brault}, \&
  {Testerman}}]{Atlas}
{Kurucz}, R.~L., {Furenlid}, I., {Brault}, J., \& {Testerman}, L. 1984, {Solar
  flux atlas from 296 to 1300 nm} (New Mexico: National Solar Observatory)

\bibitem[{{Lawler} \& {Dakin}(1989)}]{1989JOSAB...6.1457L}
{Lawler}, J.~E., \& {Dakin}, J.~T. 1989, Journal of the Optical Society of
  America B Optical Physics, 6, 1457

\bibitem[{{Lawler} {et~al.}(2001){Lawler}, {Wickliffe}, {den Hartog}, \&
  {Sneden}}]{Lawler_Eu}
{Lawler}, J.~E., {Wickliffe}, M.~E., {den Hartog}, E.~A., \& {Sneden}, C. 2001,
  \apj, 563, 1075

\bibitem[{{Lind} {et~al.}(2009){Lind}, {Asplund}, \&
  {Barklem}}]{2009A&A...503..541L}
{Lind}, K., {Asplund}, M., \& {Barklem}, P.~S. 2009, \aap, 503, 541

\bibitem[{{Ljung} {et~al.}(2006){Ljung}, {Nilsson}, {Asplund}, \&
  {Johansson}}]{LNAJ}
{Ljung}, G., {Nilsson}, H., {Asplund}, M., \& {Johansson}, S. 2006, \aap, 456,
  1181

\bibitem[{{Lodders} {et~al.}(2009){Lodders}, {Plame}, \& {Gail}}]{Lodders2009}
{Lodders}, K., {Plame}, H., \& {Gail}, H.-P. 2009, in Landolt-B{\"o}rnstein -
  Group VI Astronomy and Astrophysics Numerical Data and Functional
  Relationships in Science and Technology Volume 4B: Solar System. Edited by
  J.E. Tr{\"u}mper, 2009, 4.4., 44--54

\bibitem[{{Maeda} \& {Nomoto}(2003)}]{2003ApJ...598.1163M}
{Maeda}, K., \& {Nomoto}, K. 2003, \apj, 598, 1163

\bibitem[{{Mashonkina}(2013)}]{mash_mg13}
{Mashonkina}, L. 2013, \aap, 550, A28

\bibitem[{{Mashonkina} {et~al.}(2016){Mashonkina}, {Belyaev}, \&
  {Shi}}]{mash_al2016}
{Mashonkina}, L., {Belyaev}, A.~K., \& {Shi}, J.-R. 2016, Astronomy Letters,
  42, 366

\bibitem[{{Mashonkina} \& {Gehren}(2000)}]{mash_eu}
{Mashonkina}, L., \& {Gehren}, T. 2000, \aap, 364, 249

\bibitem[{{Mashonkina} \& {Gehren}(2001)}]{Mashonkina2001sr}
---. 2001, \aap, 376, 232

\bibitem[{{Mashonkina} {et~al.}(1999){Mashonkina}, {Gehren}, \&
  {Bikmaev}}]{Mashonkina1999}
{Mashonkina}, L., {Gehren}, T., \& {Bikmaev}, I. 1999, \aap, 343, 519

\bibitem[{{Mashonkina} {et~al.}(2011){Mashonkina}, {Gehren}, {Shi}, {Korn}, \&
  {Grupp}}]{mash_fe}
{Mashonkina}, L., {Gehren}, T., {Shi}, J.-R., {Korn}, A.~J., \& {Grupp}, F.
  2011, \aap, 528, A87

\bibitem[{{Mashonkina} {et~al.}(2003){Mashonkina}, {Gehren}, {Travaglio}, \&
  {Borkova}}]{2003A&A...397..275M}
{Mashonkina}, L., {Gehren}, T., {Travaglio}, C., \& {Borkova}, T. 2003, \aap,
  397, 275

\bibitem[{{Mashonkina} {et~al.}(2007{\natexlab{a}}){Mashonkina}, {Korn}, \&
  {Przybilla}}]{mash_ca}
{Mashonkina}, L., {Korn}, A.~J., \& {Przybilla}, N. 2007{\natexlab{a}}, \aap,
  461, 261

\bibitem[{{Mashonkina} \& {Zhao}(2006)}]{2006A&A...456..313M}
{Mashonkina}, L., \& {Zhao}, G. 2006, \aap, 456, 313

\bibitem[{{Mashonkina} {et~al.}(2008){Mashonkina}, {Zhao}, {Gehren}, {Aoki},
  {Bergemann}, {Noguchi}, {Shi}, {Takada-Hidai}, \& {Zhang}}]{Mashonkina2008}
{Mashonkina}, L., {Zhao}, G., {Gehren}, T., {et~al.} 2008, \aap, 478, 529

\bibitem[{{Mashonkina} {et~al.}(2007{\natexlab{b}}){Mashonkina}, {Vinogradova},
  {Ptitsyn}, {Khokhlova}, \& {Chernetsova}}]{Mashonkina2007}
{Mashonkina}, L.~I., {Vinogradova}, A.~B., {Ptitsyn}, D.~A., {Khokhlova},
  V.~S., \& {Chernetsova}, T.~A. 2007{\natexlab{b}}, Astronomy Reports, 51, 903

\bibitem[{{Matteucci} \& {Francois}(1989)}]{1989MNRAS.239..885M}
{Matteucci}, F., \& {Francois}, P. 1989, \mnras, 239, 885

\bibitem[{{McWilliam} {et~al.}(1995){McWilliam}, {Preston}, {Sneden}, \&
  {Shectman}}]{1995AJ....109.2736M}
{McWilliam}, A., {Preston}, G.~W., {Sneden}, C., \& {Shectman}, S. 1995, \aj,
  109, 2736

\bibitem[{{Mel{\'e}ndez} {et~al.}(2007){Mel{\'e}ndez}, {Bautista}, \&
  {Badnell}}]{ca2_bautista}
{Mel{\'e}ndez}, M., {Bautista}, M.~A., \& {Badnell}, N.~R. 2007, \aap, 469,
  1203

\bibitem[{{Mishenina} \& {Kovtyukh}(2001)}]{2001A&A...370..951M}
{Mishenina}, T.~V., \& {Kovtyukh}, V.~V. 2001, \aap, 370, 951

\bibitem[{{Mishenina} {et~al.}(2013){Mishenina}, {Pignatari}, {Korotin},
  {Soubiran}, {Charbonnel}, {Thielemann}, {Gorbaneva}, \&
  {Basak}}]{2013A&A...552A.128M}
{Mishenina}, T.~V., {Pignatari}, M., {Korotin}, S.~A., {et~al.} 2013, \aap,
  552, A128

\bibitem[{{Nissen} \& {Schuster}(1997)}]{1997A&A...326..751N}
{Nissen}, P.~E., \& {Schuster}, W.~J. 1997, \aap, 326, 751

\bibitem[{{Nissen} \& {Schuster}(2010)}]{2010A&A...511L..10N}
---. 2010, \aap, 511, L10

\bibitem[{{Nomoto} {et~al.}(2013){Nomoto}, {Kobayashi}, \&
  {Tominaga}}]{2013ARA&A..51..457N}
{Nomoto}, K., {Kobayashi}, C., \& {Tominaga}, N. 2013, \araa, 51, 457

\bibitem[{{Nomoto} {et~al.}(2006){Nomoto}, {Tominaga}, {Umeda}, {Kobayashi}, \&
  {Maeda}}]{2006NuPhA.777..424N}
{Nomoto}, K., {Tominaga}, N., {Umeda}, H., {Kobayashi}, C., \& {Maeda}, K.
  2006, Nuclear Physics A, 777, 424

\bibitem[{{Pakhomov}(2015)}]{2015ARep...59..952P}
{Pakhomov}, Y.~V. 2015, Astronomy Reports, 59, 952

\bibitem[{{Pakhomov} \& {Zhao}(2013)}]{2013AJ....146...97P}
{Pakhomov}, Y.~V., \& {Zhao}, G. 2013, \aj, 146, 97

\bibitem[{{Pignatari} {et~al.}(2010){Pignatari}, {Gallino}, {Heil}, {Wiescher},
  {K{\"a}ppeler}, {Herwig}, \& {Bisterzo}}]{2010ApJ...710.1557P}
{Pignatari}, M., {Gallino}, R., {Heil}, M., {et~al.} 2010, \apj, 710, 1557

\bibitem[{{Prochaska} {et~al.}(2000){Prochaska}, {Naumov}, {Carney},
  {McWilliam}, \& {Wolfe}}]{2000AJ....120.2513P}
{Prochaska}, J.~X., {Naumov}, S.~O., {Carney}, B.~W., {McWilliam}, A., \&
  {Wolfe}, A.~M. 2000, \aj, 120, 2513

\bibitem[{{Ralchenko} {et~al.}(2010){Ralchenko}, {Kramida}, {Reader}, \&
  Team}]{NIST08}
{Ralchenko}, Y.~A., {Kramida}, E., {Reader}, J., \& Team, N.~A. 2010, NIST
  Atomic Spectra Database (version 3.1.5) (USA)

\bibitem[{{Ram{\'{\i}}rez} {et~al.}(2012){Ram{\'{\i}}rez}, {Fish}, {Lambert},
  \& {Allende Prieto}}]{2012ApJ...756...46R}
{Ram{\'{\i}}rez}, I., {Fish}, J.~R., {Lambert}, D.~L., \& {Allende Prieto}, C.
  2012, \apj, 756, 46

\bibitem[{{Reader} {et~al.}(1980){Reader}, {Corliss}, {Wiese}, \&
  {Martin}}]{1980wtpa.book.....R}
{Reader}, J., {Corliss}, C.~H., {Wiese}, W.~L., \& {Martin}, G.~A. 1980,
  {Wavelengths and transition probabilities for atoms and atomic ions: Part 1.
  Wavelengths, part 2. Transition probabilities}

\bibitem[{{Reddy} {et~al.}(2003){Reddy}, {Tomkin}, {Lambert}, \& {Allende
  Prieto}}]{2003MNRAS.340..304R}
{Reddy}, B.~E., {Tomkin}, J., {Lambert}, D.~L., \& {Allende Prieto}, C. 2003,
  \mnras, 340, 304

\bibitem[{Reetz(1991)}]{Reetz}
Reetz, J.~K. 1991, Diploma Thesis (Universit\"at M\"unchen)

\bibitem[{{Romano} {et~al.}(2010){Romano}, {Karakas}, {Tosi}, \&
  {Matteucci}}]{2010A&A...522A..32R}
{Romano}, D., {Karakas}, A.~I., {Tosi}, M., \& {Matteucci}, F. 2010, \aap, 522,
  A32, (R10)

\bibitem[{{Ro{\v s}kar} {et~al.}(2008){Ro{\v s}kar}, {Debattista}, {Quinn},
  {Stinson}, \& {Wadsley}}]{2008ApJ...684L..79R}
{Ro{\v s}kar}, R., {Debattista}, V.~P., {Quinn}, T.~R., {Stinson}, G.~S., \&
  {Wadsley}, J. 2008, \apjl, 684, L79

\bibitem[{{Ryabchikova} {et~al.}(2015){Ryabchikova}, {Piskunov}, {Kurucz},
  {Stempels}, {Heiter}, {Pakhomov}, \& {Barklem}}]{2015PhyS...90e4005R}
{Ryabchikova}, T., {Piskunov}, N., {Kurucz}, R.~L., {et~al.} 2015, \physscr,
  90, 054005

\bibitem[{{Ryabchikova} {et~al.}(2016){Ryabchikova}, {Piskunov}, {Pakhomov},
  {Tsymbal}, {Titarenko}, {Sitnova}, {Alexeeva}, {Fossati}, \&
  {Mashonkina}}]{Ryabchikova2015}
{Ryabchikova}, T., {Piskunov}, N., {Pakhomov}, Y., {et~al.} 2016, \mnras, 456,
  1221

\bibitem[{{Rybicki} \& {Hummer}(1991)}]{rh91}
{Rybicki}, G.~B., \& {Hummer}, D.~G. 1991, \aap, 245, 171

\bibitem[{{Rybicki} \& {Hummer}(1992)}]{rh92}
---. 1992, \aap, 262, 209

\bibitem[{{Salvadori} {et~al.}(2007){Salvadori}, {Schneider}, \&
  {Ferrara}}]{2007MNRAS.381..647S}
{Salvadori}, S., {Schneider}, R., \& {Ferrara}, A. 2007, \mnras, 381, 647

\bibitem[{{Samland}(1998)}]{1998ApJ...496..155S}
{Samland}, M. 1998, \apj, 496, 155, (S98)

\bibitem[{{Sansonetti} {et~al.}(1995){Sansonetti}, {Richou}, {Engleman}, \&
  {Radziemski}}]{1995PhRvA..52.2682S}
{Sansonetti}, C.~J., {Richou}, B., {Engleman}, Jr., R., \& {Radziemski}, L.~J.
  1995, \pra, 52, 2682

\bibitem[{{Shi} {et~al.}(2008){Shi}, {Gehren}, {Butler}, {Mashonkina}, \&
  {Zhao}}]{Shi_si_sun}
{Shi}, J.~R., {Gehren}, T., {Butler}, K., {Mashonkina}, L.~I., \& {Zhao}, G.
  2008, \aap, 486, 303

\bibitem[{{Shi} {et~al.}(2009){Shi}, {Gehren}, {Mashonkina}, \&
  {Zhao}}]{2009A&A...503..533S}
{Shi}, J.~R., {Gehren}, T., {Mashonkina}, L., \& {Zhao}, G. 2009, \aap, 503,
  533

\bibitem[{{Shi} {et~al.}(2014){Shi}, {Gehren}, {Zeng}, {Mashonkina}, \&
  {Zhao}}]{cu1_nlte_shi}
{Shi}, J.~R., {Gehren}, T., {Zeng}, J.~L., {Mashonkina}, L., \& {Zhao}, G.
  2014, \apj, 782, 80

\bibitem[{{Shi} {et~al.}(2007){Shi}, {Gehren}, {Zhang}, {Zeng}, \&
  {Zhao}}]{2007A&A...465..587S}
{Shi}, J.~R., {Gehren}, T., {Zhang}, H.~W., {Zeng}, J.~L., \& {Zhao}, G. 2007,
  \aap, 465, 587

\bibitem[{{Siqueira-Mello} {et~al.}(2015){Siqueira-Mello}, {Andrievsky},
  {Barbuy}, {Spite}, {Spite}, \& {Korotin}}]{2015A&A...584A..86S}
{Siqueira-Mello}, C., {Andrievsky}, S.~M., {Barbuy}, B., {et~al.} 2015, \aap,
  584, A86

\bibitem[{{Sitnova} {et~al.}(2015){Sitnova}, {Zhao}, {Mashonkina}, {Chen},
  {Liu}, {Pakhomov}, {Tan}, {Bolte}, {Alexeeva}, {Grupp}, {Shi}, \&
  {Zhang}}]{2015ApJ...808..148S}
{Sitnova}, T., {Zhao}, G., {Mashonkina}, L., {et~al.} 2015, \apj, 808, 148

\bibitem[{{Sitnova} {et~al.}(2013){Sitnova}, {Mashonkina}, \&
  {Ryabchikova}}]{sitnova_o}
{Sitnova}, T.~M., {Mashonkina}, L.~I., \& {Ryabchikova}, T.~A. 2013, Astronomy
  Letters, 39, 126

\bibitem[{{Sitnova} {et~al.}(2016){Sitnova}, {Mashonkina}, \&
  {Ryabchikova}}]{sitnova_ti}
---. 2016, \mnras, 461, 1000

\bibitem[{{Smith}(1981)}]{1981A&A...103..351S}
{Smith}, G. 1981, \aap, 103, 351

\bibitem[{{Smith}(1988)}]{1988JPhB...21.2827S}
---. 1988, Journal of Physics B Atomic Molecular Physics, 21, 2827

\bibitem[{{Smith} \& {O'Neill}(1975)}]{1975A&A....38....1S}
{Smith}, G., \& {O'Neill}, J.~A. 1975, \aap, 38, 1

\bibitem[{{Smith} \& {Raggett}(1981)}]{1981JPhB...14.4015S}
{Smith}, G., \& {Raggett}, D.~S.~J. 1981, Journal of Physics B Atomic Molecular
  Physics, 14, 4015

\bibitem[{{Sneden} {et~al.}(2016){Sneden}, {Cowan}, {Kobayashi}, {Pignatari},
  {Lawler}, {Den Hartog}, \& {Wood}}]{2016ApJ...817...53S}
{Sneden}, C., {Cowan}, J.~J., {Kobayashi}, C., {et~al.} 2016, \apj, 817, 53

\bibitem[{{Spergel} {et~al.}(2007){Spergel}, {Bean}, {Dor{\'e}}, {Nolta},
  {Bennett}, {Dunkley}, {Hinshaw}, {Jarosik}, {Komatsu}, {Page}, {Peiris},
  {Verde}, {Halpern}, {Hill}, {Kogut}, {Limon}, {Meyer}, {Odegard}, {Tucker},
  {Weiland}, {Wollack}, \& {Wright}}]{2007ApJS..170..377S}
{Spergel}, D.~N., {Bean}, R., {Dor{\'e}}, O., {et~al.} 2007, \apjs, 170, 377

\bibitem[{{Spite} \& {Spite}(1978)}]{1978A&A....67...23S}
{Spite}, M., \& {Spite}, F. 1978, \aap, 67, 23

\bibitem[{{Spite} \& {Spite}(1982)}]{1982Natur.297..483S}
---. 1982, \nat, 297, 483

\bibitem[{{Spite} {et~al.}(2011){Spite}, {Caffau}, {Andrievsky}, {Korotin},
  {Depagne}, {Spite}, {Bonifacio}, {Ludwig}, {Cayrel}, {Fran{\c c}ois}, {Hill},
  {Plez}, {Andersen}, {Barbuy}, {Beers}, {Molaro}, {Nordstr{\"o}m}, \&
  {Primas}}]{2011A&A...528A...9S}
{Spite}, M., {Caffau}, E., {Andrievsky}, S.~M., {et~al.} 2011, \aap, 528, A9

\bibitem[{{Spite} {et~al.}(2012){Spite}, {Andrievsky}, {Spite}, {Caffau},
  {Korotin}, {Bonifacio}, {Ludwig}, {Fran{\c c}ois}, \&
  {Cayrel}}]{2012A&A...541A.143S}
{Spite}, M., {Andrievsky}, S.~M., {Spite}, F., {et~al.} 2012, \aap, 541, A143

\bibitem[{{Steenbock} \& {Holweger}(1984)}]{Steenbock1984}
{Steenbock}, W., \& {Holweger}, H. 1984, \aap, 130, 319

\bibitem[{{Takeda} {et~al.}(2002){Takeda}, {Zhao}, {Chen}, {Qiu}, \&
  {Takada-Hidai}}]{2002PASJ...54..275T}
{Takeda}, Y., {Zhao}, G., {Chen}, Y.-Q., {Qiu}, H.-M., \& {Takada-Hidai}, M.
  2002, \pasj, 54, 275

\bibitem[{{Thielemann} {et~al.}(1996){Thielemann}, {Nomoto}, \&
  {Hashimoto}}]{1996ApJ...460..408T}
{Thielemann}, F.-K., {Nomoto}, K., \& {Hashimoto}, M.-A. 1996, \apj, 460, 408

\bibitem[{{Timmes} {et~al.}(1995){Timmes}, {Woosley}, \&
  {Weaver}}]{1995ApJS...98..617T}
{Timmes}, F.~X., {Woosley}, S.~E., \& {Weaver}, T.~A. 1995, \apjs, 98, 617

\bibitem[{{Travaglio} {et~al.}(1999){Travaglio}, {Galli}, {Gallino}, {Busso},
  {Ferrini}, \& {Straniero}}]{Travaglio1999}
{Travaglio}, C., {Galli}, D., {Gallino}, R., {et~al.} 1999, \apj, 521, 691

\bibitem[{{Travaglio} {et~al.}(2004){Travaglio}, {Gallino}, {Arnone}, {Cowan},
  {Jordan}, \& {Sneden}}]{2004ApJ...601..864T}
{Travaglio}, C., {Gallino}, R., {Arnone}, E., {et~al.} 2004, \apj, 601, 864

\bibitem[{{Velichko} {et~al.}(2010){Velichko}, {Mashonkina}, \&
  {Nilsson}}]{Velichko2010_zr}
{Velichko}, A.~B., {Mashonkina}, L.~I., \& {Nilsson}, H. 2010, Astronomy
  Letters, 36, 664

\bibitem[{{Wallerstein}(1962)}]{1962ApJS....6..407W}
{Wallerstein}, G. 1962, \apjs, 6, 407

\bibitem[{{Wood} {et~al.}(2013){Wood}, {Lawler}, {Sneden}, \&
  {Cowan}}]{2013_gf_ti2}
{Wood}, M.~P., {Lawler}, J.~E., {Sneden}, C., \& {Cowan}, J.~J. 2013, \apjs,
  208, 27

\bibitem[{{Woosley} \& {Weaver}(1995)}]{1995ApJS..101..181W}
{Woosley}, S.~E., \& {Weaver}, T.~A. 1995, \apjs, 101, 181

\bibitem[{{Yan} {et~al.}(2016){Yan}, {Shi}, {Nissen}, \&
  {Zhao}}]{2016A&A...585A.102Y}
{Yan}, H.~L., {Shi}, J.~R., {Nissen}, P.~E., \& {Zhao}, G. 2016, \aap, 585,
  A102

\bibitem[{{Yan} {et~al.}(2015){Yan}, {Shi}, \& {Zhao}}]{2015ApJ...802...36Y}
{Yan}, H.~L., {Shi}, J.~R., \& {Zhao}, G. 2015, \apj, 802, 36

\bibitem[{{Zhang} {et~al.}(2006{\natexlab{a}}){Zhang}, {Butler}, {Gehren},
  {Shi}, \& {Zhao}}]{2006A&A...453..723Z}
{Zhang}, H.~W., {Butler}, K., {Gehren}, T., {Shi}, J.~R., \& {Zhao}, G.
  2006{\natexlab{a}}, \aap, 453, 723

\bibitem[{{Zhang} {et~al.}(2006{\natexlab{b}}){Zhang}, {Gehren}, {Butler},
  {Shi}, \& {Zhao}}]{2006A&A...457..645Z}
{Zhang}, H.~W., {Gehren}, T., {Butler}, K., {Shi}, J.~R., \& {Zhao}, G.
  2006{\natexlab{b}}, \aap, 457, 645

\bibitem[{{Zhang} {et~al.}(2008){Zhang}, {Gehren}, \& {Zhao}}]{Zhang2008_sc}
{Zhang}, H.~W., {Gehren}, T., \& {Zhao}, G. 2008, \aap, 481, 489

\bibitem[{{Zhao} {et~al.}(1998){Zhao}, {Butler}, \&
  {Gehren}}]{1998A&A...333..219Z}
{Zhao}, G., {Butler}, K., \& {Gehren}, T. 1998, \aap, 333, 219

\bibitem[{{Zhao} \& {Gehren}(2000)}]{2000A&A...362.1077Z}
{Zhao}, G., \& {Gehren}, T. 2000, \aap, 362, 1077

\bibitem[{{Zhao} \& {Magain}(1990{\natexlab{a}})}]{1990A&A...238..242Z}
{Zhao}, G., \& {Magain}, P. 1990{\natexlab{a}}, \aap, 238, 242

\bibitem[{{Zhao} \& {Magain}(1990{\natexlab{b}})}]{1990A&AS...86...85Z}
---. 1990{\natexlab{b}}, \aaps, 86, 85

\end{thebibliography}
\bibliographystyle{aasjournal}

\end{document}